\documentclass[english,12pt]{article}

\usepackage[pdftex]{graphicx}
\usepackage{verbatim}
\usepackage[width=\textwidth]{caption}
\usepackage{amsmath,amssymb,bm}
\usepackage{a4wide}
\usepackage[]{float}
\usepackage[]{placeins}
\usepackage{flafter}
\usepackage[]{longtable}
\usepackage{csvsimple}
\usepackage{verbatim}
\usepackage{amsopn}
\usepackage{dsfont}
\usepackage{color}
\usepackage{pdflscape}
\usepackage{array}
\usepackage{natbib}
\usepackage{xr}
\usepackage{booktabs}
\usepackage{subcaption}
\usepackage{hyperref}
\makeatletter

\newtheorem{thm}{Theorem}
\newtheorem{cor}{Corollary}

\newtheorem{res}{Result}

\newtheorem{appxthm}{Theorem}[section]

\newtheorem{appxres}{Result}[section]
\newtheorem{appxlem}{Lemma}[section]
\newtheorem{appxhyp}{Assumption}[section]

\newtheorem{hyp}{Assumption}

\newcommand{\abs}[1]{\left\vert#1\right\vert}

\newcommand{\indep}{\perp \!\!\! \perp}
\newcommand{\convP}{\stackrel{\mathbb{P}}{\longrightarrow}}

\newcommand{\E}{\mathbb{E}}
\renewcommand{\section}{\@startsection{section}{2}{0mm}{-1.5\baselineskip}{1\baselineskip}{\normalfont\large\bfseries}}
\renewcommand{\subsection}{\@startsection{subsection}{2}{0mm}{-1.2\baselineskip}{1\baselineskip}{\normalfont\normalsize\bfseries}}
\renewcommand{\subsubsection}{\@startsection{subsubsection}{3}{0mm}{-0.8\baselineskip}{0.4\baselineskip}{\normalfont\normalsize\itshape}}

\makeatletter
\def\widebreve{\mathpalette\wide@breve}
\def\wide@breve#1#2{\sbox\z@{$#1#2$}%
     \mathop{\vbox{\m@th\ialign{##\crcr
\kern0.08em\brevefill#1{0.8\wd\z@}\crcr\noalign{\nointerlineskip}%
                    $\hss#1#2\hss$\crcr}}}\limits}
\def\brevefill#1#2{$\m@th\sbox\tw@{$#1($}%
  \hss\resizebox{#2}{\wd\tw@}{\rotatebox[origin=c]{90}{\upshape(}}\hss$}
\makeatletter

\date{\today}
\begin{document}

\title{A Framework for Using Value-Added in Regressions}

\author{Antoine Deeb%
\thanks{University of California at Santa Barbara, antoinedib@ucsb.edu. I thank Cl\'{e}ment de Chaisemartin, Allegra Cockburn, Jaime Ramirez-Cuellar, Peter Kuhn, Dick Startz, Doug Steigerwald, Gonzalo Vazquez-Bare, and members of the UCSB econometrics research group for helpful comments and suggestions.}%
} \maketitle ~\vspace{-1cm}
\maketitle
%

\begin{abstract}

As increasingly popular metrics of worker and institutional quality, estimated value-added (VA) measures are now widely used as dependent or explanatory variables in regressions. For example, VA is used as an explanatory variable when examining the relationship between teacher VA and students' long-run outcomes. Due to the multi-step nature of VA estimation, the standard errors (SEs) researchers routinely use when including VA measures in OLS regressions are incorrect. In this paper, I show that the assumptions underpinning VA models naturally lead to a generalized method of moments (GMM) framework.  Using this insight,  I construct correct SEs' for regressions that use VA as an explanatory variable and for regressions where VA is the outcome. In addition, I identify the causes of incorrect SEs when using OLS, discuss the need to adjust SEs under different sets of assumptions, and propose a more efficient estimator for using VA as an explanatory variable. Finally, I illustrate my results using data from North Carolina, and show that correcting SEs results in an increase that is larger than the impact of clustering SEs.

\end{abstract}
\thispagestyle{empty}
\clearpage
\setcounter{page}{1}
\section{Introduction}

\medskip

Assessing the quality of workers and institutions and the impact of this quality on  various outcomes is a common topic in economics. For instance, it is common to assess the impact of teachers' value added (VA) on long-run student outcomes, often by regressing those outcomes on the estimated VA measures \citep{chetty2014measuring,jackson2018test,canaan2019VA}.
Furthermore, some studies ask whether observable characteristics of teachers can predict VA and regress VA measures on those characteristics. In this paper, I start by showing that when VA is the outcome or the explanatory variable in a regression, the regression's robust standard errors researchers routinely use to draw inference are invalid. I then construct consistent standard errors for the estimators used in such studies. Finally, for models using VA as an explanatory variable I propose a more efficient estimator, construct optimal instruments under certain assumptions, and discuss a specification test. \medskip

For ease of exposition, I present my results  in the context of teacher test-score value added, estimated using leave-year out measures. However, the results can be extended to  different VA measures, and to different methods of estimating VA. While I consider the case of linear regressions, the framework discussed in this paper can also be extended to non-linear models. \medskip

The main insight underlying the results in this paper is that the assumptions underpinning VA models naturally lead to a generalized method of moments (GMM) framework. After demonstrating this, I show how one can use that framework for estimation and inference in models that employ VA measures in  regressions. Specifically, in models using VA measures as explanatory variables, I show that the estimation of the VA measures, and correlations between the observable characteristics used to construct them and true teacher quality, lead to incorrect inference due to inconsistent standard error estimators. I then propose corrected standard errors from a GMM framework, and discuss other possible solutions under stronger assumptions.\medskip

Next I show that models using VA measures as explanatory variables are often overidentified systems of moment conditions resembling instrumental variables systems, and use that fact to propose a more efficient estimator of the impact of having a high test-score VA teacher on long-run outcomes. Finally, I also provide corrected standard errors for models using VA measures as a dependent variable. \medskip

\medskip

My main theoretical findings are as follows. First, I describe the current practices used to estimate the impact of teacher test-score VA on student earnings. This is typically done using a multi-step procedure where the effect of observable characteristics is removed from test-scores and earnings, the best linear predictor of current year VA is constructed using the measures from other years, and residualized earnings are regressed on the best linear predictor. I set up the treatment effect model underlying those practices, and show how the  assumptions for the residuals of that model imply the higher level assumptions typically made in the literature, and lead to identification through a system of moment equations. Specifically,  
I show that these models rest on three important assumptions for the residuals of the treatment effect model. Teachers' VA has to be mean independent the unobserved determinants of students' test scores and earnings, and the average unobserved determinants of test-scores and earnings of students matched to a given teacher have to be uncorrelated across years. These assumptions imply the forecast unbiasedness assumption used to justify the use of VA measures as explanatory variables.\medskip

Second, I show that the aforementioned assumptions lead to the identification of the parameters of interest using a system that contains four sets of moment equations that map to the steps researchers routinely use to estimate VA's effect on long-run outcomes: a first set of moments to remove the effect of covariates from the test-scores, a second set of moments  used to construct the best linear predictor of current year VA using the measures from other years, an optional third set of moments to remove the effect of covariates from the outcome, and a fourth set to get the impact of VA on the residualized outcome.\footnote{Whether the outcome needs to be residualized or not depends on the context of each specific application. In some VA studies, the outcome is residualized, but in other studies it is not.} \medskip

Third, I use the fact that the asymptotic results for GMM estimators in this context naturally capture the aforementioned multi-step procedure to derive the asymptotic distribution of the estimator of the impact of true teacher VA on earnings used in the literature. I show that while an OLS regression will yield a consistent estimate of the coefficient, the associated standard error estimators will be inconsistent. Indeed, these standard error estimators  take into  account neither the correlations between the observable characteristics of students and true teacher VA, nor the construction of the best linear predictor. Given that most VA papers rely on a base assumption of selection on observables, the aforementioned correlations are likely to be strong, and thus accounting for them and the estimation of the best linear predictor is important.\medskip

Fourth, I show that if a researcher is interested in constructing confidence intervals using the OLS coefficients and doesn't wish to estimate a system by GMM, then corrected standard errors obtained from the GMM formula need to be calculated using the estimated parameters.\footnote{Another possibility is bootstrapping the estimation of the system. In practice this can be done without using GMM by bootstrapping the entire analysis starting with the estimation of value added.} I also show that in settings with random assignment of students to teachers, correct standard errors can be obtained by a 2SLS regression of earnings on current year VA while instrumenting the current year's VA by other year's VAs. \medskip

Fifth, I demonstrate that simply regressing the outcome on the best linear predictor of VA fails to take advantage of overidentification. I show that the previous system of moments is nested in a more general and overidentified system. Indeed, under the same assumptions required to use the best linear predictor of current year VA as an explanatory variable, we have that the VA measures  in years $s \neq t$ are valid instruments for year $t$'s VA. Then when we have more than two years of data, we have more than one valid instrument for the endogenous year $t$ measure. I use this to show that one can then obtain a more precise estimator of the effect of VA on earnings using an optimal weighting matrix. Furthermore, in a constant effect framework, one can test the validity of the model using an overidentification test. \medskip

Sixth, I  consider the construction of optimal instruments under the stronger assumption of random assignment of students to teachers, when researchers also choose to include covariates in their analysis. I find that if one is willing to assume that the conditional mean of the current year's VA given other years' VA is linear and errors are homoskedastic, a 3SLS estimator using the best linear predictor can be optimal.  \medskip

Finally, I focus on cases where VA is used as a dependent variable. I derive corrected standard errors for coefficients from such regressions using a GMM framework and provide a testable condition under which using OLS with unadjusted standard errors can also lead to valid inference. \medskip

The preceding theoretical results are confirmed by simulations and an application. The simulations focus on the importance of adjusting standard errors when estimating an OLS regression using VA measures as explanatory variables. Indeed, even in a simple model with constant true value added over time,  $95\%$ confidence intervals constructed using OLS coefficients and standard errors only adjusted for clustering have a coverage rate of $72.4\%$. On the other hand, using the multi-step OLS estimator with the proposed corrected standard errors from GMM yields correct inference with a coverage rate of $94.2\%$. Furthermore, the proposed optimal GMM estimator has a variance that is $1.3\%$ lower than the multi-step OLS estimator.\medskip

My application uses data on third graders in North Carolina public schools. I document the presence of correlations between the variables used to predict VA and true teacher VA. I find an effect of sorting on lagged test scores at the student level that is similar in magnitude to \citet{chetty2014measuring1}, but I also find that these correlations are much stronger for the classroom and school-year level lagged test scores used as controls to estimate VA. Indeed, sorting on lagged test scores at the classroom and school-year level is respectively five and three times larger than at the individual level.\footnote{\citet{rothstein2017measuring} also finds significant sorting at the school level.} I then illustrate my theoretical findings by showing the impact that this sorting has on the standard errors of coefficients in regressions using value added. The GMM standard errors I propose are between $37$ and $70\%$ larger than the standard errors currently used in the literature. Notably, in this application, the increase in standard errors resulting from my adjustment is larger on average than the impact of clustering standard errors. \medskip

This paper contributes to various literatures. First, for applied researchers, this paper provides a simple-to-implement framework to correctly and efficiently draw inference when using VA measures in regressions. Papers using value added as an explanatory variable are quite common in the economics of education literature, with notable examples being  \citet{chetty2014measuring} and \citet{jackson2018test} which use VA to highlight the importance of school teachers in improving students' adult outcomes. \citet{rose2019} use  VA to study the link between teacher quality and future student criminal behavior. \citet{canaan2019VA} use VA to show that advisors who raise short-run student achievement, such as GPA, improve subsequent long-run outcomes such as graduation. \citet{mulhern2019beyond} uses VA to show that good high-school counselors tend to improve all measures of educational attainment for students. \citet{liu2021engaging} use VA to explore the link between a teacher's impact on student's attendance and their long-run outcomes. \citet{opper2019does} uses VA as an explanatory variable when testing for spillover effects to determine whether the impact of teachers extends beyond the students in their classrooms.\medskip

This paper contributes to the applied literature by providing a theoretical framework that underpins the methods used in the literature while providing guidelines for the need to correct standard errors when using VA in regressions. Indeed, this paper identifies the correlations between the student characteristics  and true teacher quality as one of the main causes of incorrect inference, and the construction of the best linear predictor as another. While most applied researchers are aware of the importance of properly accounting for these factors in order to obtain unbiased estimates of teacher VA, this paper documents the presence of these correlations in practice and stresses the importance of accounting for them when conducting inference as well. \medskip

Second, this paper is also related to the  methodological literature on VA. This literature has mostly focused on different ways to estimate and shrink VA measures. For example \citet{angrist2017leveraging} show how researchers can leverage lotteries to create better VA estimates by combining non-experimental and quasi-experimental methods to obtain VA estimates with lower MSE, while \citet{gilraine2020} propose a non-parametric method to shrink VA estimates. Finally, \citet{opper2019does} develops a method of moments estimator to construct VA measures that account for spillovers. This paper fills a different gap in the literature by providing a flexible framework for estimation and correct inference when using VA measures in regressions, as well as  providing a guide for efficient estimation when using value-added measures as explanatory variables.\medskip

Third, the study of the properties of estimators in this paper is also related to the work of \citet{pagan1986two} who provides a unified framework for the properties of two-step estimators with a focus on the possibility of invalid inference, and the work of \citet{newey1984method} and \citet{newey1994large} which shows that the properties of multi-step estimators can be obtained by framing the estimators as method of moment estimators. This paper uses the results of the latter to derive its findings. \medskip

Finally, the conceptual link between VA models and GMM outlined in this paper, and the associated causes of incorrect inference and their solutions,  can be extended to different settings using estimated measures of quality in regressions. Indeed, models resembling the VA framework described in this paper are commonly used in economics. Examples include studies of the effectiveness of healthcare providers \citep{currie2021doing}, the effectiveness of bosses \citep{lazear2015value,bertrand2003managing}, and the impact of individuals working as part of a team \citep{arcidiacono2017productivity}. While the standard error estimators derived in this paper might not directly apply in those settings because the procedures used to estimate VA measures can sometimes depart from the ones described here, researchers could draw correct inference in such cases by following the general approach outlined in this paper and jointly modeling the estimation of the VA measures and subsequent regression analysis together in a GMM framework. \medskip

The rest of this paper is organized as follows. Section 2 presents the results for using VA as an explanatory variable. Section 3 presents the results for using VA as a dependent variable. Section 4 presents the simulation results. Section 5 goes over an empirical example and section 6 concludes.

%

\section{Using Value Added as an Explanatory Variable}

Researchers are often interested in the effect of teachers on the long-run outcomes of students. For example, how does the quality of a teacher affect the adult earnings of their students? To estimate this effect, one constructs a value-added measure of teacher quality and collects the outcomes of students that were matched with each teacher.

\subsection{Setup \label{section:setup}}

To begin, we must define the value added of a teacher. The common definition is the improvement in a student's test score attributable to the teacher. Let $R_{it}$ be the test score for student $i$ in year $t$. Then the potential test score, if the student were matched to a teacher who  has value added $\mu$ is

\begin{equation*}
R^{pot}_{it}(\mu)= \mu + R^{pot}_{it}(0)
\end{equation*}

where the test score is normalized to have a mean of 0 and a variance of 1. Because the test score is normalized, the value-added measure is also normalized to have a mean of 0. The part of the test score that is not attributable to the teacher's value added is

\begin{equation*}
R^{pot}_{it}(0)= X_{it}'\beta_0+\epsilon_{it},
\end{equation*}

where $X_{it}$ captures the observed characteristics of the student and $\epsilon_{it}$ captures the unobserved characteristics of the student that determine test scores. \medskip

The value added of a teacher, while defined in terms of a short-run outcome, may also affect long-run outcomes, such as adult earnings. Let $Y_{i}$ be the adult earnings for student $i$. Consider a model in which the value added of a teacher has a linear effect on earnings that is constant for all students (I relax the constant effect assumption in Appendix \ref{het-effects}). In this setting, the potential outcome function for the adult earnings of student $i$ is given by

\begin{equation*}
Y^{pot}_{i}(\mu)= \kappa_0 \mu + Y^{pot}_i(0)
\end{equation*}

and the potential outcome for an average quality teacher is

\begin{equation*}
Y^{pot}_i(0)= X_{it}'\beta_0^Y+ \eta_{it}
\end{equation*}

The student-level characteristics that determine this relation are measured at the time of exposure to the teacher: $X_{it}$ captures the observed characteristics of the student and $\eta_{it}$ captures the unobserved characteristics of the student and teacher. \medskip

To implement this framework, I allow for the possibility that the value added of a teacher varies over time, which for teacher $j$ is denoted $\mu_{jt}$. It follows from the potential outcome framework that the observed earnings for student $i$ who was matched to teacher $j$ in year $t$ are equal to:

\begin{equation} \label{eq:obsearn}
Y^{obs}_{i}= X_{it}'\beta_0^Y+ \kappa_0 \mu_{jt} + \eta_{it}.
\end{equation}

The coefficient $\kappa_0$ measures the effect a teacher has on future earnings that arises from the teacher's contribution to the student's test scores. However, it is important to note that the effect captured in
$\kappa_0$ might not be due solely to the direct effect on student test scores. There may be an indirect effect that can arise, for example, when a high value-added teacher in year $t$ leads parents to select better teachers in years after $t$, which also positively affects earnings.\footnote{For further discussion about interpreting $\kappa_0$ when modeling earnings in a linear setting see \citet{chetty2014measuring}} \medskip

I now describe a four step procedure in terms of population quantities used to identify $\kappa_0$  which I formalize in a system of moment conditions in the next section. Each step will be followed by a brief description of how it is implemented in practice. Before doing so, I introduce some helpful notation. There are $J$ teachers and each teacher is observed for $T$ years. A balanced panel across teachers and years simplifies the presentation, but is not required for the results. Each teacher has class size $n_j$, which is assumed to be constant over time.

In step one, we remove the effect of covariates from both test-scores and earnings. To do so, we begin with observed student test scores

\begin{equation}
\label{eq:scores}
R^{obs}_{it}= X_{it}'\beta_0 + \mu_{jt} +\epsilon_{it}
\end{equation}

and remove the effect of student characteristics yielding

\begin{equation} \label{eq:k-mom}
R_{it}=\mu_{jt} + \epsilon_{it}
\end{equation}

As we will see later,  \eqref{eq:k-mom} leads to a set of $K$ moment conditions, where $K$ is the dimension of $\beta_0$. In practice $\widehat{\beta}_0$ is estimated using within teacher variation by running a regression of observed test-scores $R^{obs}_{it}$ on the covariates $X_{it}$ and teacher fixed effects, one then obtains an estimate $\widehat{R}_{it}=R^{obs}_{it}-X_{it}'\widehat{\beta}_0$. The teacher fixed effect is included to correct for possible sorting on observable characteristics. If students with highly educated parents select the most gifted teacher, then estimates of the student characteristics would be biased by the omitted teacher effect. To remove this bias, we include teacher fixed-effects. Importantly, the teacher fixed effect is only used to estimate $\beta_0$, it is not used to construct $\widehat{R}_{it}$. Note that if we removed the teacher fixed effect from $\widehat{R}_{it}$, we would effectively remove the teacher's value added.\medskip

Next we remove the effect of the same set of covariates from earnings. We have: 

\begin{equation}\label{eq:earningres2}
Y_{it}=Y^{obs}_{i} - X_{it}'\beta_0^Y
\end{equation}

such that $Y_{it}$ is the earnings residual of student $i$, matched to teacher $j$ in year $t$. This step will lead to another additional $K$ moment conditions, where $K$ is the dimension of $\beta_0^Y$. In practice $\widehat{\beta}_0^Y$ is  estimated using within teacher variation by running a regression of observed earnings $Y^{obs}_{i}$ on the covariates $X_{it}$ and teacher fixed effects, one then obtains an estimate $\widehat{Y}_{it}=Y^{obs}_{i}-X_{it}'\widehat{\beta}_0^Y$. As before the teacher fixed effect is included to correct for possible sorting on observable characteristics when estimating $\beta_0^Y$, it is not used to construct $\widehat{Y}_{it}$. 

In step two, we construct a preliminary measure of value added. Averaging $R_{it}$ over all students in a class would yield the preliminary measure of value added

\begin{equation}
\overline{R}_{jt}= \frac{1}{n_{j}}\sum_{i=1}^{n_{j}}R_{it}=\mu_{jt} +\overline{\epsilon}_{jt},
\end{equation}

with the idea that, on average, the $\overline{\epsilon}_{jt}$ are close to 0. This measure we construct contains both the value-added measure, $\mu_{jt}$, and the effect of unobserved student characteristics, $\epsilon_{it}$.\medskip

Given that to estimate the effect of teacher value added on long-run outcomes, we want to use value added as an explanatory variable, there are two issues with using this initial constructed measure $\overline{R}_{jt}$. The first is that it is a noisy measure of value added, and the second is that students with positive unobserved determinants of test scores are likely to have positive unobserved determinants of earnings. In other words if we want to use $\overline{R}_{jt}$ as an explanatory variable then we have  mechanical endogeneity from using the same students to form both the value-added measures and the outcome \citep{jacob2010persistence}. To address these concerns there are two possible solutions. The first, which is currently used in applied research and will be discussed in step three, involves constructing for each teacher the best linear predictor of preliminary value added for the current year $\overline{R}_{jt}$, from the preliminary value-added measures for all  years $s \neq t$. The second solution which I discuss in section \ref{section:overid} will involve using the preliminary  value-added measures for all  years $s \neq t$ as instruments in an overidentified system of moment equations. I will  show that there are efficiency gains from using the latter solution. \medskip

As previously mentioned, step three improves upon the preliminary measures by constructing the best linear predictor of value added for the current year $t$, from the value-added measures for all other years $s$. The best linear predictor removes the endogeneity because we assume that the unobserved factors that influence tests scores are uncorrelated over time as the classes have no students in common and all sorting of students to teacher is captured by the observable characteristics. In the context of our model it is saying that $\overline{\epsilon}_{jt}$ and $\overline{\eta}_{jt}$ are correlated but $\overline{\epsilon}_{js}$ is uncorrelated with $\overline{\epsilon}_{jt}$ and $\overline{\eta}_{jt}$ for $s \neq t$. Under the assumptions of our model the best linear predictor also shrinks the value-added measure towards a mean of zero and reduces the noise, Appendix \ref{section:shrinkage} illustrates this with a simple example. \medskip
\medskip

It is reasonable to assume that value added is stationary. This then reduces the number of parameters to be estimated. This requires that mean teacher value added does not vary across calendar years and  the correlation of value added across any pair of years depends only on the amount of time which elapses between those years. We can then define this improved measure as:

 \begin{align}
\mu^*_{jt}= \underset{\abs{s-t}\neq 0}{\sum} \phi_{0\abs{s-t}}\overline{R}_{js} \label{eq:va-shrink}.
\end{align}
 \medskip

%
%
%

where

\begin{align}\label{eq:blup}
 \boldsymbol{\phi_0}= \underset{ \phi_{\abs{s-t}}}{argmin} \E\left(\left(\overline{R}_{jt}- \sum_{\abs{s-t}\neq 0}\phi_{\abs{s-t}}\overline{R}_{js}\right)^2\right).
\end{align}


Under the assumption of stationarity, this step will lead to an additional $T-1$ moment conditions, where $T-1$ is the dimension of $\boldsymbol{\phi_0}$. In practice $\boldsymbol{\phi_0}$ can be estimated by regressing  $\widehat{R}_{jt}=\frac{1}{n_{j}}\sum_{i=1}^{n_{j}}\widehat{R}_{it}$ on $\widehat{R}_{js}$, and the estimate of $\mu^*_{jt}$ is then a fitted value from an OLS regression of $\widehat{R}_{jt}$ on the  residuals from all other years \citep{chetty2014measuring1}. \medskip

%

\medskip

In step four we consider residualized earnings. They contain both the effect of teacher value added on student earnings, $\kappa_0\mu_{jt}$, and the effect of other unobserved characteristics of the student and teacher, $\eta_{it}$: \medskip

\begin{equation} \label{eq:refover}
Y_{it}=  \kappa_0 \mu_{jt} + \eta_{it}.
\end{equation}

\medskip

Averaging over all students in a class yields:

\begin{equation} \label{eq:earningres}
\overline{Y}_{jt}=  \kappa_0 \mu_{jt} + \overline{\eta}_{jt}.
\end{equation}

which leads to one additional moment condition. In practice, the unobserved value $\mu_{jt}$ must be replaced with an estimate. \medskip

Specifically the common practice estimator of $\kappa_0$ is obtained by estimating the following  sample counterpart of \eqref{eq:earningres} using OLS: \medskip

\begin{equation} \label{eq:estimation}
\widehat{Y}_{jt} = \kappa\widehat{\mu}_{jt} + \zeta_{jt},
\end{equation}

where $\widehat{\mu}_{jt}=\sum_{\abs{s-t}\neq 0}\widehat{\phi}_{\abs{s-t}}\widehat{R}_{js}$ is an estimate of $\mu_{jt}^*$  and $\widehat{Y}_{jt}=\frac{1}{n_j}\sum_{i}Y^{obs}_{i} - X_{it}'\widehat{\beta}_0^Y$. I will refer to this estimator of $\kappa_0$ as the multi-step OLS estimator.

\medskip

\medskip

The next section formalizes the identification of $\kappa_0$ using a system of moment equations that follows the steps outlined above. \medskip

\subsection{Identification}

I now discuss  Assumptions \ref{hyp:rand-assign} and \ref{hyp:tech-id} which give interpretable primitive conditions on the residuals of the treatment effect model under which  $\kappa_0$ is identified.

\begin{hyp}\label{hyp:rand-assign}  ~ \medskip
\begin{enumerate}
\item For all $i$, $j$, $t$: $\E[\epsilon_{it}|\boldsymbol{\mu}_{j}]=\E[\epsilon_{it}]=0$ where $\boldsymbol{\mu}_{j}=\begin{pmatrix} \mu_{j1} \\ \vdots  \\  \mu_{jT} \end{pmatrix}$.
\item  For all $s \neq t$: $\E(\eta_{it}|\mu_{js})=0$.
\item  For all $s \neq t$, $\left(\overline{\epsilon}_{jt},\overline{\eta}_{jt}\right)\indep  \left(\overline{\epsilon}_{js},\overline{\eta}_{js}\right)$
\end{enumerate}
\end{hyp}

Point 1 of Assumption \ref{hyp:rand-assign} requires that students be sorted to teachers based only on observable characteristics so that teacher quality is unrelated to unobservable
determinants of student short-run outcomes. Points 2 and 3 underpin the leave-one-out procedure described in the previous section. Specifically, Point 2 requires that the unobserved determinants of earnings in year $t$ be mean independent of the true test-score value added of teacher $j$ in years $s\neq t$. The mean independence assumption in point 2 can be weakened to be $\mu_{js}$ and $\overline{\eta}_{jt}$ being uncorrelated, but if teacher value-added is constant then this assumption must hold for $s=t$. Note that Point 2 does not rule out all sorting on long-run outcomes, it only requires that any sorting of students to teacher based on long-run outcomes be independent of a teacher's test-score value added. Finally Point 3 requires that the observable characteristics used to residualize short-run outcomes be sufficiently rich such that the average unobserved determinants of short-run and long-run achievement within teacher, be independent over time.

%

I impose the following additional assumption before establishing my identification result. To state the assumption compactly, for any variable $H$, let $\overline{H}_{jt}=\frac{1}{n_j}\sum_{i}H_{it}$ and $\ddot{H}_{jt}=\overline{H}_{jt}-\frac{1}{T}\sum_{t}\overline{H}_{jt}$. For each $j$: $\boldsymbol{H}_{j}$ is a matrix stacking $\overline{H}_{jt}$  over $t$ and  $\boldsymbol{\ddot{H}}_{j}$ is a matrix stacking $\ddot{H}_{jt}$ over $t$. \medskip
 \medskip

\begin{hyp}\label{hyp:tech-id}   ~ \medskip
\begin{enumerate}
\item $\E(\boldsymbol{\ddot{X}}_{j}'\boldsymbol{\ddot{X}}_{j})$ is finite and invertible.
\item $0<Var(\mu^*_{jt})<\infty$.
\item  $\E\left(\boldsymbol{\ddot{X}_j}\boldsymbol{\ddot{\mu}_j}\right)=0$, $\E(\boldsymbol{\ddot{X}_{j}\ddot{\epsilon}_j})=0$ and $\E(\boldsymbol{\ddot{X}_{j}}\boldsymbol{\ddot{\eta}}_j)=0$.
\end{enumerate}
\end{hyp}

Assumption \ref{hyp:tech-id} contains the assumptions required for identification. Point 1 requires no perfect multicolinearity in the covariates.  Point 2 requires that the linear projection of residual test scores in the current year on other years have non-zero and finite variance. Point 3 requires that fluctuations in covariates be uncorrelated with fluctuations in unobserved shocks over time as well as fluctuations in value added over time. This third point is required to identify $\boldsymbol{\beta_0}$ and $\boldsymbol{\beta_0^Y}$ in a regression with teacher fixed effects. \medskip

I will now show $\kappa_0$ is identified by a set of four moment conditions.
%
%
%

\medskip

\begin{res} \label{prop:ID-moment}
If Assumptions \ref{hyp:rand-assign} and \ref{hyp:tech-id} hold, then $(\boldsymbol{\beta_0}, \boldsymbol{\beta_0^Y},\boldsymbol{\phi_0})$ and $\kappa_0$ are identified by the following system of moments:
\begin{align}
&\E\left(\boldsymbol{\ddot{X}}_{j}'\left(\boldsymbol{\ddot{R}}^{obs}_{j}- \boldsymbol{\ddot{X}_{j}\beta_0}\right)\right)=0  \label{eq:mom1}\\
&\E\left(\boldsymbol{\ddot{X}}_{j}'\left(\boldsymbol{\ddot{Y}}^{obs}_{j}- \boldsymbol{\ddot{X}_{j}\beta_0^Y}\right)\right)=0 \label{eq:mom3} \\
&\E\left(\boldsymbol{R^{(-t)'}_{j}}\left(\boldsymbol{R}_{j}- \boldsymbol{R^{(-t)}_{j}\phi_0}\right)\right)=0 \label{eq:mom2}\\
&\E\left(\boldsymbol{\phi_0'R^{(-t)'}_{j}}\left(\boldsymbol{Y}_{j}- \kappa_0 \boldsymbol{R^{(-t)}_{j}\phi_0} \right)\right)=0   \label{eq:mom4}
\end{align}
where $\boldsymbol{Y}_{j}$ is a vector stacking $\overline{Y}_{jt}$ for teacher $j$, $\overline{R}^{(-t)}_{j}$ is a $T-1$ row vector stacking the $\overline{R}_{js}$ excluding $\overline{R}_{jt}$, and $\boldsymbol{R^{(-t)}_{j}}$ is a $T\times (T-1)$ matrix stacking the $\overline{R}^{(-t)}_{j}$.
\end{res}

To map these moments into the four step procedure from the previous section, note that the first set of moment conditions will be used to estimate $\boldsymbol{\beta_0}$ and maps to the first step of the procedure. The second set of moment conditions will be used to estimate $\boldsymbol{\beta^Y_0}$, this also maps to the first step of the procedure. The third set of moment conditions will be used to obtain the best linear prediction of $\overline{R}_{jt}$ as a function of other years. This maps to the second and third step of the procedure with $\boldsymbol{R^{(-t)}_{j}\phi_0}=\boldsymbol{\mu_{j}}^*$.  The fourth set of moment conditions will be used to estimate $\kappa_0$ and maps to the last step.
\medskip

For further intuition on this result, Appendix \ref{sec:extraID} also presents an exposition of this identification result using variances and covariances. \medskip

Finally note  that Assumptions \ref{hyp:rand-assign} and \ref{hyp:tech-id} imply the assumptions commonly made when using value added as a regressor. To see that, consider  $\mu^*_{jt}= \sum_{\abs{s-t}\neq 0}\phi_{0\abs{s-t}} \overline{R}_{js}$ from  \eqref{eq:va-shrink}. Note that  $\mu^*_{jt}$ is the best linear predictor of $\overline{R}_{jt}=\mu_{jt} +\overline{\epsilon}_{jt}$ as a function of other years. We can write:

\begin{equation} \label{eq:decomp}
\overline{R}_{jt}= \sum_{\abs{s-t}\neq 0}\phi_{0\abs{s-t}} \overline{R}_{js} + \theta_{jt},
\end{equation}

where $\theta_{jt}$ is the error from the best linear prediction. Then under Assumptions \ref{hyp:rand-assign} and \ref{hyp:tech-id} we have the following result:

\begin{res} \label{lem:assumption}
If Assumptions \ref{hyp:rand-assign} and \ref{hyp:tech-id} hold, then:
\begin{align*}
&Cov\left(\overline{\eta}_{jt}, \mu^*_{jt}\right)=0\\
&\frac{Cov\left(\mu_{jt}, \mu^*_{jt}\right)}{Var\left(\mu^*_{jt}\right)}=1.
\end{align*}
\end{res}

Now notice that it follows from \eqref{eq:earningres} that $\kappa_0$ is identified under Result \ref{lem:assumption}:

\begin{equation*}
\frac{Cov(\overline{Y}_{jt},\mu^*_{jt})}{Var(\mu^*_{jt})}=\kappa_0\frac{Cov(\mu_{jt},\mu^*_{jt})}{Var(\mu^*_{jt})}+\frac{Cov(\overline{\eta}_{jt},\mu^*_{jt})}{Var(\mu^*_{jt})}=\kappa_0,
\end{equation*}

since the conditions in Result \ref{lem:assumption} are the population equivalent of the assumptions usually stated in the current literature. For example, \citet{chetty2014measuring} suggest that the reduced form coefficient from an OLS regression of earnings on estimated value added would identify $\kappa_0$, if  an implicit and infeasible regression of true value added on estimated value added yields a coefficient of one. They show that one can recover the parameter of interest under the following conditions, which they call forecast unbiasedness for an estimate $\widehat{\mu}_{jt}$ of $\mu_{jt}$ and selection on observables:
\begin{align*}
\frac{Cov\left(\mu_{jt}, \widehat{\mu}_{jt}\right)}{Var\left(\widehat{\mu}_{jt}\right)}=1 \ \ \text{and} \ \
Cov\left(\eta_{ijt}, \widehat{\mu}_{jt}\right)=0.
\end{align*}

Therefore Assumptions \ref{hyp:rand-assign} and \ref{hyp:tech-id} are primitive assumptions that ensure that the higher level assumptions currently made in the literature hold.

\subsection{Estimation and Inference }

This section will focus on drawing correct inference on $\kappa_0$. In many applications, neither $n_j$ (class size) nor $T$ (the number of years over which a teacher is observed) are very large, therefore I  will consider asymptotics where $J$ (the number of teachers) goes to infinity and $n_j$ and $T$ are fixed. \medskip

In practice, researchers use the multi-step OLS estimator $\widehat{\kappa}$ along with $\widehat{s}$, a consistent estimator of  $\widetilde{\sigma}^2=(G_{\kappa}^{-1})^2\E\left(g(\boldsymbol{Z})^2\right)$, in order to draw inference on $\kappa_0$.\footnote{The actual form of $\widehat{s}$ used varies depending on whether one uses heteroskedasticity-robust variance estimators or cluster-robust variance estimators.} In this section, I will show that $\widehat{s}$ will not be a consistent estimator of the true variance of $\widehat{\kappa}$, and propose an alternative and consistent estimator of the variance of $\widehat{\kappa}$. Then, I will show that one can construct an optimal GMM estimator that is more efficient than $\widehat{\kappa}$. Next, I will show that if students are randomly assigned to teachers, one can draw valid inference on $\kappa_0$ using a 2SLS procedure. Finally under the assumption of random assignment, I will show that given some distributional assumptions a 3SLS estimator of $\kappa_0$ is optimal if the researcher chooses to include covariates in their analysis.      \medskip

\subsubsection{Asymptotic Distribution of the Multi-Step OLS Estimator \label{section:estim-inf}}

The multi-step OLS estimator of $\kappa_0$, described using the four steps in section 2.1, is an exactly identified method-of-moments estimator where the moments correspond to \eqref{eq:mom1}, \eqref{eq:mom3}, \eqref{eq:mom2}, and \eqref{eq:mom4}. Thus, asymptotic results for GMM estimators in this context naturally capture the multi-step OLS estimator. Therefore, one can treat $\widehat{\kappa}$ as part of a joint GMM estimator of the system of moments in Result \ref{prop:ID-moment} in order to derive its asymptotic distribution. \medskip

To establish consistency and asymptotic normality of the GMM estimators, regularity conditions are required - they are listed and discussed in Appendix \ref{regularity} as Assumptions \ref{hyp:tech-cons} and \ref{hyp:tech-norm}. \medskip

The object of interest here is the asymptotic distribution of $\widehat{\kappa}$, the multi-step OLS estimator of the impact of value added on earnings. Theorem \ref{prop:kappa-dist} below establishes the asymptotic distribution of this estimator. One should note that Theorem \ref{prop:kappa-dist} is a subset of the more general Theorem \ref{lem:asym-norm} which establishes the asymptotic normality and consistency of the joint estimators $(\boldsymbol{\widehat{\beta},\widehat{\phi},\widehat{\beta}^Y,}\widehat{\kappa})$. Given that $(\boldsymbol{\beta_0,\phi_0,\beta^Y_0})$ are often regarded as nuisance parameters required to estimate value-added measures and lack any meaningful causal interpretation, my discussion will focus only on the asymptotic distribution of $\widehat{\kappa}$.

\begin{thm} \label{prop:kappa-dist}
If Assumptions \ref{hyp:rand-assign}, \ref{hyp:tech-id}, \ref{hyp:tech-cons}, and \ref{hyp:tech-norm} hold, then:
\begin{enumerate}
\item \begin{align}
\sqrt{J}(\widehat{\kappa}-\kappa_0)\rightsquigarrow \mathcal{N}\left(0,\sigma^2\right)
\end{align}
where $\sigma^2=(G_{\kappa}^{-1})^2\E\left(\left(g(\boldsymbol{Z})+G_{\beta^Y}\psi_3(\boldsymbol{Z})+G_{\phi}\psi_2(\boldsymbol{Z})+G_{\beta}\psi_1(\boldsymbol{z})-G_{\phi}M_{2\phi}^{-1}M_{2\beta}\psi_1(\boldsymbol{Z})\right)^2\right)$ and $\boldsymbol{Z_j}=\left(\boldsymbol{X_j,R_j^{obs},Y_j^{obs}}\right)$, $\boldsymbol{Z}$ stacks the $\boldsymbol{Z_j}$, $g(\boldsymbol{Z})=\boldsymbol{\phi_0'R^{(-t)'}_{j}}\left(\boldsymbol{Y}_{j}- \kappa_0 \boldsymbol{R^{(-t)}_{j}\phi_0} \right)$ is the moment function used to estimate $\kappa_0$, $G_{\kappa}=\E[\nabla_{\kappa}g(\boldsymbol{Z},\boldsymbol{\beta_0,\phi_0,\beta^Y_0},\kappa_0)]$  and $G_{\beta}$, $G_{\phi}$, $G_{\beta^Y}$ are defined analogously. The remaining terms are defined in the proof.
\item Let $\widehat{\sigma}^2$  correspond to  an estimator of $\sigma^2$, constructed by replacing the population moments in $\sigma^2$ by averages and the parameters by the GMM estimators. Then: 
    \begin{equation*}
    \widehat{\sigma}^2 \convP \sigma^2.
    \end{equation*}
\end{enumerate}
\end{thm}

Theorem \ref{prop:kappa-dist} provides the asymptotic distribution of $\widehat{\kappa}$, and notably its variance $\sigma^2$. Crucially this variance depends on $G_{\beta}$, $G_{\phi}$, $G_{\beta^Y}$, the expected values of the gradient of the moment conditions used to identify $\kappa_0$ with respect to $\boldsymbol{\beta, \phi, \beta^Y}$ evaluated at $(\boldsymbol{\beta_0,\phi_0,\beta^Y_0,}\kappa_0)$. Intuitively, we need to consistently estimate  $(\boldsymbol{\beta_0,\phi_0,\beta^Y_0})$ in order to consistently estimate $\kappa_0$, therefore the  variance of $\kappa_0$ must reflect the fact that the parameters used to construct the dependent variable and the value-added measure are estimated \citep{newey1994large}. \medskip

The first implication of Theorem \ref{prop:kappa-dist} is that  GMM estimation of the system of moments from Result \ref{prop:ID-moment} will yield a consistent estimator of $\kappa_0$ and correct standard errors for that estimator. These standard errors can be used for hypothesis testing and confidence intervals. This is especially convenient as GMM estimation and the associated variance estimator are standard routines in most statistical software. \medskip

Another implication of Theorem \ref{prop:kappa-dist} is that $\widehat{s}$, the variance estimator routinely used in practice, will be an inconsistent estimator of $\sigma^2$. Indeed,  $\widehat{s}$ is a consistent estimator  of $\widetilde{\sigma}^2$, and it follows from point one of the theorem that $\widetilde{\sigma}^2\neq \sigma^2$. It is then worth examining what terms cause the difference between the two variances. \medskip

The first term I consider arises from the need to consistently estimate $\mu_{jt}^*$ and will depend on $G_{\phi}$. As discussed in section 2.1, the construction of $\mu_{jt}^*$ when using the multi-step OLS estimator plays a crucial role in circumventing endogeneity issues arising from correlations between  $\overline{\epsilon}_{jt}$ and $\overline{\eta}_{jt}$. Mathematically, we can show that $G_{\phi}$ is only zero in trivial cases, so $\widehat{s}$ can only be consistent in trivial cases. Indeed:

\begin{align*}
G_{\phi}&=-\kappa_0\boldsymbol{\phi_0'}\E\left(\boldsymbol{R_j^{(-t)'}R_j^{(-t)}}\right)
\end{align*}

and under point 3 of Assumption \ref{hyp:tech-norm} we have $\E\left(\boldsymbol{R_j^{(-t)'}R_j^{(-t)}}\right)\neq 0$ and $\boldsymbol{\phi_0}\neq 0$. Therefore $G_{\phi}$ is only equal to zero in the trivial case where $\kappa_0=0$. Then we have the following corollary:

\begin{cor} \label{cor:no-OLS}
If Assumptions \ref{hyp:rand-assign}, \ref{hyp:tech-id}, \ref{hyp:tech-cons}, and \ref{hyp:tech-norm} hold, then $\widetilde{\sigma}^2$ cannot be equal to $\sigma^2$ unless $\kappa_0=0$.
\end{cor}
 \medskip

A direct implication of Corollary \ref{cor:no-OLS} is that under our assumptions, $\widehat{s}$ will be an inconsistent estimator of $\sigma^2$ in all non-trivial cases. Therefore, to correctly draw inference on $\kappa_0$ when estimating it using the multi-step OLS estimator, one needs to manually construct the variance estimator following the formula for $\sigma^2$ in Theorem \ref{prop:kappa-dist}. This can be done by replacing population moments with sample averages and parameters with their estimator.  For example $G_{\phi}$ can be replaced with $\widehat{G}_{\phi}=-\widehat{\kappa}\boldsymbol{\widehat{\phi}'}\left(\frac{1}{J}\sum_{j=1}^{J}\boldsymbol{\widehat{R}_j^{(-t)'}\widehat{R}_j^{(-t)}}\right)$. \medskip

To further examine the difference, I consider the remaining two terms. The second term arises from the need to consistently estimate the relationship between covariates and test scores. It will depend on:

\begin{align*}
G_{\beta}&=\E\left[-\left(\boldsymbol{Y_j'-2\kappa_0\phi_0'R_j^{(-t)'}}\right)A\right]
\end{align*}

where $A$ is a function of the covariates in years $s\neq t$.\footnote{$A$ is a $T\times K$ matrix such that each row consists of ($\phi_1\overline{X}_{jt-1}+\phi_2\overline{X}_{jt-2}+...$).} Crucially, $G_{\beta}$ depends on the covariance between the unobserved determinants of earnings $\overline{\eta}_{jt}$ and the covariates $\overline{X}_{js}$ for $s \neq t$, the covariance between covariates $\overline{X}_{jt}$ and true value added $\mu_{jt}$  in all years including $t$, and the covariance between $\overline{X}_{jt}$ and the unobserved determinants of test scores $\overline{\epsilon}_{jt}$ in all years including $t$. Given that most value-added models rest on assumptions of selection on observables, it is reasonable to expect that the covariance between covariates and value added is non-zero and thus $G_{\beta} \neq 0$.\footnote{Note that point 3 of Assumption \ref{hyp:tech-id} only rules out correlations between $\boldsymbol{\ddot{X}_j}$ and $\boldsymbol{\ddot{\eta}_j}$, and correlations between $\boldsymbol{\ddot{X}_j}$ and $\boldsymbol{\ddot{\epsilon}_j}$. So under the assumptions of the model it need not be that the remaining terms in $G_{\beta}$ are zero either.} The application to North Carolina data in section \ref{section:sorting} and previous studies \citep{rothstein2017measuring} show that these correlations are large and significant.  \medskip

The final term that causes the difference between the two variances arises from the need to consistently estimate the relationship between earnings and covariates, and will depend on:

\begin{align*}
G_{\beta^Y}=\E\left[-\left(\boldsymbol{\phi_0'R_j^{(-t)'}X_j}\right)\right].
\end{align*}

Similarly to $G_{\beta}$, $G_{\beta^Y}$ will depend on the covariance between the covariates $\overline{X}_{jt}$ and test-score value added $\mu_{js}$ for $ s\neq t$, and the covariance between $\overline{X}_{jt}$  and the unobserved determinants of test scores $\overline{\epsilon}_{js}$ for $s \neq t$. Again since value-added models rest on assumptions of selection on observables, one can expect that $G_{\beta^Y}\neq 0$. \medskip

Therefore, the multi-step nature of the procedure used to estimate $\kappa_0$, and the correlations between true value added and the observables used to estimate it, make the standard errors that researchers routinely use  inconsistent estimators of $\sigma^2$.  Appendix \ref{section:variance-compare} presents a comparison between $\sigma^2$ and $\widetilde{\sigma}^2$, which suggests that one could expect $\widetilde{\sigma}^2$ to be smaller. The next section will focus on the most efficient way to estimate $\kappa_0$. \medskip


\subsubsection{Overidentification and Efficiency \label{section:overid}}

If one is  interested in estimating $\kappa_0$, then the multi-step OLS estimator  will not be the estimator with the lowest variance. Indeed, under Assumptions \ref{hyp:rand-assign} and \ref{hyp:tech-id}, the exactly identified system consisting of \eqref{eq:mom1}, \eqref{eq:mom3}, \eqref{eq:mom2}, and \eqref{eq:mom4} masks a more general overidentified system when $T>2$. To see that note that \eqref{eq:mom4} only requires that a linear combination of $\E\left(\boldsymbol{\overline{R}^{(-t)'}_{j}}\left(\boldsymbol{Y}_{j}- \kappa_0\boldsymbol{R_{j}}\right)\right)$  be equal to zero.\footnote{See section \ref{section:proofid} for a proof.} However, under Assumptions \ref{hyp:rand-assign} and \ref{hyp:tech-id} we have the stronger conditions:

\begin{align*}
\E\left(\boldsymbol{\overline{R}^{(-t)'}_{j}}\left(\boldsymbol{Y}_{j}-\kappa_0\boldsymbol{R}_{j}\right)\right)=0.
\end{align*}

Given that $\E\left(\boldsymbol{R^{(-t)'}_{j}}\left(\boldsymbol{Y}_{j}- \kappa_0\boldsymbol{R_{j}}\right)\right)=0$, any linear combination of these moments will also be zero which makes estimating $\boldsymbol{\phi_0}$ and thus \eqref{eq:mom2} redundant. Then,  by replacing \eqref{eq:mom2} and \eqref{eq:mom4} with $\E\left(\boldsymbol{R^{(-t)'}_{j}}\left(\boldsymbol{Y}_{j}- \kappa_0\boldsymbol{R_{j}}\right)\right)=0$ we instead have the following system of moment conditions:

\begin{align}
&\E\left(\boldsymbol{\ddot{X}}_{j}'\left(\boldsymbol{\ddot{R}}^{obs}_{j}- \boldsymbol{\ddot{X}_{j}\beta_0}\right)\right)=0 \nonumber \\
&\E\left(\boldsymbol{\ddot{X}}_{j}'\left(\boldsymbol{\ddot{Y}}^{obs}_{j}- \boldsymbol{\ddot{X}_{j}\beta_0^Y}\right)\right)=0 \nonumber \\&\E\left(\boldsymbol{R^{(-t)'}_{j}}\left(\boldsymbol{Y}_{j}-\kappa_0\boldsymbol{R}_{j}\right)\right)=0 \label{eq:mom-over}
\end{align}

with $2K+1$ parameters to estimate and $2K+(T-1)$ moments to estimate them with. Therefore when $T>2$, the system is overidentified. Intuitively, the system consisting of \eqref{eq:mom1}, \eqref{eq:mom3}, and \eqref{eq:mom-over} is akin to an instrumental variables system where $\kappa_0$ is a scalar parameter that can be identified from $T-1$  moments. In essence, under the same set of assumptions required to use the multi-step OLS estimator of $\kappa_0$, we have that the preliminary measures of value added in years $s \neq t$ are valid instruments for the preliminary measure of value added in year $t$ from $\eqref{eq:k-mom}$. As such when we have more than two years of data, we have more than one valid instrument for the endogenous current year measure. As a result, we can rewrite the system as an overidentified systems of moment conditions resembling an instrumental variables framework. The first two set of moments are taken from the previous setup and are used to create the dependent variable (residualized earnings), the instruments (preliminary measures of value added in years $s \neq t$), and the endogenous variable (preliminary measure of value added in year $t$). The third set of moments provide the overidentifying information.\medskip

Given that we have shown that the system of moments in Result \ref{prop:ID-moment} is nested in the more general overidentified system consisting of  \eqref{eq:mom1}, \eqref{eq:mom3}, and \eqref{eq:mom-over}, we can now establish that the GMM estimator with the optimal weighting matrix will be a more efficient estimator of $\kappa_0$. The optimal GMM estimator will then have a variance that is no greater than the multi-step OLS estimator.\footnote{The GMM system consisting of \eqref{eq:mom1}, \eqref{eq:mom3}, and \eqref{eq:mom-over} weighed using \begin{equation*}
W_1=\begin{pmatrix} I_{K\times K} & 0 & 0 \\  0 & I_{K\times K} & 0 \\ 0& 0 &\boldsymbol{\phi_0\phi_0'} \end{pmatrix}.
\end{equation*}
where $\boldsymbol{\phi_0}=\E\left(\boldsymbol{R^{(-t)'}_{j}R^{(-t)}_{j}}\right)^{-1}\E\left(\boldsymbol{\overline{R}^{(-t)'}_{j}R_{j}}\right)$ would have the same objective function as the multi-step OLS.} \medskip

Let $\widehat{W}^*$ be an estimate of the optimal weighting matrix $W^*=\E[\widetilde{g}_1(\boldsymbol{Z},\boldsymbol{\beta_0,\beta^Y_0,}\kappa_0)\widetilde{g}(\boldsymbol{Z},\boldsymbol{\beta_0,\beta^Y_0,}\kappa_0)']^{-1}$ that replaces population moments with sample averages and parameters with their estimator. Then we have that the estimators resulting from GMM minimization with an optimal weighting matrix are consistent and asymptotically normal. Theorem  \ref{prop:kappa-optimal} formalizes the result for the optimal estimator $\widehat{\kappa}^*$. It is a subset of the more general Theorem \ref{prop:optimal} which formalizes the result for the joint estimators $(\boldsymbol{\widehat{\beta}^*,\widehat{\beta}^{Y*}},\widehat{\kappa}^*)$. \medskip

\begin{thm} \label{prop:kappa-optimal}
If Assumptions \ref{hyp:rand-assign}, \ref{hyp:tech-id}, and \ref{hyp:tech-opt} hold, then
\begin{equation*}
\sqrt{J}\left[\widehat{\kappa}^*-\kappa_0\right] \rightsquigarrow \mathcal{N}\left(0,\sigma_*^{2}\right).
\end{equation*}

where $\widehat{\kappa}^*$ is the estimate resulting from a GMM minimization using $\widehat{W}^*$ as a weighting matrix. One has that $\sigma_*^{2} \leq \sigma^2$.
\end{thm}

Theorem \ref{prop:kappa-optimal} shows that researchers can use an optimal GMM procedure to obtain more precise estimates of $\kappa_0$, such that $\sigma_*^{2}\leq \sigma^2 $ without imposing any additional assumptions. Furthermore, the validity of Assumptions \ref{hyp:rand-assign}, \ref{hyp:tech-id}, and \ref{hyp:tech-opt}, and the specifications of  \eqref{eq:scores}, \eqref{eq:earningres2}, and \eqref{eq:earningres} is falsifiable using Hansen's overidentification test under the assumption of constant treatment effects. This is formalized in Result \ref{prop:over-id}.

\subsubsection{Estimation and Inference in the Presence of Random Assignment}


I now turn to the case where students are randomly assigned to teachers. In such settings,  one would not need to adjust for covariates in the analysis and drawing inference on $\kappa_0$ is a simpler problem. Indeed, while  $\widehat{s}$ is still an inconsistent estimator of $\sigma^2$, one needs to account only for the estimation of the linear projection $\mu_{jt}^*$. Then one can use a 2SLS regression of the outcome on the preliminary value-added measure in year $t$, $\overline{R}_{jt}$, while instrumenting it by the preliminary measures in years $s\neq t$. \medskip

To formalize this result and ease exposition, consider the following assumption similar to \citet{canaan2019VA}:

\begin{hyp}\label{hyp:no-cov}  ~ \medskip
In every year students are randomly assigned to teachers such that one can model observed scores and earnings as:
\begin{align*}
&R^{obs}_{it}= \alpha_t + \mu_{jt} +\epsilon_{it}\\
&Y^{obs}_{i}= \alpha^Y_t+ \kappa_0 \mu_{jt} + \eta_{it}.
\end{align*}
\end{hyp}

Under Assumption \ref{hyp:no-cov}, given that students are randomly assigned to teachers, there is no need to control for covariates aside from year fixed effects. Indeed, \eqref{eq:mom1} and \eqref{eq:mom3} are now redundant since the time effects $\alpha_t$ and $\alpha^Y_t$ can be removed by subtracting the averages of $R^{obs}_{it}$ and $Y^{obs}_{i}$ for year $t$. Then we are left with \eqref{eq:mom-over} and the system of moments becomes:

\begin{align*}
&\E\left(\boldsymbol{R^{(-t)'}_{j}}\left(\boldsymbol{Y}_{j}- \kappa_0 \boldsymbol{R_{j}} \right)\right)=0,
\end{align*}

which now aligns with a traditional instrumental variables problem. We can then estimate and draw inference on $\kappa_0$ using the following 2SLS regression:

\begin{align}
\widehat{R}_{jt}= \sum_{\abs{s-t}\neq 0}\phi_{\abs{s-t}} \widehat{R}_{js} + \iota_{jt} \label{eq:2SLS-1}\\
\widehat{Y}_{jt}=  \kappa \sum_{\abs{s-t}\neq 0}\widehat{\phi}_{\abs{s-t}} \widehat{R}_{js}  + \zeta_{jt} \label{eq:2SLS-2}
\end{align}
where $\widehat{R}_{jt}=R^{obs}_{jt} - \frac{1}{J} \sum_{j=1}^{J}R^{obs}_{jt}$ where $R^{obs}_{jt}=\frac{1}{n_j}\sum_{i=1}^{n_{j}}R^{obs}_{it}$, and $\widehat{Y}_{jt}$ is defined analogously. This corresponds to a regression of the outcome on current year value added while instrumenting year $t$ value added by value added in years $s\neq t$. Furthermore, let $\widehat{s}_{2SLS}$ be the variance estimator that is computed by statistical softwares when using 2SLS.

This leads to the following result:

\begin{res} \label{cor:2sls}
If Assumptions  \ref{hyp:rand-assign}, \ref{hyp:tech-id}, \ref{hyp:no-cov},  \ref{hyp:tech-cons}, and \ref{hyp:tech-norm} hold:
Valid inference can be drawn on $\kappa_0$ by estimating the 2SLS regression defined by \eqref{eq:2SLS-1} and \eqref{eq:2SLS-2} while using  $\widehat{s}_{2SLS}$ as a variance estimator.
\end{res}

The intuition behind Result \ref{cor:2sls} is simple, if the covariates are uninformative of teacher value added or the unobserved determinants, then they are not required for a consistent estimation of the residualized outcome or the value-added measures. In that case, standard errors need only account for the fact that the current year preliminary value added measure was instrumented by the preliminary value added measures in other years. However, given that Assumption \ref{hyp:no-cov} is unlikely to hold in most applied settings, constructing a consistent estimator of $\sigma^2$ or GMM estimation of the system in Result \ref{prop:ID-moment} is required to construct confidence intervals using $\widehat{\kappa}$. \medskip

Finally, it follows from standard results that this 2SLS estimator will be efficient under homoskedasticity. The next section will show that if a researcher wants to include covariates in their analysis, then this 2SLS estimator is no longer efficient under homoskedasticity. \medskip

\subsubsection{Optimal Instruments in the Presence of Random Assignment}

Even in the presence of random assignment, researchers sometimes include covariates in the analysis, often to improve statistical precision. For such cases, I will now construct the optimal instruments for the system defined by \eqref{eq:mom1}, \eqref{eq:mom3}, and \eqref{eq:mom-over}, and show that the traditional 3SLS estimator is optimal under a homoscedasticity assumption and if $\E[\boldsymbol{R}_{j}|\boldsymbol{R^{(-t)}_{j}}]$ is linear.\footnote{I use the term traditional 3SLS to refer to the 3SLS estimator that uses the linear projections of the endogenous variables as instruments. This is in contrast with the GMM 3SLS estimator which is optimal under homoskedasticity without further assumptions. The traditional 3SLS estimator is equivalent to GMM 3SLS when all equations use the same instruments \citep{wooldridge2010econometric}.}

For the remainder of this section I assume that students are randomly sorted to teachers and the following assumption holds:

\begin{hyp}\label{hyp:exog-cov}  ~ \medskip
\begin{enumerate}
\item  $\boldsymbol{X}_j \indep (\boldsymbol{\mu_j,\epsilon^{(-t)}_j,\eta^{(-t)}_j})$.
\item $\boldsymbol{\mu_j} \indep (\boldsymbol{\epsilon_j,\eta_j})$.
\item $\E(\boldsymbol{\epsilon_j}|\boldsymbol{\ddot{X}_j})$=0.
\end{enumerate}
\end{hyp}

Point 1 of Assumption \ref{hyp:exog-cov} requires that the observable characteristics of students matched to teacher $j$ in year $t$ be independent of teacher $j$'s value-added in year $t$ and the unobservable characteristics of students assigned to teacher $j$ in years $s \neq t$. Point 2 requires that the value-added of teacher $j$ be independent of all student unobservables. These conditions should hold by design if students are randomly assigned to teachers. The third point slightly strengthens point 3 of Assumption \ref{hyp:tech-id} to require that the unobservable determinants of test-scores be mean independent of within teacher fluctuations in covariates.\medskip
We can now construct the optimal instruments for the system:
\begin{align*}
&\E\left(\boldsymbol{\ddot{X}}_{j}'\left(\boldsymbol{\ddot{R}}^{obs}_{j}- \boldsymbol{\ddot{X}_{j}\beta_0}\right)\right)=0 \\
&\E\left(\boldsymbol{\ddot{X}}_{j}'\left(\boldsymbol{\ddot{Y}}^{obs}_{j}- \boldsymbol{\ddot{X}_{j}\beta_0^Y}\right)\right)=0  \\
&\E\left(\boldsymbol{R^{(-t)'}_{j}}\left(\boldsymbol{Y}_{j}- \kappa_0 \boldsymbol{R_{j}} \right)\right)=0.
\end{align*}

Let $\boldsymbol{u}=\begin{pmatrix} \boldsymbol{\ddot{\epsilon}_{j}} +\boldsymbol{\ddot{\mu}_{j}} \\ \boldsymbol{\ddot{\eta}_{j}} + \kappa_0\boldsymbol{\ddot{\mu}_{j}} \\  \boldsymbol{\eta_{j}} -\kappa_0 \boldsymbol{\epsilon_{j}} \end{pmatrix}$, then we have the following result:

\begin{res} \label{res:optimal-mom}
If Assumptions \ref{hyp:rand-assign}, \ref{hyp:tech-id}, and \ref{hyp:exog-cov}  hold, then:

If  $\E[\boldsymbol{R}_{j}|\boldsymbol{R^{(-t)}_{j}}]$ is linear such that $\E[\boldsymbol{R}_{j}|\boldsymbol{R^{(-t)}_{j}}]=\boldsymbol{R^{(-t)}_{j}\phi_0}$ and $\boldsymbol{\E[uu'|\ddot{X}_{j},R^{(-t)}_{j}]}=\boldsymbol{\E[uu']}$ then the optimal moment conditions are:
\begin{equation*}
\E\left[\left(\boldsymbol{\E[uu']}^{-1}\begin{pmatrix} \boldsymbol{\ddot{X}_{j}} &0 & 0 \\ 0 & \boldsymbol{\ddot{X}_{j}} & 0\\ 0 & 0 & \boldsymbol{R^{(-t)}_{j}\phi_0}\end{pmatrix}\right)'\boldsymbol{u}\right]=0.
\end{equation*}
Those are the moment conditions satisfied by the 3SLS estimator. Then in this case the optimal estimator is the 3SLS estimator which first estimates $(\boldsymbol{\beta_0}, \boldsymbol{\beta_0^Y})$ using OLS and constructs $\boldsymbol{R}_{j}$ and $\boldsymbol{Y}_{j}$, then estimates $\kappa_0$ by a 2SLS regression of $\boldsymbol{Y}_{j}$ on $\boldsymbol{R}_{j}$ while instrumenting $\boldsymbol{R}_{j}$ with $\boldsymbol{R^{(-t)}_{j}}$, uses those estimates to construct an estimate of $\boldsymbol{\E[uu']}$, and finally estimates the entire system again using GLS. \medskip
\end{res}

%

Result \ref{res:optimal-mom} states that under random assignment, if the errors are homoskedastic and the conditional expectation of the preliminary value-added measures $\boldsymbol{R}_{j}$ in year $t$ given preliminary value-added measures $\boldsymbol{R^{(-t)}_{j}}$ in years $s\neq t$ is actually linear, then estimating the system composed of \eqref{eq:mom1}, \eqref{eq:mom3}, and \eqref{eq:mom-over} by 3SLS is efficient. Result \ref{res:optimal-mom}  also rules out the system underlying the multi-step OLS estimator  being optimal, as that would require the three components in $\boldsymbol{u}$ to be uncorrelated.\medskip 

It is then useful to consider under what distributional assumptions the conditions from Result \ref{res:optimal-mom}  hold. The first condition to consider is  homoskedasticity, namely the possibility that $\boldsymbol{\E[uu'|\ddot{X}_{j},R^{(-t)}_{j}]}=\boldsymbol{\E[uu']}$. Homoskedasticity requires that the variances and covariance of the unobservable determinants and value added $\boldsymbol{\epsilon_{j}}, \boldsymbol{\eta_{j}},\boldsymbol{\mu_{j}}$ do not vary across teachers with different levels of the preliminary measures of value added in years $s\neq t$, $\boldsymbol{R^{(-t)}_{j}}=\boldsymbol{\mu^{(-t)}_{j}}+\boldsymbol{\epsilon^{(-t)}_{j}}$, and with different levels of covariate fluctuations $\boldsymbol{\ddot{X}_{j}}$. \medskip

The second condition is $\E[\boldsymbol{R}_{j}|\boldsymbol{R^{(-t)}_{j}}]$ being linear. Suppose we have:
\begin{equation} \label{eq:norm-dist}
\begin{bmatrix}
\boldsymbol{\mu_{j}}\\
\boldsymbol{\epsilon_j}\\
\end{bmatrix}
\sim \mathcal{N} \begin{pmatrix}
0, & \begin{pmatrix}
\boldsymbol{\Sigma}_{\boldsymbol{\mu}}\otimes I_J & 0 \\
0&\sigma^2_{\epsilon}I_{JT}
\end{pmatrix} \end{pmatrix}
\end{equation}

such that teacher value added and the average unobservable determinants of student test scores are joint normally distributed. Under  \eqref{eq:norm-dist} teacher value added is independent across teachers but correlated within teacher, and it is independent of the average unobservable determinants of student test scores which are i.i.d across teacher-years (this second requirement is likely to hold under random assignment).  \medskip

Then we have:
\begin{align}
&\E[\boldsymbol{R}_{j}|\boldsymbol{R^{(-t)}_{j}}] \nonumber \\
=&\E[\boldsymbol{\mu_{j}}+\boldsymbol{\epsilon_j}|\boldsymbol{\mu^{(-t)}_{j}}+\boldsymbol{\epsilon^{(-t)}_j}] \nonumber \\
=&\E[\boldsymbol{\mu_{j}}|\boldsymbol{\mu^{(-t)}_{j}}+\boldsymbol{\epsilon^{(-t)}_j}]+\E[\boldsymbol{\epsilon_j}|\boldsymbol{\mu^{(-t)}_{j}}+\boldsymbol{\epsilon^{(-t)}_j}] \nonumber  \\
=&\E[\boldsymbol{\mu_{j}}|\boldsymbol{\mu^{(-t)}_{j}}+\boldsymbol{\epsilon^{(-t)}_j}] \nonumber  \\
=&(\boldsymbol{\mu^{(-t)}_{j}}+\boldsymbol{\epsilon^{(-t)}_j})\boldsymbol{\Sigma}^{-1}_{\boldsymbol{\mu^{(-t)}_{j}+\epsilon^{(-t)}_j}}\boldsymbol{\Sigma}_{\boldsymbol{\mu\mu^{(-t)}}} \nonumber \\
=&\boldsymbol{R^{(-t)}_{j}}\boldsymbol{\Sigma}^{-1}_{\boldsymbol{R^{(-t)}_{j}}}\boldsymbol{\Sigma}_{\boldsymbol{R}_{j}\boldsymbol{R^{(-t)}_{j}}}\\
=& \boldsymbol{R^{(-t)}_{j}\phi_0}\label{eq:is-linear}
\end{align}

where $\boldsymbol{\Sigma}_{\boldsymbol{\mu\mu^{(-t)}}}$ is the covariance matrix of $\boldsymbol{\mu}$ and $\boldsymbol{\mu^{(-t)}}$, and $\boldsymbol{\Sigma}_{\boldsymbol{R^{(-t)}_{j}}}$ is the covariance matrix of $\boldsymbol{R^{(-t)}_{j}}$. The third equality follows from the fact that $\boldsymbol{\epsilon_j}$ is independent of $\boldsymbol{\mu^{(-t)}_{j}}+\boldsymbol{\epsilon^{(-t)}_j}$ by  \eqref{eq:norm-dist}. The fourth equality follows from the conditional expectation formula for multivariate normal distributions, and the fifth equality follows from the fact that the covariance matrix of $\boldsymbol{R_{j}}$ and $\boldsymbol{R^{(-t)}_{j}}$ is the same as $\boldsymbol{\Sigma}_{\boldsymbol{\mu\mu^{(-t)}}}$ by  \eqref{eq:norm-dist}.

\medskip

Then if teacher value added and the average unobservable determinants of student test scores are joint normally distributed following  \eqref{eq:norm-dist}, then
$\E[\boldsymbol{R}_{j}|\boldsymbol{R^{(-t)}_{j}}]$ is linear.  To gain some intuition for  \eqref{eq:is-linear}, consider the case with $T=2$ where
 \begin{equation*}
\E[\boldsymbol{R}_{j}|\boldsymbol{R^{(-t)}_{j}}]=\frac{Cov(\mu_{jt},\mu_{j(t-1)})}{Var(\mu)+Var(\epsilon)}\overline{R}_{j(t-1)}.
\end{equation*} \medskip

The distributional assumption in   \eqref{eq:norm-dist} could be plausible in a setting with random assignment of students to teachers, a large number of teachers $J$, a large number of students per teachers $n_j$, and where value added is either constant over time for every teacher or $\mu_{jt}=\mu_{j}+\omega_{jt}$ where $\mu_j$ and $\boldsymbol{\omega_{j}}$ are joint normal and independent. Overall, if one is willing to assume that students are randomly assigned to teachers then the 3SLS estimator is efficient if one assumes homoskedasticity and joint normality of teacher value added and unobserved determinants.

\section{Using Value Added as a Dependent Variable \label{sec:outcome}}

Researchers are also often interested in examining whether the observable characteristics of teachers predict their value added. For example, do teachers with National Board Certification or more experience have higher value added? One can also look into whether the implementation of a given policy is linked to an increase in teacher value added. For instance, does a training program for teachers raise their value added? To answer such  questions, researchers regress their estimated value-added measures on a set of explanatory variables. \medskip

In our setting, suppose that one is interested in the relationship between some teacher-year level variables $D_{jt}$ and teacher value-added $\mu_{jt}$. The parameters of interest in this section are the coefficients from  the linear projection of $\mu_{jt}$ on $D_{jt}$:

\begin{equation} \label{eq:VA-outcome}
\mu_{jt}=D_{jt}'\boldsymbol{\alpha_0} + \zeta_{jt}.
\end{equation}

Given that $\mu_{jt}$ is unobserved, one cannot estimate \eqref{eq:VA-outcome} as it is. In practice,  researchers often replace $\mu_{jt}$ by the estimated measure $\widehat{\mu}_{jt}=\sum_{\abs{s-t}\neq 0}\widehat{\phi}_{\abs{s-t}}\widehat{R}_{js}$ and proceed by estimating \eqref{eq:VA-outcome} using OLS. \medskip

This section will go over how to identify, consistently estimate, and draw inference on $\boldsymbol{\alpha_0}$ using an exactly identified GMM procedure. The asymptotic result for this GMM estimator will again naturally capture the cases in which the steps are separately estimated using OLS. \medskip

I impose the following assumption with additional regularity conditions in Assumption \ref{hyp:as-outcome-tech} in Appendix \ref{regularity}:
\begin{hyp}\label{hyp:as-outcome}  ~ \medskip
\begin{enumerate}
\item $\E(\boldsymbol{D}_{j}'\boldsymbol{D}_{j})$ is finite and invertible.
\item $\E(\boldsymbol{D}_{j}'\boldsymbol{\epsilon}_{j})=0$.
\end{enumerate}
\end{hyp}

Point 1 requires no perfect multicolinearity in the $\boldsymbol{D}_{j}$. Point 2 requires that the  average shocks be uncorrelated with the variables $D_{jt}$.  Similarly to Assumption \ref{hyp:rand-assign}, this assumption requires that the observable characteristics used to residualize short-run outcomes be sufficiently rich such that the remaining unobservables, excluding value added, be uncorrelated with our variables of interest. Note that point 2 can be weakened to require that the average unobserved determinants of test scores in years $s\neq t$ be uncorrelated with $D_{jt}$, such that $\E(\boldsymbol{D}_{j}'\boldsymbol{\epsilon^{(-t)}}_{j})=0$. \medskip

We can then show that $\boldsymbol{\alpha_0}$ is identified by the following system of moment conditions:
\begin{res} \label{lem:as-outcome}
If Assumptions \ref{hyp:rand-assign}, \ref{hyp:tech-id}, and \ref{hyp:as-outcome}  hold, then $(\boldsymbol{\beta_0,\alpha_0})$ are uniquely identified by the following system of moments:
\begin{align}
&\E\left(\boldsymbol{\ddot{X}}_{j}'\left(\boldsymbol{\ddot{R}}^{obs}_{j}- \boldsymbol{\ddot{X}_{j}\beta_0}\right)\right)=0 \\
&\E\left(\boldsymbol{D_{j}}'\left(\boldsymbol{R}_j-\boldsymbol{D_{j}\alpha_0}\right)\right)=0
\end{align}
\end{res}

Result \ref{lem:as-outcome} shows that the coefficient from a linear projection of a set of variables $D_{jt}$ on teacher value-added $\mu_{jt}$ is identified by two sets of moment conditions. The first set of moments is used to identify $\boldsymbol{\beta_0}$ and construct preliminary measures of value added. They are discussed in sections 2.1 and 2.2. The second set of moments is used to identify $\boldsymbol{\alpha_0}$ . Note that unlike the previous sections, under point 2 of Assumption \ref{hyp:as-outcome}, we do not need to create a leave-year out measure of value added to recover the relationship between $\mu_{jt}$ and $\boldsymbol{D}_{jt}$. Instead, we can use the preliminary measures of value added $\overline{R}_{jt}=\mu_{jt}+\overline{\epsilon}_{jt}$ as the outcome since the unobserved determinants of student test scores are assumed to be uncorrelated with the variables of interest $D_{jt}$. If one suspects that point 2 of Assumption \ref{hyp:as-outcome} is unlikely to hold, and is instead willing to assume that $\E(\boldsymbol{D}_{j}'\boldsymbol{\epsilon^{(-t)}}_{j})=0$, then  $\overline{R}_{jt}$ can be replaced by a simple leave-year out average $\widetilde{\mu}_{jt}=\frac{1}{T-1}\sum_{s\neq t}\overline{R}_{js}$ with the idea that $\boldsymbol{D}_{jt}$ is unlikely to be correlated with the unobservable determinants of test scores of students in different years. \medskip

To estimate $(\boldsymbol{\beta_0,\alpha_0})$, let the GMM weighting matrix be the identity matrix $W=I$. Then Theorem \ref{lem:as-outcome-dist} shows that we can consistently estimate and draw inference on $\boldsymbol{\alpha_0}$ using GMM. Using partitioned inversion and Theorem \ref{lem:as-outcome-dist}, we can obtain the asymptotic variance of $\boldsymbol{\widehat{\alpha}}$. \medskip

\begin{thm} \label{prop:alpha-dist}
If Assumptions \ref{hyp:rand-assign}, \ref{hyp:tech-id}, and \ref{hyp:as-outcome}  hold, then
\begin{align}
\sqrt{J}(\widehat{\alpha}-\alpha_0)\rightsquigarrow \mathcal{N}\left(0,V_1\right)
\end{align}
where $V_1=\E(\boldsymbol{D}_{j}'\boldsymbol{D}_{j})^{-1}\E\left(\Gamma\Gamma'\right)\E(\boldsymbol{D}_{j}'\boldsymbol{D}_{j})^{-1'}$, and \\
$\Gamma=\left(\boldsymbol{D_{j}}'\left(\boldsymbol{R}_j-\boldsymbol{D_{j}\alpha_0}\right)-\E\left(\boldsymbol{D_j'X_j}\right)\E\left(\boldsymbol{\ddot{X}}_{j}'\boldsymbol{\ddot{X}}_{j}\right)^{-1}\boldsymbol{\ddot{X}}_{j}'\left(\boldsymbol{\ddot{R}}^{obs}_{j}- \boldsymbol{\ddot{X}_{j}\beta_0}\right)\right)$
\end{thm}

Theorem \ref{prop:alpha-dist} shows that a GMM estimation of the system in Result \ref{lem:as-outcome} will yield a consistent and asymptotically normal estimator of $\boldsymbol{\alpha_0}$, and the standard errors from that procedure can be used to correctly draw inference. Furthermore, because the GMM estimator captures the estimator of $\boldsymbol{\alpha_0}$ obtained from an OLS regression of an estimate of $\boldsymbol{R}_j$ on $D_{jt}$, one has to construct a consistent estimator of  $V_1$ in order to construct confidence interval using the OLS estimator of $\boldsymbol{\alpha_0}$. Indeed since the uncorrected OLS variance estimator is a consistent estimator of $$\E(\boldsymbol{D}_{j}'\boldsymbol{D}_{j})^{-1}\E\left(\boldsymbol{D_{j}}'\left(\boldsymbol{R}_j-\boldsymbol{D_{j}\alpha_0}\right)\left(\boldsymbol{R}_j-\boldsymbol{D_{j}\alpha_0}\right)'\boldsymbol{D_{j}}\right)\E(\boldsymbol{D}_{j}'\boldsymbol{D}_{j})^{-1'},$$ it will only consistently estimate $V_1$ if the covariates used to create the value-added measures and the characteristics of interest are uncorrelated such that $\E\left(\boldsymbol{D_j'X_j}\right)=0$. Given that both $\boldsymbol{D_j}$ and $\boldsymbol{X_j}$ are observable, this condition can be tested using the sample equivalent of $\E\left(\boldsymbol{D_j'X_j}\right)$. A key point here is that $\boldsymbol{R}_j$  is a noisy measure of value added containing true teacher value added $\boldsymbol{\mu_{j}}$ that is correlated with $\boldsymbol{X_j}$ and the noise term $\boldsymbol{\epsilon_j}$. If $\boldsymbol{D_j}$ and $\boldsymbol{X_j}$ are correlated, consistent estimation of the first step which removes the effect of $\boldsymbol{X_j}$ is required to obtain a consistent estimator of $\boldsymbol{\alpha_0}$. Since ignoring the first steps in calculating standard errors is valid only if inconsistency in the first steps doesn't lead to inconsistency in later steps \citep{newey1994large}, one needs to account for the estimation of current year value added if $\boldsymbol{D_j}$ and $\boldsymbol{X_j}$ are correlated.

\section{Simulations \label{simul}}

I now illustrate my findings using a simulation study. This simulation will show that using unadjusted standard errors with the multi-step OLS estimator will lead to coverage rates that are too low for $95\%$ confidence intervals. It will also show that coverage deteriorates further as correlations between the covariates and true value added increase. On the other hand, the $95\%$ confidence intervals from using the multi-step OLS estimator with corrected standard errors from GMM perform well in all cases.  To demonstrate, I draw  1000 replication samples with a total sample size of $n=900,000$ observations holding the number of students per class $n_j$ and classes per teacher $T$  constant at 30 and 10 respectively, so that the sample contains $J=3,000$ distinct teachers who each teach 1 class a year for 10 years. The parameters of the simulations are as follows to allow the covariate to be correlated with value added:
\begin{enumerate}
\item $\mu_{jt} \sim N(0,0.01)$.
\item $\rho=0.5$.
\item $X=\frac{\rho\mu_{jt} + (1-\rho)\mathcal{N}(0,0.01)}{\sqrt{\rho^2+(1-\rho)^2}}$.
\item $R^{obs}_{it}= X_{it} +\mu_{jt} + \epsilon_{it}*U[0,2]$, where  $\epsilon \sim \mathcal{N}(0,0.81)$.
\item $Y^{obs}_{i}= 5+ 10X_{it} + \kappa_0\mu_{jt} + \eta_{it}*U[0,2]$, where  $\eta \sim \mathcal{N}(0,100)$.
\item $\kappa_0=100$.
\end{enumerate}

$\mu_{jt}$ has a standard deviation of 0.1, which is in line with estimates in \citet{chetty2014measuring1} and the simulation in \citet{chettyreply}. Test scores are constructed to be mean zero and standard deviation one, with $U[0,2]$ allowing for heteroskedasticity. Earnings and $\kappa_0$ are chosen to resemble the simulation in \citet{chettyreply} while allowing for heteroskedasticity. Finally, given that the simulation only uses one covariate, the correlation of this covariate with VA is set to be high at $0.5$ to obtain results that are in line with this paper's empirical application.

Each replication first estimates value added as described above and then estimates  \eqref{eq:estimation}, clustering standard errors at the teacher level. \medskip

Column 1 of Table \ref{table:simul2} presents the results when the true effect of test score value added on earnings is set to be 100, and  $\kappa_0$ is estimated using the multi-step OLS estimator and the standard errors are not corrected. Results show that the standard errors obtained from simply clustering and making no other adjustments when using OLS are incorrect. The estimated standard errors were far too small - on average, the standard error estimates were less than two thirds of the correct value. The inconsistent standard errors lead to incorrect coverage rates for $95\%$ confidence intervals, with these intervals  containing the true value $\kappa_0$ (100) only $72.4\%$ of the time.

\medskip

Column 2 of Table \ref{table:simul2} presents the results when the true effect of test score value added on earnings is set to be 100, and $\kappa_0$ is estimated using the multi-step OLS estimator with the corrected standard errors obtained from the GMM formula. Confidence intervals constructed using this estimator provide correct coverage. The corrected GMM standard errors account for the added variability resulting from the correlations between covariates and true value added, as such the estimated variance is close to the true variance of the estimator.
\medskip

\begin{figure}
\section{Figures}
\begin{center}
\caption{Coverage Rate using OLS SEs and GMM SEs with Different Correlation Levels}
\label{fig:cov}
		\includegraphics[width=0.7\linewidth]{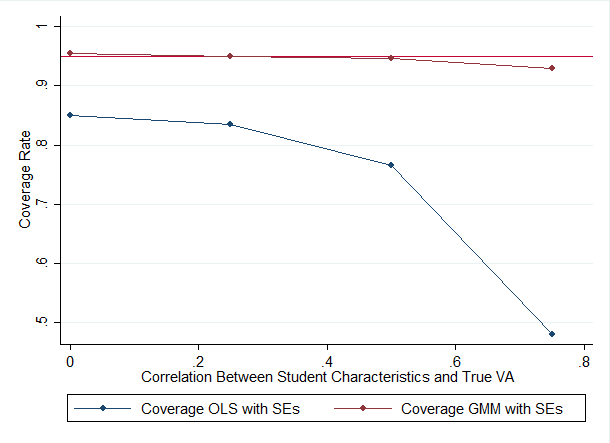}
		\end{center}
			\begin{itemize}
		\small\item[]{Each dot represents a coverage rate obtained by taking the average of an indicator of whether the true value $\kappa_0$ is in the estimated confidence interval for a 1000 replications with standard errors clustered at the teacher level. The number of students per class $n_j$ and classes per teacher $T$ are held constant at 30 and 10 respectively in all simulations, the correlation between student characteristics and true VA is increased by increasing the parameter $\rho$ from the data generating process described in section \ref{simul}. $\rho$ is set to be 0, 0.25 , 0.5, and 0.75. }
		\end{itemize}
		\end{figure}

\begin{figure}

\begin{center}
\caption{Actual Standard Deviation of $\widehat{\kappa}$ vs Standard Error Obtained from OLS}
 \label{fig:sd}
		\includegraphics[width=0.7\linewidth]{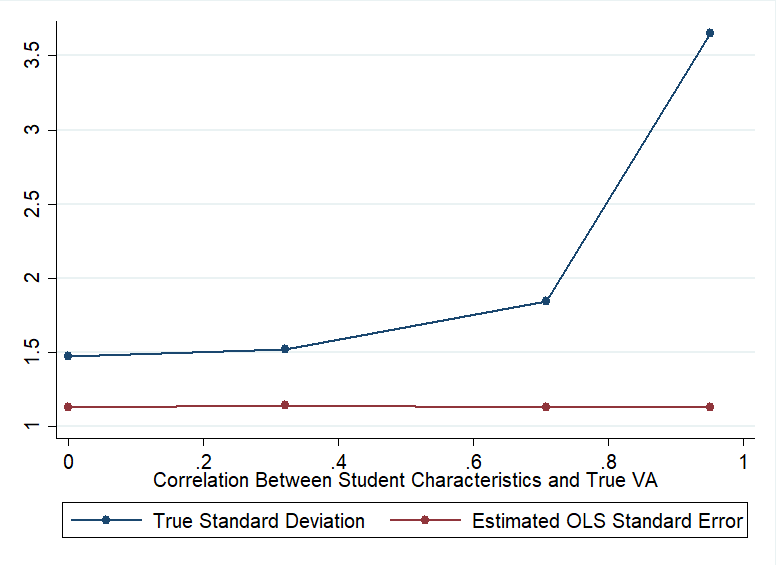}
		\end{center}
			\begin{itemize}
		\small\item[]{Each dot of the blue line represents the standard deviation of $\widehat{\kappa}$ obtained from a Monte-Carlo using a 1000 replications. Each dot of the red line represents the average of the standard errors estimated using OLS from 1000 replications with standard errors clustered at the teacher level. The number of students per class $n_j$ and classes per teacher $T$ are held constant at 30 and 10 respectively in all simulations, the correlation between student characteristics and true VA is increased by increasing the parameter $\rho$ from the data generating process described in section \ref{simul}. $\rho$ is set to be 0, 0.25 , 0.5, and 0.75. }
		\end{itemize}
		\end{figure}

To better illustrate the role of the correlations between the observable characteristics of students used to estimate value added and true teacher quality,  I let $\rho$ be equal 0 , 0.25, 0.5, and 0.75 and draw a 1000 replications for each value. For each set of replications, I repeat the previous exercise and obtain the coverage rate of the estimated confidence intervals from  using the multi-step OLS estimator with the incorrect and corrected standard errors. Figure \ref{fig:cov} presents the results of this exercise, while the actual coverage rate of confidence intervals obtained from OLS deteriorates as the correlation between observable characteristics and true quality increases, the confidence intervals constructed using the from the GMM standard errors perform well even when the correlation is set to be unrealistically high with $\rho=0.75$. To understand why the confidence intervals from OLS provide coverage well below their nominal rates as the correlation increases, I plot for each set of replications the actual standard deviation of $\widehat{\kappa}_{OLS}$ from the Monte-Carlo and the average standard errors obtained from OLS. Figure \ref{fig:sd} presents the results. While the increasing correlation leads the actual standard deviation of $\widehat{\kappa}_{OLS}$ to increase, the average standard errors obtained from OLS do not change since they do not take into account the correlation. This results in a larger drop of coverage for confidence intervals as the correlation increases. \medskip

Finally, Table \ref{table:simul3} presents the monte-carlo variances of the multi-step OLS estimator of $\kappa_0$ and the optimal GMM estimator of $\kappa_0$ from section \ref{section:overid}. The optimal GMM  has a slightly a lower variance than the multi-step OLS estimator, specifically it is  $1.3\%$ lower.\footnote{The small magnitude of the  difference could be due to the fact that the simulations have only a moderate amount of heteroskedasticity, no correlations between $\epsilon$ and $\eta$,  and constant VA over time which implies $\E[\boldsymbol{R}_{j}|\boldsymbol{R^{(-t)}_{j}}]$ is linear.}

\section{Application}

To further illustrate my results, I draw on administrative data for students in the North Carolina public schools, in the years 2000-2005. Specifically, I focus on third grade students in those years.\footnote{I limit my focus to third grade students from 2000 to 2005 in order to remain as close as possible to the theoretical setting of this paper. I am able to match over $70\%$ of those students to their teacher. Of those matched, none are missing covariates used for VA estimation, and I can observe long-run outcomes for over half of them. Furthermore, the results found in this paper are comparable to those reported in \citet{rothstein2017measuring} who considers students in grades 3 through 5 for the years 1997-2011.}  Students in grade three in North Carolina take end-of-grade tests in math as well as pre-tests in the Fall which will be used as lagged test scores. I standardize all scores within year-grade cell. \medskip

I start with 611,870 distinct students, in 1,313 schools. After excluding students with missing test scores, special education classes, classes with fewer than 10 students, and students that are not matched to their teachers, I am left with 444,262 distinct students matched to 8,210 teachers in 22,295 distinct classrooms. Given that the procedure described in section \ref{section:setup} is a leave-year out procedure, it excludes all teachers who only teach for a year. As such my final sample consists of 388,191 students matched to 5,266 teachers in 19,351 classrooms. I draw long-run outcomes for these students from high-school transcripts (graduation, GPA, class rank), end-of-course algebra scores in high-school, and exit surveys (college plans).  \medskip

The summary statistics for the sample are presented in Table \ref{tab:sumstat}. Half of the students in my sample are female, around $34\%$ of students are Black or Hispanic, and around $61\%$ of students are white. Only $3\%$ of students are English language learners, whereas $10\%$ are special education students. Of students matched to their long-run outcomes, about $91\%$ graduate from high school, $78\%$ plan on attending college, and $41\%$ plan on attending a 4-year college. \medskip

The remainder of this section is organized as follows. First, I will illustrate the results of section $\ref{section:estim-inf}$ by showing that that the  standard errors routinely used in the literature are incorrect. Next, to understand why the unadjusted standard errors are incorrect, I will show that there is evidence of unconditional sorting i.e strong correlations between the variables used to predict VA and teacher VA. In doing so I will also illustrate the results of section \ref{sec:outcome}.   \medskip

\subsection{Correcting Inference}

To begin, I estimate value-added measures following the procedure laid out in section \ref{section:setup}, notably estimating $\beta_0$ using only within teacher variation. I use a rich vector of student controls that is similar to \citet{rothstein2017measuring} and contains: cubic polynomials in prior scores, gender, age, indicators for special education, limited English, year, lunch eligibility, ethnicity, as well as class- and school-year means of those variables. Table \ref{tab:va-stats} describes the generated measures, there are 19,351 distinct measures corresponding to the 19,351 classrooms. The measures are mean zero and have a standard deviation of $0.177$ such that a one standard deviation increase in estimated VA corresponds to a $17.7\%$ increase in test scores.\footnote{Assuming that the VA measures are forecast unbiased.} \medskip

 It then follows from Theorem \ref{prop:kappa-dist} that when using the estimated VA measures in a regression on long-run outcomes, we must adjust standard errors to obtain the true variance of $\kappa_0$. I focus on estimating the impact of teacher VA on a set of long-run outcomes for the third grade students, namely:  high-school algebra scores, high-school graduation, plan to attend college, plan to attend a 4 year college, high-school GPA,  and high-school class rank. \medskip

I estimate the effect of teacher VA on those outcomes following the methodology in section \ref{section:setup}. The results are found in Table \ref{tab:long-run}. I find that a one standard deviation increase in teacher VA in third grade leads to a $3.84$ percent of a standard deviation increase in high-school algebra scores, a $0.5$ percentage point increase in high-school graduation, a $0.88$ percentage point increase in college enrollment plans, a $1.76$ percentage point increase in 4-year college enrollment plans, a $0.036$ point increase in weighted high-school GPA, and a $0.93$ percentage point increase in high-school class rank. The magnitude of these impacts are similar to the ones obtained by  \citet{rothstein2017measuring} using a larger sample of students from North Carolina, and the ones of  \citet{chetty2014measuring} in New York. \medskip

Importantly, the unadjusted standard errors obtained from simply running an OLS regression and clustering standard errors at the teacher level are incorrect. Indeed, comparing the unadjusted standard errors to those   obtained from a GMM estimation of the system shows that the unadjusted standard errors are too small. The clustered standard errors from GMM are $37\%$ to $70\%$ larger than their unadjusted clustered OLS counterparts. Given that the magnitude of the effects are relatively large, the coefficients remain statistically significant even though their t-values decrease substantially. For example, the t-statistic for graduation drops from 7.14 to 4.16, and the t-statistic for planning to attend college drops from 8.18 to 5. In applications for which estimated effects are not so large, this drop could mean the difference between statistically significant and insignificant results. To put this change into perspective, Table \ref{tab:long-run} also gives the heteroskedasticity robust standard errors obtained from running an OLS regression. For all but one outcome, the difference between clustered standard errors obtained by running GMM and clustered standard errors obtained by running OLS is at least as large as the difference between clustered and heteroskedasticity robust standard errors obtained by running OLS. \medskip

To summarize, adjusting standard errors to account for correlations between teacher VA and the controls, and for the estimation of VA is likely to be important in practice. In this application on data from North Carolina, the increase in standard errors resulting from the adjustment is on average larger than the impact of clustering standard errors.\footnote{In this application, I cluster standard errors at the teacher level. If one were to cluster standard errors at a higher level, say school or school-year, the results should be similar. Indeed,  \citet{rothstein2017measuring} finds that correlations between VA and controls is stronger between schools than within schools, stating that schools with higher VA teachers have much higher prior year test scores and better socioeconomic conditions.}

\subsection{Presence and Effects of Unconditional Sorting \label{section:sorting}}

The previous section has shown that the unadjusted standard errors obtained from simply running OLS are too small. The theoretical results imply that this is likely due to significant correlations between teacher VA and the vector of covariates used to estimate the measures. To examine that, I empirically test for the presence of these correlations. \medskip

Similar to \citet{chetty2014measuring1}, given that my VA measures are estimated using within teacher variation in the controls, I can use these measures to estimate the correlations between true VA and the controls. \citet{chetty2014measuring1} do so by running a univariate regression of the VA measures on lagged test-scores and other covariates.\footnote{To account for the attenuation resulting from the fact that the VA measures are shrunk, they multiply their coefficients by an estimate $\frac{SD(\mu_{jt})}{SD(\widehat{\mu_{jt}})}=1.56$ in their data. This ratio is 1.17 in my data and I adjust my estimates accordingly when running the analysis.} They find positive and statistically significant but small evidence of unconditional sorting on lagged test-scores, better students are assigned slightly better teachers. Given that their point estimate for unconditional sorting is small, and that they obtain VA measures estimated using within and between teacher variation that are highly correlated (0.979) with their original measures estimated using only within teacher variation, they conclude that unconditional sorting is relatively minimal in practice.\medskip

I conduct a similar analysis. I find that the degree of unconditional sorting is not minimal but occurs at the group rather than individual student level. As such the class and school-year level means included in the estimation of VA  are strong predictors of teacher VA, these findings are consistent with \citet{rothstein2017measuring}. Furthermore, I show that although VA measures estimated using both within and between teachers variation can be highly correlated with the baseline measures using only within teacher variation, they significantly understate the effect of certain teachers and are poor predictors of long-run outcomes. Thus I find that using within teacher variation as proposed by \citet{chetty2014measuring1} is important in practice.\medskip

I estimate the degree of unconditional sorting in two ways. First, I follow \citet{chetty2014measuring1} by regressing the VA measures on different controls then adjusting the coefficient to account for shrinkage, but correct inference by obtaining the standard errors by bootstrapping the GMM system.\footnote{I use the bootstrap on the GMM system since the standard errors in Theorem \ref{prop:alpha-dist} are for a system that regresses preliminary VA measures, not the shrunk measures,x on covariates.} Second, I make use of Theorem \ref{prop:alpha-dist} and run a regression of the preliminary VA measures on controls by GMM estimation of the system in Result \ref{lem:as-outcome}. The results of this are presented in Table \ref{tab:correls}. \medskip

Panel A presents the results from the regressions of the VA measures on three different controls: student level lagged test scores, classroom mean lagged test score, and school year mean lagged test score. Surprisingly, the highly significant estimate for unconditional sorting on student level lagged test scores of 0.010 (Column (1), Table  \ref{tab:correls} Panel A)  is similar to the estimate found by \citet{chetty2014measuring1} of 0.012, even though the two analyses use different data sets. This suggests a small degree of unconditional sorting on student level past test scores; better students are matched with higher VA teachers. I then further examine this by also regressing the VA measures on the class and school-year average of lagged test scores. I find that sorting on test scores at the class and school-year level is significantly stronger than at the individual level. Notably the point estimate of 0.056 for classroom level lagged test scores is highly significant and five times larger than the one at the individual level. Finally, one can see that the unadjusted standard errors from OLS are incorrect, and that the standard errors obtained from bootstrapping the GMM system are about $50\%$ larger.  \medskip

I confirm these findings with a direct application of Theorem \ref{prop:alpha-dist}. I first estimate a regression of preliminary VA, $\widehat{R}_{jt}$, on classroom mean lagged test score, then re-estimate the parameter using the GMM system as described in Theorem \ref{prop:alpha-dist}. The results correspond to Panel B of  Table  \ref{tab:correls}.  The point estimate of 0.073 is larger than the one in Panel A, but they are statistically indistinguishable. Again, the unadjusted standard errors are smaller than the ones obtained by GMM, with the GMM standard errors being approximately $80\%$ larger. The difference between the standard errors is directly explained by Theorem \ref{prop:alpha-dist}. Here the vector of variables $\boldsymbol{D}_{j}$ is a subset of the covariates used to construct VA, $\boldsymbol{X}_{j}$, as such it must be that $\E\left(\boldsymbol{D_j'X_j}\right)\neq 0$. \medskip

In summation, the results in Table \ref{tab:correls} point to a strong correlation between the covariates used to estimate VA and true teacher VA. Consequently, this means that estimating $\beta_0$ without teacher fixed effects, using  both between and within teacher variation, will lead to biased estimates and therefore incorrect VA measures and point estimates for long run impacts. To show this, I estimate another set of VA measures without including teacher fixed effects. At first glance, it seems that these measures are very similar to the ones obtained using within teacher variation only. Indeed, Table \ref{tab:cor-is-bull} shows that the measures have a very high correlation of 0.965.\medskip

\begin{figure}

\begin{center}
\caption{Baseline VA vs VA Using All Variation}
 \label{fig:vas}
		\includegraphics[width=0.7\linewidth]{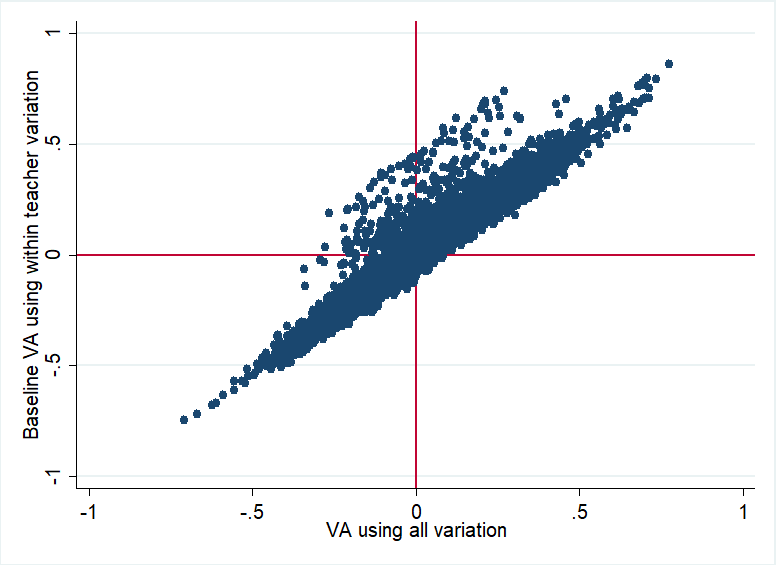}
		\end{center}
			\begin{itemize}
		\small\item[]{This graphs plots the baseline VA measures, constructed using an estimator of $\beta_0$ that was estimated using teacher fixed-effects, against VA measures constructed using an estimator of $\beta_0$ without fixed effects. Controls used are: cubic polynomials in prior scores; gender; age; indicators for special education, limited English, year, lunch eligibility, ethnicity; as well as class- and school-year means of those variables.}
		\end{itemize}
		\end{figure}

\begin{figure}

\begin{center}
\caption{Estimates of Long-Run Impacts}
 \label{fig:wrong-long}
		\includegraphics[width=0.7\linewidth]{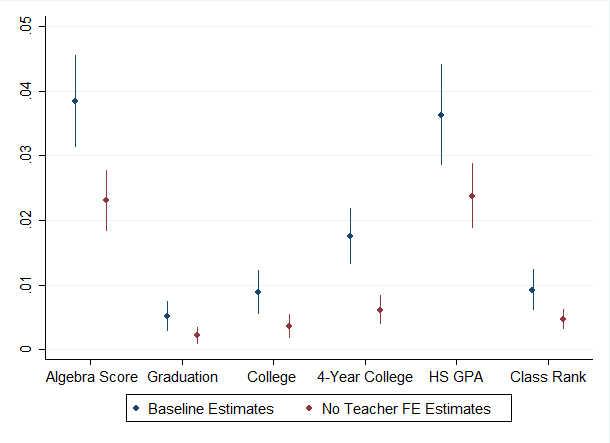}
		\end{center}
			\begin{itemize}
		\small\item[]{This graphs plots the effect of a one standard deviation increase in teacher VA on different long-run outcomes. They are obtained by a univariate regression of the residualized outcome on teacher VA, following the methodology of section \ref{section:setup}.  The estimates in blue are the estimates of $\kappa_0$ for different outcomes using the exact methodology in section \ref{section:setup}, while the estimates in red are the estimates of $\kappa_0$ without including teacher fixed effects to estimate $\beta_0$ and $\beta_0^Y$. Standard errors are clustered at the teacher level. The standard errors for the baseline estimates are obtained by GMM, the standard errors for the other estimates are unadjusted. Controls for estimation of VA and residualization of outcome include cubic polynomials in prior scores; gender; age; indicators for special education, limited English, year, lunch eligibility, ethnicity; as well as class- and school-year means of those variables.}
		\end{itemize}
		\end{figure}

However, further examination shows that the measures estimated without teacher fixed effects will dramatically understate the effectiveness of certain teachers. As shown in the upper left quadrant of Figure \ref{fig:vas}, these measures are sometimes negative for teachers that have positive VA when estimated using within teacher variation. This in turn will lead to biased estimates of $\kappa_0$, Figure \ref{fig:wrong-long} illustrates this point. It presents the results of regressions of a variety of long-run outcomes on teacher VA, following the procedure laid out in section \ref{section:setup}. One set of results is obtained by estimating $\beta_0$ and $\beta_0^Y$ using teacher fixed effects, and the other set is obtained by estimating $\beta_0$ and $\beta_0^Y$ without fixed effects. The estimates in blue are the estimates of $\kappa_0$ for different outcomes using the exact methodology in section \ref{section:setup}, while the estimates in red are the estimates of $\kappa_0$ without including teacher fixed effects to estimate $\beta_0$ and $\beta_0^Y$. It is clear that although the measures obtained without fixed effects are highly correlated with the baseline measures, they systematically underestimate $\kappa_0$.

\section{Conclusion}

In this paper, I consider how to correctly and efficiently draw inference in models using value-added measures in regressions. Starting with models using value-added measures as an explanatory variable, I show that they can be reframed as GMM systems, and use that to construct corrected standard errors for regressions with value-added on the right hand side. I then show that these models can also be written as  systems resembling instrumental variables where the preliminary value-added measures for years $s\neq t$ serve as instruments for the preliminary measure in year $t$. Then when one has more than two years of data, there are multiple instruments available for the measure in year $t$. I use this overidentifying information to propose a more efficient estimator for the impact of value-added on long-run outcomes using optimal GMM, and to propose a specification test for these models. For regressions using value-added measures as an outcome, I derive corrected standard errors from GMM and provide a testable condition under which unadjusted standard errors can also lead to valid inference. The theoretical results of this paper are checked using a simulation study. Finally, I demonstrate the practical implications of my results in an application on data from North Carolina public schools. I first document the presence of correlations between teacher test-score value added and student observables, and then show that adjusting  standard errors to account for those correlations and the estimation of VA measures is relevant in practice.
\newpage
\bibliography{biblio}

\begin{thebibliography}{}

\bibitem[Angrist et~al., 2017]{angrist2017leveraging}
Angrist, J.~D., Hull, P.~D., Pathak, P.~A., and Walters, C.~R. (2017).
\newblock Leveraging lotteries for school value-added: Testing and estimation.
\newblock {\em The Quarterly Journal of Economics}, 132(2):871--919.

\bibitem[Arcidiacono et~al., 2017]{arcidiacono2017productivity}
Arcidiacono, P., Kinsler, J., and Price, J. (2017).
\newblock Productivity spillovers in team production: Evidence from
  professional basketball.
\newblock {\em Journal of Labor Economics}, 35(1):191--225.

\bibitem[Bertrand and Schoar, 2003]{bertrand2003managing}
Bertrand, M. and Schoar, A. (2003).
\newblock Managing with style: The effect of managers on firm policies.
\newblock {\em The Quarterly journal of economics}, 118(4):1169--1208.

\bibitem[Canaan et~al., 2021]{canaan2019VA}
Canaan, S., Deeb, A., and Mouganie, P. (2021).
\newblock Advisor value-added and student outcomes: Evidence from randomly
  assigned college advisors.
\newblock {\em American Economic Journal: Economic Policy}.

\bibitem[Chamberlain, 1987]{chamberlain1987asymptotic}
Chamberlain, G. (1987).
\newblock Asymptotic efficiency in estimation with conditional moment
  restrictions.
\newblock {\em Journal of econometrics}, 34(3):305--334.

\bibitem[Chetty et~al., 2014a]{chetty2014measuring1}
Chetty, R., Friedman, J.~N., and Rockoff, J.~E. (2014a).
\newblock Measuring the impacts of teachers i: Evaluating bias in teacher
  value-added estimates.
\newblock {\em American Economic Review}, 104(9):2593--2632.

\bibitem[Chetty et~al., 2014b]{chetty2014measuring}
Chetty, R., Friedman, J.~N., and Rockoff, J.~E. (2014b).
\newblock Measuring the impacts of teachers ii: Teacher value-added and student
  outcomes in adulthood.
\newblock {\em American economic review}, 104(9):2633--79.

\bibitem[Chetty et~al., 2017]{chettyreply}
Chetty, R., Friedman, J.~N., and Rockoff, J.~E. (2017).
\newblock Measuring the impacts of teachers: Reply.
\newblock {\em American Economic Review}, 107(6):1685--1717.

\bibitem[Currie and Zhang, 2021]{currie2021doing}
Currie, J. and Zhang, J. (2021).
\newblock Doing more with less: Predicting primary care provider effectiveness.
\newblock Technical report, National Bureau of Economic Research.

\bibitem[Gilraine et~al., 2020]{gilraine2020}
Gilraine, M., Gu, J., and McMillan, R. (2020).
\newblock A new method for estimating teacher value-added.
\newblock Working Paper 27094, National Bureau of Economic Research.

\bibitem[Hansen, 1982]{hansen1982large}
Hansen, L.~P. (1982).
\newblock Large sample properties of generalized method of moments estimators.
\newblock {\em Econometrica: Journal of the Econometric Society}, pages
  1029--1054.

\bibitem[Jackson, 2018]{jackson2018test}
Jackson, C.~K. (2018).
\newblock What do test scores miss? the importance of teacher effects on
  non--test score outcomes.
\newblock {\em Journal of Political Economy}, 126(5):2072--2107.

\bibitem[Jacob et~al., 2010]{jacob2010persistence}
Jacob, B.~A., Lefgren, L., and Sims, D.~P. (2010).
\newblock The persistence of teacher-induced learning.
\newblock {\em Journal of Human resources}, 45(4):915--943.

\bibitem[Kane and Staiger, 2008]{kane2008estimating}
Kane, T.~J. and Staiger, D.~O. (2008).
\newblock Estimating teacher impacts on student achievement: An experimental
  evaluation.
\newblock Technical report, National Bureau of Economic Research.

\bibitem[Lazear et~al., 2015]{lazear2015value}
Lazear, E.~P., Shaw, K.~L., and Stanton, C.~T. (2015).
\newblock The value of bosses.
\newblock {\em Journal of Labor Economics}, 33(4):823--861.

\bibitem[Liu and Loeb, 2021]{liu2021engaging}
Liu, J. and Loeb, S. (2021).
\newblock Engaging teachers measuring the impact of teachers on student
  attendance in secondary school.
\newblock {\em Journal of Human Resources}, 56(2):343--379.

\bibitem[Mulhern, 2019]{mulhern2019beyond}
Mulhern, C. (2019).
\newblock Beyond teachers: Estimating individual guidance counselors’ effects
  on educational attainment.

\bibitem[Newey and McFadden, 1994]{newey1994large}
Newey, K. and McFadden, D. (1994).
\newblock Large sample estimation and hypothesis.
\newblock {\em Handbook of Econometrics, IV, Edited by RF Engle and DL
  McFadden}, pages 2112--2245.

\bibitem[Newey, 1984]{newey1984method}
Newey, W.~K. (1984).
\newblock A method of moments interpretation of sequential estimators.
\newblock {\em Economics Letters}, 14(2-3):201--206.

\bibitem[Opper, 2019]{opper2019does}
Opper, I.~M. (2019).
\newblock Does helping john help sue? evidence of spillovers in education.
\newblock {\em American Economic Review}, 109(3):1080--1115.

\bibitem[Pagan, 1986]{pagan1986two}
Pagan, A. (1986).
\newblock Two stage and related estimators and their applications.
\newblock {\em The Review of Economic Studies}, 53(4):517--538.

\bibitem[Rose et~al., 2019]{rose2019}
Rose, E., Schellenberg, J., and Shem-Tov, Y. (2019).
\newblock The effects of teacher quality on criminal behavior.

\bibitem[Rothstein, 2017]{rothstein2017measuring}
Rothstein, J. (2017).
\newblock Measuring the impacts of teachers: Comment.
\newblock {\em American Economic Review}, 107(6):1656--84.

\bibitem[Wooldridge, 2010]{wooldridge2010econometric}
Wooldridge, J.~M. (2010).
\newblock {\em Econometric analysis of cross section and panel data}.
\newblock MIT press.

\end{thebibliography}

\clearpage
\appendix

\section{Tables}

%

\clearpage

\begin{table}
		\centering
		\caption{Simulation Evidence of the Properties of $\widehat{\kappa}$ }
			\label{table:simul2}
			\begin{tabular}{l*{3}{c}}
				\toprule
				&\multicolumn{1}{c}{OLS}&\multicolumn{1}{c}{ OLS}  \\
                    & \multicolumn{1}{c}{Current Practice} &\multicolumn{1}{c}{Corrected SEs  }  \\
                 &\multicolumn{1}{c}{(1)}&\multicolumn{1}{c}{(2)}\\
				\midrule
								\addlinespace
				\multicolumn{1}{l}{\textbf{ n=900,000 and J=3,000 }} \\
	\midrule
\addlinespace
              SD from Monte-Carlo& 2.156 &2.156\\
              Average SD of $\widehat{\kappa}$& 1.265 &2.100\\
                           	\addlinespace
                Coverage Rate of $95\%$ CI& 0.724 & 0.942\\
				\toprule
			\end{tabular}\\
			\begin{minipage}{12.0cm}
				\small
				Results are based on 1000 replications. The coverage rate is obtained by taking the average of an indicator of whether the true value $\kappa_0$ is in the estimated confidence interval for a 1000 replications with standard errors clustered at the teacher level. The number of students per class $n_j$ and classes per teacher $T$ are held constant at 30 and 10 respectively in all simulations.
			\end{minipage}

	\end{table}

\clearpage
\begin{table}
		\centering
		\caption{Simulation Evidence of the Properties of $\widehat{\kappa}$ }
			\label{table:simul3}
			\begin{tabular}{l*{3}{c}}
				\toprule
				&\multicolumn{1}{c}{OLS}&\multicolumn{1}{c}{ Optimal}  \\
                    & \multicolumn{1}{c}{Current Practice} &\multicolumn{1}{c}{GMM }  \\
                 &\multicolumn{1}{c}{(1)}&\multicolumn{1}{c}{(2)}\\
				\midrule
								\addlinespace
				\multicolumn{1}{l}{\textbf{ n=900,000 and J=3,000 }} \\
	\midrule
\addlinespace
                Variance from Monte Carlo& 4.652 & 4.592 \\
                           	\addlinespace

				\toprule
			\end{tabular}\\
			\begin{minipage}{10.5cm}
				\small
				Results are based on 1000 replications. The number of students per class $n_j$ and classes per teacher $T$ are held constant at 30 and 10 respectively in all simulations.
			\end{minipage}

	\end{table}

\clearpage

\begin{table}

		\centering
		
		\caption{Summary Statistics}	
		\vspace*{0.3cm}
		
		\label{tab:sumstat}
		
		{
			\def\sym#1{\ifmmode^{#1}\else\(^{#1}\)\fi}
			\begin{tabular}{l*{6}{c}}
				\toprule
				&\multicolumn{1}{c}{Mean}&\multicolumn{1}{c}{S.D.}&\multicolumn{1}{c}{Obs.}\\
				&\multicolumn{1}{c}{(1)}&\multicolumn{1}{c}{(2)}&\multicolumn{1}{c}{(3)}\\
				
				\midrule
				\addlinespace
				
				\multicolumn{1}{l}{\textbf{A. Student Level Short-Run Variables \ \ \ \ \ \ \ \ }} \\
				\addlinespace
				Math Test Score   & 0.060 & 0.974 & 388,191   \\
				Class Size  & 21.968  &  3.379 & 388,191  \\
				Female & 0.494  & 0.500 & 388,191 \\
				Lunch Eligibility &  0.446 & 0.497 & 388,191 \\
                Black & 0.284 & 0.451 & 388,191 \\
                Hispanic & 0.053&0.223 &388,191 \\
                White & 0.608& 0.488& 388,191 \\
                English Language Learner &0.032 &0.175 & 388,191  \\
                Special Education & 0.110 & 0.312 & 388,191 \\
				\addlinespace
				
				\midrule
				\addlinespace
				\multicolumn{1}{l}{\textbf{B. Student Level Long-Run Outcomes \ \ \ \ \ \ \ \ }} \\
				\addlinespace
				High-School Algebra Score  & 0.243 & 0.924 & 303,826  \\
				Graduate High School  & 0.907 & 0.290 & 280,542  \\
				Plan College &  0.788&  0.409 & 272,990\\
				Plan 4-Year College &  0.417& 0.493 & 272,987\\
				Weighted High-School GPA & 3.072 & 0.939 & 193,927\\
				Class Rank & 0.513 & 0.286 & 193,594\\
				\addlinespace
				
				\midrule
				\addlinespace
				\toprule
			\end{tabular}\\
				\begin{minipage}{14.0cm}
\small
The sample consists of 388,191 North Carolina public schools third grade students matched to 5,266 teachers in 19,351 classrooms in the years 2000-2005.
				\end{minipage}
		}
	\end{table}

	\begin{table}
		\centering
		\caption{Summary Statistics for VA measures}
		\label{tab:va-stats}
		{
			\def\sym#1{\ifmmode^{#1}\else\(^{#1}\)\fi}
			\begin{tabular}{l*{1}{c}}
				\toprule
				\midrule
				Mean of VA &  0.013 \\
				S.D of VA &  .177 \\
				Number of Observations     &       19,351       \\
				\toprule
			\end{tabular}\\
			
				\begin{minipage}{10.0cm}
\small
Summary statistics for the VA measures of 5,266 teachers in 19,351 classrooms in the years 2000-2005. They are estimated use within teacher variation following the procedure described in section \ref{section:setup}. Controls include cubic polynomials in prior scores, gender, age, indicators for special education, limited English, year, lunch eligibility, ethnicity, as well as class- and school-year means of those variables
				\end{minipage}
		}
	\end{table}

\begin{landscape}
			
			\begin{table}
				\centering
				\footnotesize
				\caption{Estimates of Long-Run Impacts}
				\label{tab:long-run}
				
				{
					\def\sym#1{\ifmmode^{#1}\else\(^{#1}\)\fi}
					\begin{tabular}{l*{7}{c}}
						\toprule
						&\multicolumn{1}{c}{Algebra Score }&\multicolumn{1}{c}{Graduation }&\multicolumn{1}{c}{Plan College }&\multicolumn{1}{c}{ Plan 4-Year College}&\multicolumn{1}{c}{HS GPA}&\multicolumn{1}{c}{Class Rank}\\
						\midrule
						\addlinespace
                        Teacher VA   & 0.038 &  0.005 & 0.009 & 0.018 &0.036&0.009\\
                        OLS Heteroskedasticity Robust SE&   (0.0013) &(0.0005) & (0.0007)& (0.0008)& (0.0017)& (0.0005)   \\
						OLS Clustered SE&   (0.0027) &(0.0007) & (0.0011)& (0.0015)& (0.0027)& (0.0010)   \\
                        GMM Clustered SE&   (0.0037) & (0.0012) & (0.0018)& (0.0022)& (0.0040)&(0.0016)\\
						\addlinespace
						\midrule
						$N$    &        303,733       &  280,456        &  272,907 & 272,904    &  193,867       & 193,535  \\
						\toprule
					\end{tabular}\\
					\begin{minipage}{20.5cm}
						\small
						Teacher VA is standardized. The results in this table are obtained by a univariate regression of the residualized outcome on teacher VA, following the methodology of section \ref{section:setup}.   Standard errors are clustered at the teacher level. Controls for estimation of VA and residualization of outcome include cubic polynomials in prior scores; gender; age; indicators for special education, limited English, year, lunch eligibility, ethnicity; as well as class- and school-year means of those variables.
					\end{minipage}
				}
			\end{table}
		\end{landscape}

\begin{landscape}
			\begin{table}
				\centering
				\caption{Estimates of Unconditional Sorting}
				\label{tab:correls}
				\begin{tabular}{l*{4}{c}}
					\toprule

					Dependent Variable &\multicolumn{3}{c}{Teacher VA}\\
					\midrule
					&\multicolumn{1}{c}{(1)}&\multicolumn{1}{c}{(2)}&\multicolumn{1}{c}{(3)}\\
                        \midrule
					 \multicolumn{1}{l}{\textbf{A. Regression of teacher VA measures on covariates }} \\
                    \midrule
					\addlinespace
					Lagged Test Scores  & 0.010& & \\
					OLS Standard Error & (0.0012)& & \\
                    Bootstrap Standard Error &(0.0018)& &  \\
					\addlinespace
					Classroom Mean Lagged Test Score  & & 0.056& \\
						OLS Standard Error & &(0.0065) &\\
                    Bootstrap Standard Error & &(0.0100)&\\
                    \addlinespace
                    School-Year Mean Lagged Test Score  & & & 0.041 \\
                    OLS Standard Error & & & (0.0083)\\
                    Bootstrap Standard Error & & &(0.0133)\\
					\addlinespace
                    \midrule
					$N$    & 388,191   &   388,191 &  388,191  \\
                     \midrule
                    \multicolumn{1}{l}{\textbf{B. Regression of preliminary teacher VA measures on covariates }} \\
					\midrule
                    \addlinespace
	               Classroom Mean Lagged Test Score  & 0.073\\
				    OLS Standard Error & (0.0063)\\
                    GMM Standard Error & (0.0114)\\
                    \addlinespace
					\midrule
					$N$ & 444,018\\
					\toprule
				\end{tabular}\\
				\begin{minipage}{20cm}
					\small
					Standard errors are clustered at the teacher level. Panel A presents the results of regressions of the estimated teacher VA measures on different controls. Bootstrapped standard errors are calculated by block bootstrapping the sample at the teacher level then estimating VA followed by the regressions, this is equivalent to bootstrapping the GMM system. Following \citet{chetty2014measuring1}, the coefficients and standard errors in panel A are multiplied by 1.17 to offset shrinkage of the dependent variable. Panel B presents the results of a regression of preliminary (unshrunk VA measure) on  classroom mean lagged test score following Theorem \ref{prop:alpha-dist}. OLS standard errors are the unadjusted estimators produced by statistical softwares, GMM standard errors are obtained  following Theorem \ref{prop:alpha-dist}. Observations are slightly higher in panel B because teachers who only teach one year can be included when following this methodology.
				\end{minipage}
				
			\end{table}
\end{landscape}

		\begin{table}
			\centering
			\caption{ Correlations Between Different VA measures}
			\label{tab:cor-is-bull}
			
			{
				\def\sym#1{\ifmmode^{#1}\else\(^{#1}\)\fi}
				\begin{tabular}{l*{3}{c}}
					\toprule
					&\multicolumn{1}{c}{Baseline VA }&\multicolumn{1}{c}{VA using all variation }\\
					\midrule
					\addlinespace
					Baseline VA     &  1.00 & 0.965 \\
					\addlinespace
					VA using all variation &   0.965  &  1.00   \\
					\addlinespace
					\toprule
				\end{tabular}\\
				\begin{minipage}{14.0cm}
					\small
					This table presents the two-way correlation coefficient between the baseline VA measures estimated using only within teacher variation, and the VA measures estimated using between and within teacher variation.
				\end{minipage}
			}
		\end{table}

\clearpage

\section{Heterogeneous Treatment Effects \label{het-effects}}

Suppose  that the potential outcome function for the adult earnings of student $i$ is given by

\begin{equation*}
Y^{pot}_{i}(\mu)= \kappa_i \mu + Y^{pot}_i(0)
\end{equation*}

where $Y^{pot}_i(0)$ is the same as before but $\kappa_i$ is stochastic varies across students. Then the true teacher-year level residual earnings are:

\begin{equation} \label{eq:res-het}
\overline{Y}_{jt}=  \kappa_{jt} \mu_{jt} + \overline{\eta}_{jt},
\end{equation}

where $\kappa_{jt}=\frac{1}{n_j}\sum_{i=1}^{n_j} \kappa_i$. Let $\boldsymbol{\kappa_j}$ be a vector stacking the $\kappa_{jt}$. I impose the following assumptions:

\begin{hyp}\label{hyp:het-effects} Heterogeneous Treatment Effects \\
\begin{enumerate}
\item $\E(\boldsymbol{\kappa_j})=\kappa^*<\infty$.
\item $\kappa_{jt} \indep \left(\boldsymbol{\mu}_j,\boldsymbol{\epsilon^{(-t)}_j}\right)$.
\end{enumerate}
\end{hyp}

Point 1 requires that the mean of the individual level effects be finite. Point 2 requires that the treatment effects in year $t$ be independent of a teacher's true VA and of the unobserved determinants of the teacher's other students in years $s\neq t$. \medskip

Consider the linear projection of $\overline{Y}_{jt}$ from  \eqref{eq:res-het} on $\mu_{jt}^*=\sum_{k\neq t}\phi_{0k} \overline{R}_{jk}$:

\begin{equation}
\overline{Y}_{jt}=  \kappa_0 \mu_{jt}^* + \overline{u}_{jt},
\end{equation}

where $\kappa_0=\frac{Cov(\overline{Y}_{jt}, \mu_{jt}^*)}{Var( \mu_{jt}^*)}$. Now to show that $\kappa_0=\E(\kappa_{jt})$:

\begin{align*}
\kappa_0&=\frac{Cov(\overline{Y}_{jt}, \mu_{jt}^*)}{Var( \mu_{jt}^*)}\\
&=\frac{Cov(\kappa_{jt} \mu_{jt} + \overline{\eta}_{jt}, \mu_{jt}^*)}{Var( \mu_{jt}^*)}\\
&=\frac{\E(\kappa_{jt} \mu_{jt} \mu_{jt}^*) + \E(\overline{\eta}_{jt}\mu_{jt}^*)}{\E( \mu_{jt}^*)}\\
&=\frac{\E(\kappa_{jt} \mu_{jt} \mu_{jt}^*)}{\E( \mu_{jt}^{*2})}\\
&=\frac{\E(\kappa_{jt})\E(\mu_{jt} \mu_{jt}^*)}{\E( \mu_{jt}^{*2})}\\
&=\E(\kappa_{jt})\\
\end{align*}

where the second equality follows from  \eqref{eq:res-het}. The third equality follows from the fact that $\mu_{jt}$ and $\mu_{jt}^*$ are mean zero. The fourth equality follows from Assumption \ref{hyp:rand-assign}. The fifth equality follows from Point 2 of Assumption \ref{hyp:het-effects}. The last equality follows from the fact that $\frac{\E(\mu_{jt} \mu_{jt}^*)}{\E( \mu_{jt}^*)}=1$ from Result \ref{lem:assumption}. \medskip

Then we still have $\E\left(\boldsymbol{\phi_0'\overline{R}^{(-t)'}_{j}}\left(\boldsymbol{Y}_{j}- \kappa_0 \boldsymbol{R^{(-t)}_{j}\phi_0} \right)\right)=0$ with $\kappa_0=\E(\kappa_{jt})$, then the non-optimal GMM estimator is robust to heterogeneous treatment effects under Assumption \ref{hyp:het-effects}. \medskip

%
%

\section{Shrinkage \label{section:shrinkage}}

Consider a simple case where value added is constant over time such that:

\begin{equation}
R_{it}=R^{obs}_{it} -  X_{it}'\beta_0=\mu_{j} + \epsilon_{it}
\end{equation}

and
\begin{equation}
\overline{R}_{jt}= \frac{1}{n_{j}}\sum_{i=1}^{n_{j}}R_{it}=\mu_{j} +\overline{\epsilon}_{jt}.
\end{equation}

Consider the best linear predictor of  $\overline{R}_{jt}$ using one other year $\overline{R}_{jt'}$:

\begin{equation}
\mu_{jt}^*=\phi_{0}\overline{R}_{jt'}
\end{equation}

where:
\begin{equation}
\phi_{0}=\frac{Cov(\overline{R}_{jt},\overline{R}_{jt'})}{Var(\overline{R}_{jt'})}.
\end{equation}

Then under Assumption \ref{hyp:rand-assign} we have:

\begin{align*}
\phi_{0}&=\frac{Cov(\overline{R}_{jt},\overline{R}_{jt'})}{Var(\overline{R}_{jt'})}\\
&= \frac{Cov(\mu_{j} +\overline{\epsilon}_{jt},\mu_{j} +\overline{\epsilon}_{jt'})}{Var(\mu_{j} +\overline{\epsilon}_{jt'})}\\
&= \frac{Cov(\mu_{j},\mu_{j})}{Var(\mu_{j}) +Var(\overline{\epsilon}_{jt'})}\\
&= \frac{Var(\mu_{j})}{Var(\mu_{j})+ Var(\overline{\epsilon}_{jt'})}<1
\end{align*}

where the second equality follows from the fact that $\mu_j$ is uncorrelated with $\overline{\epsilon}_{jt'}$ and $\overline{\epsilon}_{jt}$ by Point 1 of Assumption \ref{hyp:rand-assign}, and the fact that $\overline{\epsilon}_{jt'}$ and $\overline{\epsilon}_{jt}$ are uncorrelated by Point 3 of Assumption \ref{hyp:rand-assign}. \medskip

This example shows that when value added is constant, $\mu_{jt}^*$ is a shrinkage estimator similar to the one proposed by \citet{kane2008estimating}. One can show that the measures will still be shrunk towards the mean of zero when value added is not constant over time and more years are used.

\section{Alternative Identification Proof \label{sec:extraID}}

\begin{appxres} \label{prop:ID}
If Assumptions \ref{hyp:rand-assign} and \ref{hyp:tech-id} hold, then $\kappa_0$ is identified.
\end{appxres}

By Points 1 and 3 of Assumption \ref{hyp:tech-id}, $\boldsymbol{\beta}_0$ and $\boldsymbol{\beta}_0^Y$ are identified by the coefficients on $\boldsymbol{X_j}$ in a regression of $\boldsymbol{R}_j^{obs}$ and  $\boldsymbol{Y}_j^{obs}$ respectively on $\boldsymbol{X_j}$ and teacher fixed effects. Namely:
\begin{eqnarray*}
\boldsymbol{\beta}_0=\E(\boldsymbol{\ddot{X}}_{j}'\boldsymbol{\ddot{X}}_{j})^{-1}\E(\boldsymbol{\ddot{X}}_{j}'\boldsymbol{\ddot{R}}^{obs}_{j}) \\
\boldsymbol{\beta}_0^Y= \E(\boldsymbol{\ddot{X}}_{j}'\boldsymbol{\ddot{X}}_{j})^{-1}\E(\boldsymbol{\ddot{X}}_{j}'\boldsymbol{\ddot{Y}}^{obs}_{j})
\end{eqnarray*}

since the a regression on the variables using the within transform is equivalent to a regression with fixed effects.

Then starting with  \eqref{eq:scores}:
\begin{equation*}
R^{obs}_{it}= \boldsymbol{ X_{it}'\beta_0} + \mu_{jt} +\epsilon_{it},
\end{equation*}

and let:
\begin{equation}
R_{it}=R^{obs}_{it} - \boldsymbol{ X_{it}'\beta_0}=\mu_{jt} + \epsilon_{it}
\end{equation}

be the actual residual score, which is then collapsed to the teacher year level:
\begin{equation}
\overline{R}_{jt}= \frac{1}{n_{j}}\sum_{i=1}^{n_{j}}R_{it}=\mu_{jt} +\overline{\epsilon}_{jt},
\end{equation}

we can write the best linear prediction of $\overline{R}_{jt}$ as a function of other years as:
\begin{equation}
\overline{R}_{jt}= \sum_{\abs{s-t}\neq 0}\phi_{0\abs{s-t}} \overline{R}_{j\abs{s-t}} + \theta_{jt}
\end{equation}

so that:
\begin{equation}\label{eq:plug-in}
\mu_{jt}=\sum_{\abs{s-t}\neq 0}\phi_{0\abs{s-t}} \overline{R}_{j\abs{s-t}} + \theta_{jt} - \overline{\epsilon}_{jt}.
\end{equation}

Plugging  \eqref{eq:plug-in} into  \eqref{eq:earningres} and letting  $\mu_{jt}^*=\sum_{s\neq 0}\phi_{0s} \overline{R}_{js}$ yields:
\begin{equation} \label{eq:for-mom4}
\overline{Y}_{jt} = \kappa_0 \mu_{jt}^* + \kappa_0 \theta_{jt} - \kappa_0 \overline{\epsilon}_{jt} +\overline{\eta}_{jt},
\end{equation}

where $\mu_{jt}^*$ is identified. Therefore:

\begin{eqnarray}
\frac{Cov\left(\overline{Y}_{jt},\mu_{jt}^*\right)}{Var\left(\mu_{jt}^*\right)}&=& \kappa_0 + \kappa_0 \frac{Cov\left(\theta_{jt},\mu_{jt}^*\right)}{Var\left(\mu_{jt}^*\right)} - \kappa_0 \frac{Cov\left(\overline{\epsilon}_{jt},\mu_{jt}^*\right)}{Var\left(\mu_{jt}^*\right)}+
\frac{Cov\left(\overline{\eta}_{jt},\mu_{jt}^*\right)}{Var\left(\mu_{jt}^*\right)} \nonumber \\
&=& \kappa_0
\end{eqnarray}
where the second equality holds because $\theta_{jt}$ is the error from  \eqref{eq:decomp} and the variables are mean zero, and:
\begin{align*}
Cov\left(\overline{\epsilon}_{jt},\mu_{jt}^*\right)&=Cov\left(\overline{\epsilon}_{jt},\sum_{\abs{s-t}\neq 0}\phi_{0\abs{s-t}} \overline{R}_{js}\right) \\
&=Cov\left(\overline{\epsilon}_{jt},\sum_{\abs{s-t}\neq 0}\phi_{0\abs{s-t}}(\mu_{js} +\overline{\epsilon}_{js})\right)\\
&=Cov\left(\overline{\epsilon}_{jt},\sum_{\abs{s-t}\neq 0}\phi_{0\abs{s-t}}\mu_{js}\right)+Cov\left(\overline{\epsilon}_{jt},\sum_{\abs{s-t}\neq 0}\phi_{0\abs{s-t}}\overline{\epsilon}_{js}\right)\\
&=0
\end{align*}
where  both terms are 0 by Assumption \ref{hyp:rand-assign}.   \medskip

To show $Cov\left(\overline{\eta}_{jt},\mu_{jt}^*\right)=0$:
\begin{align*}
Cov\left(\overline{\eta}_{jt},\mu_{jt}^*\right)&=Cov\left(\overline{\eta}_{jt},\sum_{\abs{s-t}\neq 0}\phi_{0\abs{s-t}} \overline{R}_{js}\right) \\
&=Cov\left(\overline{\eta}_{jt},\sum_{\abs{s-t}\neq 0}\phi_{0\abs{s-t}}(\mu_{js} +\overline{\epsilon}_{js})\right)\\
&=Cov\left(\overline{\eta}_{jt},\sum_{\abs{s-t}\neq 0}\phi_{0\abs{s-t}}\mu_{js}\right)+Cov\left(\overline{\epsilon}_{jt},\sum_{\abs{s-t}\neq 0}\phi_{0\abs{s-t}}\overline{\epsilon}_{js}\right)\\
&=0
\end{align*}

where both terms are 0 by Assumption \ref{hyp:rand-assign}. \medskip

\textbf{QED.} \medskip

\section{Regularity Conditions \label{regularity}}

\begin{appxhyp}\label{hyp:tech-cons} Technical Assumptions for Consistency \\
\begin{enumerate}
\item  $(\boldsymbol{\beta_0'}, \boldsymbol{\beta_0^{Y'}},\boldsymbol{\phi_0'},\kappa_0)$ is an element in the interior of $\boldsymbol{\Theta}$ and $\boldsymbol{\Theta}$ is a compact subset of $\mathcal{R}^{2K+T}$.
\item\begin{flalign*}
&\E\left(sup_{(\boldsymbol{\beta}, \boldsymbol{\beta^Y},\boldsymbol{\phi},\kappa) \in \Theta}\left|\left|\boldsymbol{\ddot{X}}_{j}'\left(\boldsymbol{\ddot{R}}^{obs}_{j}- \boldsymbol{\ddot{X}_{j}\beta}\right)\right|\right|\right) < \infty\\
&\E\left(sup_{(\boldsymbol{\beta}, \boldsymbol{\beta^Y},\boldsymbol{\phi},\kappa) \in \Theta}\left|\left|\boldsymbol{\widetilde{R}^{(-t)'}_{j}}\left(\boldsymbol{\widetilde{R}}_{j}- \boldsymbol{\widetilde{R}^{(-t)}_{j}\phi}\right)\right|\right|\right) <\infty\\
&\E\left(sup_{(\boldsymbol{\beta}, \boldsymbol{\beta^Y},\boldsymbol{\phi},\kappa) \in \Theta}\left|\left|\boldsymbol{\ddot{X}}_{j}'\left(\boldsymbol{\ddot{Y}}^{obs}_{j}- \boldsymbol{\ddot{X}_{j}\beta^Y}\right)\right|\right|\right)< \infty\\
&\E\left( sup_{(\boldsymbol{\beta}, \boldsymbol{\beta^Y},\boldsymbol{\phi},\kappa) \in \Theta}\left|\boldsymbol{\phi'\widetilde{R}^{(-t)'}_{j}}\left(\boldsymbol{\widetilde{Y}}_{j}- \kappa \boldsymbol{\widetilde{R}^{(-t)}_{j}\phi} \right)\right|\right)<\infty
\end{flalign*}
\item $\left(\boldsymbol{X}_{j},\boldsymbol{\mu}_j, \boldsymbol{\epsilon}_j, \boldsymbol{Y}^{obs}_{j}, \boldsymbol{\eta}_j\right)$ are i.i.d across $j$.
\end{enumerate}
where $\widetilde{R}_{jt}=\overline{R}^{obs}_{jt}-\boldsymbol{\overline{X}_{jt}'\beta}$, $\boldsymbol{\widetilde{R}}_{j}$ is a vector stacking all $\widetilde{R}_{jt}$ for teacher $j$, $\boldsymbol{\widetilde{R}^{(-t)}_{j}}$ is  a matrix stacking $T$ row vectors (of dimension $1\times(T-1)$) with each row (indexed by $t$) containing $T-1$ different $\widetilde{R}_{jk}$ for $k\neq t$.
\end{appxhyp}

\medskip \medskip

\begin{appxhyp}\label{hyp:tech-norm} Technical Assumptions for Asymptotic Normality and Consistent Variance Estimation \\
\begin{enumerate}
\item For all $(\boldsymbol{\beta'}, \boldsymbol{\beta^{Y'}},\boldsymbol{\phi'},\kappa) \in \mathcal{R}^{2K+T}$: \begin{align*}
&E\left(\boldsymbol{\ddot{X}}_{j}'\left(\boldsymbol{\ddot{R}}^{obs}_{j}- \boldsymbol{\ddot{X}_{j}\beta}\right)\left(\boldsymbol{\ddot{R}}^{obs}_{j}- \boldsymbol{\ddot{X}_{j}\beta}\right)'\boldsymbol{\ddot{X}}_{j}\right) < \infty \\
&\E\left(\boldsymbol{\widetilde{R}^{(-t)'}_{j}}\left(\boldsymbol{\widetilde{R}_j}- \boldsymbol{\widetilde{R}^{(-t)}_{j}\phi}\right)\left(\boldsymbol{\widetilde{R}_j}- \boldsymbol{\widetilde{R}^{(-t)}_{j}\phi}\right)'\boldsymbol{\widetilde{R}^{(-t)}_{j}}\right)<\infty\\
&\E\left(\boldsymbol{\ddot{X}}_{j}'\left(\boldsymbol{\ddot{Y}}^{obs}_{j}- \boldsymbol{\ddot{X}_{j}\beta^Y}\right)\left(\boldsymbol{\ddot{Y}}^{obs}_{j}- \boldsymbol{\ddot{X}_{j}\beta^Y}\right)'\boldsymbol{\ddot{X}}_{j}\right) <\infty \\
&\E\left(\boldsymbol{\phi'\widetilde{R}^{(-t)'}_{j}}\left(\boldsymbol{Y}_{j}- \kappa \boldsymbol{\widetilde{R}^{(-t)}_{j}\phi} \right)\left(\boldsymbol{Y}_{j}- \kappa \boldsymbol{\widetilde{R}^{(-t)}_{j}\phi} \right)'\boldsymbol{\widetilde{R}^{(-t)}_{j}\phi}\right)<\infty.
\end{align*}

\item  $\E\left|\underset{(\boldsymbol{\beta,\phi,\beta^Y,}\kappa)\in \boldsymbol{\Theta}}{sup}\left(\boldsymbol{\phi'\widetilde{R}_j^{(-t)'}\widetilde{R}_j^{(-t)}\phi}\right)\right| <\infty$ and
$\E\left|\underset{(\boldsymbol{\beta,\phi,\beta^Y,}\kappa)\in \boldsymbol{\Theta}}{sup}\left(\boldsymbol{\widetilde{R}_j^{(-t)'}\widetilde{R}_j^{(-t)}}\right)\right| <\infty$.
\item $\E\left[\left(\boldsymbol{\phi_0'R_j^{(-t)'}R_j^{(-t)}\phi_0}\right)\right], \E\left[\left(\boldsymbol{R_j^{(-t)'}R_j^{(-t)}}\right)\right]$ are invertible.
\item\begin{align*}
&\E\left(sup_{(\boldsymbol{\beta}, \boldsymbol{\beta^Y},\boldsymbol{\phi},\kappa) \in \Theta}\left|\left|\boldsymbol{\ddot{X}}_{j}'\left(\boldsymbol{\ddot{R}}^{obs}_{j}- \boldsymbol{\ddot{X}_{j}\beta}\right)\right|\right|^2\right) < \infty\\
&\E\left(sup_{(\boldsymbol{\beta}, \boldsymbol{\beta^Y},\boldsymbol{\phi},\kappa) \in \Theta}\left|\left|\boldsymbol{\widetilde{R}^{(-t)'}_{j}}\left(\boldsymbol{\widetilde{R}}_{j}- \boldsymbol{\widetilde{R}^{(-t)}_{j}\phi}\right)\right|\right|^2\right) <\infty\\
&\E\left(sup_{(\boldsymbol{\beta}, \boldsymbol{\beta^Y},\boldsymbol{\phi},\kappa) \in \Theta}\left|\left|\boldsymbol{\ddot{X}}_{j}'\left(\boldsymbol{\ddot{Y}}^{obs}_{j}- \boldsymbol{\ddot{X}_{j}\beta^Y}\right)\right|\right|^2\right)< \infty\\
&\E\left( sup_{(\boldsymbol{\beta}, \boldsymbol{\beta^Y},\boldsymbol{\phi},\kappa) \in \Theta}\left|\boldsymbol{\phi'\widetilde{R}^{(-t)'}_{j}}\left(\boldsymbol{\widetilde{Y}}_{j}- \kappa \boldsymbol{\widetilde{R}^{(-t)}_{j}\phi} \right)\right|^2\right)<\infty
\end{align*}
\end{enumerate}
\end{appxhyp}

\begin{appxhyp}\label{hyp:tech-opt} Technical Assumptions for Optimal GMM \\
\begin{enumerate}
\item  $(\boldsymbol{\beta_0'}, \boldsymbol{\beta_0^{Y'}},\kappa_0)$  is an element in the interior of $\boldsymbol{\Theta_1}$ and $\boldsymbol{\Theta_1}$ is a compact subset of $\mathcal{R}^{2K+1}$.
\item \begin{flalign*}
&\E\left(sup_{(\boldsymbol{\beta}, \boldsymbol{\beta^Y},\kappa) \in \Theta_1}\left|\left|\boldsymbol{\ddot{X}}_{j}'\left(\boldsymbol{\ddot{R}}^{obs}_{j}- \boldsymbol{\ddot{X}_{j}\beta}\right)\right|\right|^2\right) < \infty\\
&\E\left(sup_{(\boldsymbol{\beta}, \boldsymbol{\beta^Y},\kappa) \in \Theta_1}\left|\left|\boldsymbol{\ddot{X}}_{j}'\left(\boldsymbol{\ddot{Y}}^{obs}_{j}- \boldsymbol{\ddot{X}_{j}\beta^Y}\right)\right|\right|^2\right) < \infty\\
&\E\left(sup_{(\boldsymbol{\beta}, \boldsymbol{\beta^Y},\kappa) \in \Theta_1}\left|\left|\boldsymbol{\widetilde{R}^{(-t)'}_{j}}\left(\boldsymbol{\widetilde{Y}}_{j}- \kappa \boldsymbol{\widetilde{R}_j} \right)\right|\right|^2\right) <\infty
\end{flalign*}
\item $\left(\boldsymbol{X}_{j},\boldsymbol{\mu}_j, \boldsymbol{\epsilon}_j, \boldsymbol{Y}^{obs}_{j}, \boldsymbol{\eta}_j\right)$ are i.i.d across $j$.
\item For all $(\boldsymbol{\beta}, \boldsymbol{\beta^Y},\kappa) \in \mathcal{R}^{2K+1}$:
\begin{align*}
&E\left(\boldsymbol{\ddot{X}}_{j}'\left(\boldsymbol{\ddot{R}}^{obs}_{j}- \boldsymbol{\ddot{X}_{j}\beta}\right)\left(\boldsymbol{\ddot{R}}^{obs}_{j}- \boldsymbol{\ddot{X}_{j}\beta}\right)'\boldsymbol{\ddot{X}}_{j}\right) < \infty \\
&\E\left(\boldsymbol{\ddot{X}}_{j}'\left(\boldsymbol{\ddot{Y}}^{obs}_{j}- \boldsymbol{\ddot{X}_{j}\beta^Y}\right)\left(\boldsymbol{\ddot{Y}}^{obs}_{j}- \boldsymbol{\ddot{X}_{j}\beta^Y}\right)'\boldsymbol{\ddot{X}}_{j}\right) <\infty \\
&\E\left(\boldsymbol{\widetilde{R}^{(-t)'}_{j}}\left(\boldsymbol{\widetilde{Y}}_{j} - \kappa \boldsymbol{\widetilde{R}_j} \right)\left(\boldsymbol{\widetilde{Y}}_{j} - \kappa \boldsymbol{\widetilde{R}_j} \right)'\boldsymbol{\widetilde{R}^{(-t)}_{j}}\right)<\infty.
\end{align*}
\item $\E\left|\underset{(\boldsymbol{\beta,\beta^Y,}\kappa)\in \boldsymbol{\Theta_1}}{sup}\left(\boldsymbol{\widetilde{R}_j^{(-t)'}\widetilde{R}_j^{(-t)}}\right)\right| <\infty$.
\end{enumerate}
\end{appxhyp}

Assumption \ref{hyp:tech-opt} reframes Assumptions \ref{hyp:tech-cons} and \ref{hyp:tech-norm} for the new set of moments. \medskip

\begin{appxhyp}\label{hyp:as-outcome-tech} Assumptions for Consistency, and Asymptotic Normality \\
\begin{enumerate}
\item  $(\boldsymbol{\beta_0',\alpha_0'})$ is an element of the interior of $\boldsymbol{\Theta_2}$ and $\boldsymbol{\Theta_2}$ is a compact subset of  of $\mathcal{R}^{K+K_D}$.
\item
\begin{flalign*}
&\E\left(sup_{(\boldsymbol{\beta}, \boldsymbol{\alpha}) \in \Theta_2}\left|\left|\boldsymbol{\ddot{X}}_{j}'\left(\boldsymbol{\ddot{R}}^{obs}_{j}- \boldsymbol{\ddot{X}_{j}\beta}\right)\right|\right|^2\right) < \infty\\
&\E\left(sup_{(\boldsymbol{\beta}, \boldsymbol{\alpha}) \in \Theta_2}\left|\left|\boldsymbol{D_{j}}'\left(\boldsymbol{\widetilde{R}_j}-\boldsymbol{D_{j}\alpha }\right)\right|\right|^2\right)< \infty\\
\end{flalign*}
\item $\left(\boldsymbol{X}_{j},\boldsymbol{\mu}_j, \boldsymbol{\epsilon}_j, \boldsymbol{D}_{j}\right)$ are i.i.d across $j$.
\item For all $(\boldsymbol{\beta}, \boldsymbol{\alpha}) \in \mathcal{R}^{K+K_D}$:
\begin{align*}
&E\left(\boldsymbol{\ddot{X}}_{j}'\left(\boldsymbol{\ddot{R}}^{obs}_{j}- \boldsymbol{\ddot{X}_{j}\beta}\right)\left(\boldsymbol{\ddot{R}}^{obs}_{j}- \boldsymbol{\ddot{X}_{j}\beta}\right)'\boldsymbol{\ddot{X}}_{j}\right) < \infty \\
&\E\left(\boldsymbol{D_{j}}'\left(\boldsymbol{\widetilde{R}_j}-\boldsymbol{D_{j}\alpha }\right)\left(\boldsymbol{\widetilde{R}_j}-\boldsymbol{D_{j}\alpha }\right)'\boldsymbol{D_{j}}\right) <\infty
\end{align*}
\end{enumerate}
\end{appxhyp}
\section{Consistency, Asymptotic Normality, and Consistent Variance Estimation of the GMM Estimators \label{section:gmm-stuff}}

\begin{appxlem} \label{prop:cons}
If Assumptions \ref{hyp:rand-assign}, \ref{hyp:tech-id}, and \ref{hyp:tech-cons} hold, then $(\boldsymbol{\widehat{\beta},\widehat{\phi},\widehat{\beta}^Y,}\widehat{\kappa})\overset{p}{\rightarrow}(\boldsymbol{\beta_0,\phi_0,\beta^Y_0,}\kappa_0)$, where $(\boldsymbol{\widehat{\beta},\widehat{\phi},\widehat{\beta}^Y,}\widehat{\kappa})$ are the estimators obtained from a GMM minimization of the moment system of  \eqref{eq:mom1}, \eqref{eq:mom2}, \eqref{eq:mom3}, and \eqref{eq:mom4} with the identity matrix as a weighting matrix.
\end{appxlem} \medskip

\begin{appxthm} \label{lem:asym-norm}
If Assumptions \ref{hyp:rand-assign}, \ref{hyp:tech-id}, \ref{hyp:tech-cons}, and \ref{hyp:tech-norm} hold, then
\begin{equation*}
\sqrt{J}\left[(\boldsymbol{\widehat{\beta},\widehat{\phi},\widehat{\beta}^Y,}\widehat{\kappa})-(\boldsymbol{\beta_0,\phi_0,\beta^Y_0,}\kappa_0)\right] \rightsquigarrow N\left(0,\Omega\right).
\end{equation*}
where $\Omega=\widetilde{G}^{-1}\E[\widetilde{g}(\boldsymbol{Z},\boldsymbol{\beta_0,\phi_0,\beta^Y_0,}\kappa_0)\widetilde{g}(\boldsymbol{Z},\boldsymbol{\beta_0,\phi_0,\beta^Y_0,}\kappa_0)']\widetilde{G}^{-1'}$ and $\widetilde{G}=\E\left[\triangledown_{(\boldsymbol{\beta,\phi,\beta^Y,}\kappa)}\widetilde{g}(\boldsymbol{Z},\boldsymbol{\beta_0,\phi_0,\beta^Y_0,}\kappa_0)\right]$,  and $\widetilde{g}(\boldsymbol{Z},\boldsymbol{\beta_0,\phi_0,\beta^Y_0,}\kappa_0)$ is defined in  \eqref{eq:system} in the proof. \medskip

Let $\widehat{\Omega}$ correspond to  an estimator of $\Omega$, constructed by replacing the population moments in $\Omega$ by averages and the parameters by the GMM estimators. Then:
\begin{align}
\widehat{\Omega}\overset{p}{\rightarrow} \Omega
\end{align}
\end{appxthm}
\medskip

\begin{appxlem} \label{lem:asym-step1}
If Assumptions \ref{hyp:rand-assign}, \ref{hyp:tech-id}, and \ref{hyp:tech-opt} hold, then
\begin{enumerate}
\item \begin{equation*}
\sqrt{J}\left[(\boldsymbol{\widehat{\beta},\widehat{\beta}^Y,}\widehat{\kappa})-(\boldsymbol{\beta_0,\beta^Y_0,}\kappa_0)\right] \rightsquigarrow N\left(0,\Omega_1\right).
\end{equation*}
where $(\boldsymbol{\widehat{\beta},\widehat{\beta}^Y,}\widehat{\kappa})$ are the estimators obtained from a GMM minimization and \\ $\Omega_1=\left(\widetilde{G}_1'\widetilde{G}_1\right)^{-1}\widetilde{G}_1'\E[\widetilde{g}_1(\boldsymbol{Z},\boldsymbol{\beta_0,\beta^Y_0,}\kappa_0)\widetilde{g}(\boldsymbol{Z},\boldsymbol{\beta_0,\beta^Y_0,}\kappa_0)']\widetilde{G}_1\left(\widetilde{G}_1'\widetilde{G}_1\right)^{-1}$, \\ and $\widetilde{G}_1=\E\left[\triangledown_{(\boldsymbol{\beta,\beta^Y,}\kappa)}\widetilde{g}_1(\boldsymbol{Z},\boldsymbol{\beta_0,\beta^Y_0,}\kappa_0)\right]$,  where $\widetilde{g}_1(\boldsymbol{Z},\boldsymbol{\beta_0,\beta^Y_0,}\kappa_0)$ is defined in  \eqref{eq:system2} in the proof.
\item  $\widehat{\Omega}_1\overset{p}{\rightarrow} \Omega_1$  where $\widehat{\Omega}_1$ corresponds to the sample equivalent of $\Omega_1$ replacing moments by sample moments and parameters by $(\boldsymbol{\widehat{\beta},\widehat{\beta}^Y,}\widehat{\kappa})$.
\end{enumerate}
\end{appxlem}

\begin{appxthm} \label{prop:optimal}
If Assumptions \ref{hyp:rand-assign}, \ref{hyp:tech-id}, and \ref{hyp:tech-opt} hold, then
\begin{equation*}
\sqrt{J}\left[(\boldsymbol{\widehat{\beta}^*,\widehat{\beta}^{Y*}},\widehat{\kappa}^*)-(\boldsymbol{\beta_0,\beta^Y_0,}\kappa_0)\right] \rightsquigarrow N\left(0,\Omega^*\right).
\end{equation*}

where $(\boldsymbol{\widehat{\beta}^*,\widehat{\beta}^{Y*}},\widehat{\kappa}^*)$ are the estimates resulting from a GMM minimization using $\widehat{W}^*$ as a weighting matrix,   \medskip

$\Omega^*=\left(\widetilde{G}_1'W^*\widetilde{G}_1\right)^{-1}$,
and $\Omega^* \leq \Omega_2$ where $\Omega_2$ is the submatrix of $\Omega$ that corresponds to the variance covariance matrix of $(\boldsymbol{\widehat{\beta},\widehat{\beta}^{Y}},\widehat{\kappa})$.
\end{appxthm}

\begin{appxres} \label{prop:over-id}
If Assumptions \ref{hyp:rand-assign}, \ref{hyp:tech-id}, and \ref{hyp:tech-opt} hold, then
\begin{equation*}
\widehat{\widetilde{g}}_1(\boldsymbol{Z},\boldsymbol{\widehat{\beta}^*,\widehat{\beta}^{Y*}},\widehat{\kappa}^*)'\widehat{W}^*\widehat{\widetilde{g}}_1(\boldsymbol{Z},\boldsymbol{\widehat{\beta}^*,\widehat{\beta}^{Y*}},\widehat{\kappa}^*)\rightsquigarrow \chi^2_{T-2}.
\end{equation*}
\end{appxres}

\begin{appxthm} \label{lem:as-outcome-dist}
If Assumptions \ref{hyp:rand-assign}, \ref{hyp:tech-id}, \ref{hyp:as-outcome}, \ref{hyp:as-outcome-tech} and   hold, then
\begin{equation*}
\sqrt{J}\left[(\boldsymbol{\widehat{\beta},\widehat{\alpha}})-(\boldsymbol{\beta_0,\alpha_0})\right] \rightsquigarrow N\left(0,\Omega_2\right).
\end{equation*}
where $\Omega_2=\widetilde{G}_2^{-1}\E[\widetilde{g}_2(\boldsymbol{Z}_2,\boldsymbol{\beta_0,\alpha_0})\widetilde{g}_2(\boldsymbol{Z}_2,\boldsymbol{\beta_0,\alpha_0}')]\widetilde{G}_2^{-1'}$ and $\widetilde{G}_2=\E\left[\triangledown_{(\boldsymbol{\beta,\alpha})}\widetilde{g}_2(\boldsymbol{Z}_2,\boldsymbol{\beta_0,\alpha_0})\right]$, where $\boldsymbol{Z_j}=\left(\boldsymbol{X_j,R_j^{obs},D_j}\right)$, $\boldsymbol{Z}_2$ stacks the $\boldsymbol{Z_j}$, and $\widetilde{g}_2(\boldsymbol{Z}_2,\boldsymbol{\beta_0,\alpha_0})$ is defined in  \eqref{eq:system3} in the proof.
\end{appxthm}

\section{Variance Comparison \label{section:variance-compare}}

Theorem \ref{prop:kappa-dist} also allows us to compare $\sigma^2$ to the variance obtained from OLS:
\begin{align*}
&\sigma^2-(G_{\kappa}^{-1})^2\E\left(g(\boldsymbol{Z})^2\right)\\
=&(G_{\kappa}^{-1})^2\left[\E\left(\left(g(\boldsymbol{Z})+G_{\beta^Y}\psi_3(\boldsymbol{Z})+G_{\phi}\psi_2(\boldsymbol{Z})+G_{\beta}\psi_1(\boldsymbol{z})-G_{\phi}M_{2\phi}^{-1}M_{2\beta}\psi_1(\boldsymbol{Z})\right)^2-\E\left(\left(g(\boldsymbol{Z})^2\right)\right)\right)\right]\\
=&(G_{\kappa}^{-1})^2\E\left(\left(2g(\boldsymbol{Z})\right)\left(G_{\beta^Y}\psi_3(\boldsymbol{Z})+G_{\phi}\psi_2(\boldsymbol{Z})+G_{\beta}\psi_1(\boldsymbol{z})-G_{\phi}M_{2\phi}^{-1}M_{2\beta}\psi_1(\boldsymbol{Z})\right)\right)\\
&+(G_{\kappa}^{-1})^2\E\left(\left(G_{\beta^Y}\psi_3(\boldsymbol{Z})+G_{\phi}\psi_2(\boldsymbol{Z})+G_{\beta}\psi_1(\boldsymbol{z})-G_{\phi}M_{2\phi}^{-1}M_{2\beta}\psi_1(\boldsymbol{Z})\right)^2\right)
\end{align*}

where the equality follows from factoring $a^2-b^2$ and the linearity of the expectation operator. Given that the second term is always positive, the difference between the two variances is going to depend on the covariances between the moment used to estimate $\kappa_0$ and the other three sets moments. Then the covariances of interest are:

\begin{align*}
&\E\left(\boldsymbol{\ddot{X}}_{j}'\left(\boldsymbol{\ddot{\mu}_j} + \boldsymbol{\ddot{\epsilon}_j}\right)\left(\kappa_0 \boldsymbol{\theta_{j}} - \kappa_0 \boldsymbol{\epsilon_{j}} +\boldsymbol{\eta_{j}}\right)'\boldsymbol{R^{(-t)}_{j}\phi_0}\right)\\
& \E\left(\boldsymbol{\overline{R}^{(-t)'}_{j}\theta_{j}}\left(\kappa_0 \boldsymbol{\theta_{j}} - \kappa_0 \boldsymbol{\epsilon_{j}} +\boldsymbol{\eta_{j}}\right)'\boldsymbol{R^{(-t)}_{j}\phi_0}\right)\\
&\E\left(\boldsymbol{\ddot{X}}_{j}' \boldsymbol{\ddot{\eta}_j}\left(\kappa_0 \boldsymbol{\theta_{j}} - \kappa_0 \boldsymbol{\epsilon_{j}} +\boldsymbol{\eta_{j}}\right)'\boldsymbol{R^{(-t)}_{j}\phi_0}\right).
\end{align*}

For ease of exposition of the difference between the two variances, assume that the covariances of the errors are homoskedastic with respect to $\boldsymbol{R^{(-t)}_{j}}$ and $\boldsymbol{\ddot{X}_{j}}$ such that:

\begin{align*}
&\E\left(\left(\boldsymbol{\ddot{\mu}_j} + \boldsymbol{\ddot{\epsilon}_j}\right)\left(\kappa_0 \boldsymbol{\theta_{j}} - \kappa_0 \boldsymbol{\epsilon_{j}} +\boldsymbol{\eta_{j}}\right)'| \boldsymbol{R^{(-t)}_{j}},\boldsymbol{\ddot{X}_{j}}\right)=\E\left(\left(\boldsymbol{\ddot{\mu}_j} + \boldsymbol{\ddot{\epsilon}_j}\right)\left(\kappa_0 \boldsymbol{\theta_{j}} - \kappa_0 \boldsymbol{\epsilon_{j}} +\boldsymbol{\eta_{j}}\right)'\right) \\
&\E\left(\boldsymbol{\theta_{j}}\left(\kappa_0 \boldsymbol{\theta_{j}} - \kappa_0 \boldsymbol{\epsilon_{j}} +\boldsymbol{\eta_{j}}\right)'|\boldsymbol{R^{(-t)}_{j}},\boldsymbol{\ddot{X}_{j}}\right)=\E\left(\boldsymbol{\theta_{j}}\left(\kappa_0 \boldsymbol{\theta_{j}} - \kappa_0 \boldsymbol{\epsilon_{j}} +\boldsymbol{\eta_{j}}\right)'\right)\\
&\E\left(\left(\boldsymbol{\kappa_0\ddot{\mu}_j}+ \boldsymbol{\ddot{\eta}_j}\right)\left(\kappa_0 \boldsymbol{\theta_{j}} - \kappa_0 \boldsymbol{\epsilon_{j}} +\boldsymbol{\eta_{j}}\right)'| \boldsymbol{R^{(-t)}_{j}},\boldsymbol{\ddot{X}_{j}}\right)=\E\left(\left(\boldsymbol{\kappa_0\ddot{\mu}_j}+ \boldsymbol{\ddot{\eta}_j}\right)\left(\kappa_0 \boldsymbol{\theta_{j}} - \kappa_0 \boldsymbol{\epsilon_{j}} +\boldsymbol{\eta_{j}}\right)'\right)
\end{align*}

 we get:
\begin{align*}
&\E\left(\boldsymbol{\ddot{X}}_{j}'\E\left(\left(\boldsymbol{\ddot{\mu}_j} + \boldsymbol{\ddot{\epsilon}_j}\right)\left(\kappa_0 \boldsymbol{\theta_{j}} - \kappa_0 \boldsymbol{\epsilon_{j}} +\boldsymbol{\eta_{j}}\right)'\right)\boldsymbol{R^{(-t)}_{j}\phi_0}\right)\\
& \E\left(\boldsymbol{\overline{R}^{(-t)'}_{j}}\E\left(\boldsymbol{\theta_{j}}\left(\kappa_0 \boldsymbol{\theta_{j}} - \kappa_0 \boldsymbol{\epsilon_{j}} +\boldsymbol{\eta_{j}}\right)'\right)\boldsymbol{R^{(-t)}_{j}\phi_0}\right)\\
&\E\left(\boldsymbol{\ddot{X}}_{j}' \E\left(\left(\boldsymbol{\kappa_0\ddot{\mu}_j}+ \boldsymbol{\ddot{\eta}_j}\right)\left(\kappa_0 \boldsymbol{\theta_{j}} - \kappa_0 \boldsymbol{\epsilon_{j}} +\boldsymbol{\eta_{j}}\right)'\right)\boldsymbol{R^{(-t)}_{j}\phi_0}\right).
\end{align*}

Note that the first and third set of covariances will depend on $\E\left(\ddot{X}_{jt}\overline{R}_{js}\right)$ for $s\neq t$. Simplifying further, it is reasonable to assume that those covariances are close to zero since they depend on the covariance between within teacher fluctuations in covariates in year $t$ and teacher value added in years $s\neq t$, and  average unobserved determinants of test scores in years $s\neq t$, $\E\left(\ddot{X}_{jt}\mu_{js}\right)$, and $\E\left(\ddot{X}_{jt}\overline{\epsilon}_{js}\right)$.  \medskip

Then for the variance from OLS to be larger than $\sigma^2$ it would have to be that $$2G_{\phi}\E\left(\boldsymbol{R^{(-t)'}_{j}}\E\left(\boldsymbol{\theta_{j}}\left(\kappa_0 \boldsymbol{\theta_{j}} - \kappa_0 \boldsymbol{\epsilon_{j}} +\boldsymbol{\eta_{j}}\right)'\right)\boldsymbol{R^{(-t)}_{j}\phi_0}\right)<0$$ and  $$\left|2G_{\phi}\E\left(\boldsymbol{R^{(-t)'}_{j}}\E\left(\boldsymbol{\theta_{j}}\left(\kappa_0 \boldsymbol{\theta_{j}} - \kappa_0 \boldsymbol{\epsilon_{j}} +\boldsymbol{\eta_{j}}\right)'\right)\boldsymbol{R^{(-t)}_{j}\phi_0}\right)\right| >$$ $$\E\left(\left(G_{\beta^Y}\psi_3(\boldsymbol{Z})+G_{\phi}\psi_2(\boldsymbol{Z})+G_{\beta}\psi_1(\boldsymbol{z})-G_{\phi}M_{2\phi}^{-1}M_{2\beta}\psi_1(\boldsymbol{Z})\right)^2\right).$$ In other words the contribution to $\sigma^2$ of the covariances between the moments used to estimate $\boldsymbol{\phi_0}$ and $\kappa_0$ has to be large and negative enough to outweigh the contribution to $\sigma^2$ from the variances and covariances of the moments used to estimate $\boldsymbol{\beta_0}$, $\boldsymbol{\beta^Y_0}$, and $\boldsymbol{\phi_0}$.  Therefore, it is likely that the variance from OLS be smaller than $\sigma^2$ in most cases. \medskip

\section{Proofs}

\subsection{Proof of Result \ref{prop:ID-moment}}
\begin{align*}
&\E\left(\boldsymbol{\ddot{X}}_{j}'\left(\boldsymbol{\ddot{R}}^{obs}_{j}- \boldsymbol{\ddot{X}_{j}\beta_0}\right)\right)  \\
=&\E\left(\boldsymbol{\ddot{X}}_{j}'\left(\boldsymbol{\ddot{\mu}_j} + \boldsymbol{\ddot{\epsilon}_j}\right)\right)   \\
=&0
\end{align*}

where the first equality follows from  \eqref{eq:scores}, and the second equality follows from Point 3 of Assumption \ref{hyp:tech-id}. \medskip

\begin{align*}
&\E\left(\boldsymbol{\overline{R}^{(-t)'}_{j}}\left(\boldsymbol{R}_{j}- \boldsymbol{R^{(-t)}_{j}\phi_0}\right)\right) \\
=&\E\left(\boldsymbol{\overline{R}^{(-t)'}_{j}\theta_{j}}\right) \\
=&0
\end{align*}

where the first equality follows from  \eqref{eq:decomp}, and the second equality follows from the fact that $\theta_{jt}$ is orthogonal to $\overline{R}_{jk}$ for $k\neq t$ by construction. \medskip

\begin{align*}
&\E\left(\boldsymbol{\ddot{X}}_{j}'\left(\boldsymbol{\ddot{Y}}^{obs}_{j}- \boldsymbol{\ddot{X}_{j}\beta_0^Y}\right)\right)  \\
=&\E\left(\boldsymbol{\ddot{X}_{j}}(\boldsymbol{\ddot{\eta}_{j}}+ \kappa_0\boldsymbol{\ddot{\mu}_{j}})\right)  \\
=&0
\end{align*}

where the first equality follows from  \eqref{eq:earningres} and the second from Point 3 of Assumption \ref{hyp:tech-id}.\medskip

\begin{align*}
&\E\left(\boldsymbol{\phi_0'\overline{R}^{(-t)'}_{j}}\left(\boldsymbol{Y}_{j}- \kappa_0 \boldsymbol{R^{(-t)}_{j}\phi_0} \right)\right)  \\
=&\E\left(\mu_{jt}^*\left(\boldsymbol{Y}_{j}- \kappa_0 \mu_{jt}^* \right)\right)  \\
=&\E\left(\mu_{jt}^*\left(\kappa_0 \theta_{jt} - \kappa_0 \overline{\epsilon}_{jt} +\overline{\eta}_{jt}\right)\right)  \\
=&0
\end{align*}

where the second equality follow from  \eqref{eq:for-mom4} and the third equality follows from Assumption \ref{hyp:rand-assign} and from the fact that $\theta_{jt}$ is orthogonal to $\overline{R}_{jk}$ for $k\neq t$ by construction. \medskip

\textbf{Q.E.D}

\subsection{Proof of Result \ref{lem:assumption}}

Starting with $Cov\left(\overline{\eta}_{jt}, \mu^*_{jt}\right)=0$:
\begin{align*}
&Cov\left(\overline{\eta}_{jt}, \mu^*_{jt}\right) \\
=&Cov\left(\overline{\eta}_{jt}, \sum_{\abs{s-t}\neq 0}\phi_{0\abs{s-t}} \overline{R}_{js}\right) \\
=&Cov\left(\overline{\eta}_{jt},\sum_{\abs{s-t}\neq 0}\phi_{0\abs{s-t}}\mu_{js}\right)+Cov\left(\overline{\epsilon}_{jt},\sum_{\abs{s-t}\neq 0}\phi_{0\abs{s-t}}\overline{\epsilon}_{js}\right)\\
=&0
\end{align*}

where both terms are 0 by Assumption \ref{hyp:rand-assign}. \medskip

Now for $\frac{Cov\left(\mu_{jt}, \mu^*_{jt}\right)}{Var\left(\mu^*_{jt}\right)}=1$:
\begin{align*}
&\frac{Cov\left(\mu_{jt}, \mu^*_{jt}\right)}{Var\left(\mu^*_{jt}\right)} \\
=&\frac{Cov\left(\sum_{\abs{s-t}\neq 0}\phi_{0\abs{s-t}} \overline{R}_{js} + \theta_{jt} - \overline{\epsilon}_{jt}, \mu^*_{jt}\right)}{Var\left(\mu^*_{jt}\right)} \\
=&\frac{Cov\left(\sum_{\abs{s-t}\neq 0}\phi_{0\abs{s-t}} \overline{R}_{js} + \theta_{jt} - \overline{\epsilon}_{jt}, \sum_{\abs{s-t}\neq 0}\phi_{0\abs{s-t}}\overline{R}_{js} \right)}{Var\left(\sum_{\abs{s-t}\neq 0}\phi_{0\abs{s-t}}\overline{R}_{js} \right)} \\
=&\frac{Cov\left(\sum_{\abs{s-t}\neq 0}\phi_{0\abs{s-t}} \overline{R}_{js}, \sum_{\abs{s-t}\neq 0}\phi_{0\abs{s-t}}\overline{R}_{js} \right)}{Var\left(\sum_{\abs{s-t}\neq 0}\phi_{0\abs{s-t}}\overline{R}_{js} \right)} \\
=&1
\end{align*}

where the first equality follows from  \eqref{eq:plug-in}, the third equality follows from the fact that $\theta_{jt}$ is the error from  \eqref{eq:decomp} and by Assumption \ref{hyp:rand-assign}.

\subsection{Proof of Lemma \ref{prop:cons}}
We start with the following moments:

\begin{align}
&\E\left(\boldsymbol{\ddot{X}}_{j}'\left(\boldsymbol{\ddot{R}}^{obs}_{j}- \boldsymbol{\ddot{X}_{j}\beta}\right)\right) \label{eq:fereg}\\
&\E\left(\boldsymbol{\widetilde{R}^{(-t)'}_{j}}\left(\boldsymbol{\widetilde{R}}_{j}- \boldsymbol{\widetilde{R}^{(-t)}_{j}\phi}\right)\right)\\
&\E\left(\boldsymbol{\ddot{X}}_{j}'\left(\boldsymbol{\ddot{Y}}^{obs}_{j}- \boldsymbol{\ddot{X}_{j}\beta^Y}\right)\right) \label{eq:fereg2}\\
&\E\left(\boldsymbol{\phi'\widetilde{R}^{(-t)'}_{j}}\left(\boldsymbol{\widetilde{Y}}_{j}- \kappa \boldsymbol{\widetilde{R}^{(-t)}_{j}\phi} \right)\right) \label{eq:lastmom}
\end{align}

where $\widetilde{R}_{jt}=\overline{R}^{obs}_{jt}-\boldsymbol{\overline{X}_{jt}'\beta}$, $\boldsymbol{\widetilde{R}}_{j}$ is a vector stacking all $\widetilde{R}_{jt}$ for teacher $j$, $\boldsymbol{\widetilde{R}^{(-t)}_{j}}$ is  a matrix stacking $T$ row vectors (of dimension $1\times(T-1)$) with each row (indexed by $t$) containing $T-1$ different $\widetilde{R}_{jk}$ for $k\neq t$, and $\widetilde{Y}_{jt}=\overline{Y}^{obs}_{jt}-\boldsymbol{\overline{X}_{jt}'\beta^Y}$ and $\boldsymbol{\widetilde{Y}}_{j}$ is a vector stacking them. Note that  \eqref{eq:fereg} and \eqref{eq:fereg2}  make it so that $\beta_0$ and $\beta_0^Y$ will be estimated using teacher fixed effects. \medskip

Let the sample equivalent of the moment conditions be:

\begin{align}
&\frac{1}{J}\sum_{j=1}^{J}\boldsymbol{\ddot{X}}_{j}'\left(\boldsymbol{\ddot{R}}^{obs}_{j}- \boldsymbol{\ddot{X}_{j}\beta}\right)\\
&\frac{1}{J}\sum_{j=1}^{J}\boldsymbol{\widetilde{R}^{(-t)'}_{j}}\left(\boldsymbol{\widetilde{R}}_{j}- \boldsymbol{\widetilde{R}^{(-t)}_{j}\phi}\right)\\
&\frac{1}{J}\sum_{j=1}^{J}\boldsymbol{\ddot{X}}_{j}'\left(\boldsymbol{\ddot{Y}}^{obs}_{j}- \boldsymbol{\ddot{X}_{j}\beta^Y}\right)\\
&\frac{1}{J}\sum_{j=1}^{J}\boldsymbol{\phi'\widetilde{R}^{(-t)'}_{j}}\left(\boldsymbol{\widetilde{Y}}_{j}- \kappa \boldsymbol{\widetilde{R}^{(-t)}_{j}\phi} \right)
\end{align}

Let the GMM weighting matrix be the identity matrix. 
\medskip

To show the estimator is consistent, we need to check the conditions from Theorem 2.6 of \citet{newey1994large}. Consistency of GMM when the data are i.i.d requires the following conditions:

\begin{enumerate}
\item The weighting matrix $W$ is positive semi-definite
\item $W\E[g(\boldsymbol{X,Y^{obs},R^{obs}},\boldsymbol{\beta}, \boldsymbol{\beta^Y},\boldsymbol{\phi},\kappa)]=0$ if and only if $(\boldsymbol{\beta}, \boldsymbol{\beta^Y},\boldsymbol{\phi},\kappa)=(\boldsymbol{\beta_0}, \boldsymbol{\beta_0^Y},\boldsymbol{\phi_0},\kappa_0)$, where $g(\boldsymbol{X,Y^{obs},R^{obs}},\boldsymbol{\beta}, \boldsymbol{\beta^Y},\boldsymbol{\phi},\kappa)$ is a vector stacking the moment functions.
\item $(\boldsymbol{\beta_0}, \boldsymbol{\beta_0^Y},\boldsymbol{\phi_0},\kappa_0) \in \boldsymbol{\Theta}$ and $\boldsymbol{\Theta}$ is compact.
\item $g(\boldsymbol{X,Y^{obs},R^{obs}},\boldsymbol{\beta}, \boldsymbol{\beta^Y},\boldsymbol{\phi},\kappa)$ is continuous for all $(\boldsymbol{\beta}, \boldsymbol{\beta^Y},\boldsymbol{\phi},\kappa) \in \boldsymbol{\Theta}$ with probability one.
\item $\E[sup_{(\boldsymbol{\beta}, \boldsymbol{\beta^Y},\boldsymbol{\phi},\kappa) \in \Theta} ||g(\boldsymbol{X,Y^{obs},R^{obs}},\boldsymbol{\beta}, \boldsymbol{\beta^Y},\boldsymbol{\phi},\kappa)||]<\infty$.
\end{enumerate}

%
The first condition is satisfied since the weighting matrix $W$ is the identity matrix, so it is positive semi-definite.

Furthermore by Result \ref{prop:ID-moment} all the moments are equal to 0 when evaluated at $(\boldsymbol{\beta_0}, \boldsymbol{\beta_0^Y},\boldsymbol{\phi_0},\kappa_0)$. Therefore $\E[g(\boldsymbol{X,Y^{obs},R^{obs}},\boldsymbol{\beta_0}, \boldsymbol{\beta^Y_0},\boldsymbol{\phi_0},\kappa_0)]=0$ and under Assumptions  \ref{hyp:rand-assign} and \ref{hyp:tech-id}, and  \eqref{eq:obsearn}, \eqref{eq:scores}, \eqref{eq:blup}, and \eqref{eq:decomp} we have that $(\boldsymbol{\beta_0}, \boldsymbol{\beta_0^Y},\boldsymbol{\phi_0},\kappa_0)$ are uniquely identified so that this holds if and only if $(\boldsymbol{\beta}, \boldsymbol{\beta^Y},\boldsymbol{\phi},\kappa)=(\boldsymbol{\beta_0}, \boldsymbol{\beta_0^Y},\boldsymbol{\phi_0},\kappa_0)$. Then the second condition holds. \medskip

The third condition holds by Point 1 of Assumption \ref{hyp:tech-cons}, and the fourth condition holds by inspection.

The last condition holds by Point 2 of Assumption \ref{hyp:tech-cons} and the triangle inequality applied to the euclidian norm.

\textbf{Q.E.D}

\subsection{Proof of Theorem \ref{lem:asym-norm}}

Now to show asymptotic normality we start by rewriting the moments as:
\begin{align}
&\E\left(m_1(\boldsymbol{Z},\boldsymbol{\beta})\right)=E\left(\boldsymbol{\ddot{X}}_{j}'\left(\boldsymbol{\ddot{R}}^{obs}_{j}- \boldsymbol{\ddot{X}_{j}\beta}\right)\right) \\
&\E\left(m_2(\boldsymbol{Z},\boldsymbol{\beta},\boldsymbol{\phi})\right)=\E\left(\boldsymbol{\widetilde{R}^{(-t)'}_{j}}\left(\boldsymbol{\widetilde{R}}_{j}- \boldsymbol{\widetilde{R}^{(-t)}_{j}\phi}\right)\right)\\
&\E\left(m_3(\boldsymbol{Z},\boldsymbol{\beta^Y})\right)=\E\left(\boldsymbol{\ddot{X}}_{j}'\left(\boldsymbol{\ddot{Y}}^{obs}_{j}- \boldsymbol{\ddot{X}_{j}\beta^Y}\right)\right)\\
&\E\left(g(\boldsymbol{Z},\boldsymbol{\beta,\phi,\beta^Y},\kappa)\right)=\E\left(\boldsymbol{\phi'\widetilde{R}^{(-t)'}_{j}}\left(\boldsymbol{\widetilde{Y}}_{j}- \kappa \boldsymbol{\widetilde{R}^{(-t)}_{j}\phi}\right)\right)
\end{align}

where $\boldsymbol{Z_j}=\left(\boldsymbol{X_j,R_j^{obs},Y_j^{obs}}\right)$, $\boldsymbol{Z}$ stacks the $\boldsymbol{Z}_j$, and $m_1(),m_2(),m_3(),g()$ are functions. Let
\begin{equation} \label{eq:system}
\widetilde{g}(\boldsymbol{Z},\boldsymbol{\beta,\phi,\beta^Y},\kappa)=\left[m_1(\boldsymbol{Z},\boldsymbol{\beta})',m_2(\boldsymbol{Z},\boldsymbol{\beta},\boldsymbol{\phi})',m_3(\boldsymbol{Z},\boldsymbol{\beta^Y})',g(\boldsymbol{Z},\boldsymbol{\beta,\phi,\beta^Y},\kappa)'\right]'
\end{equation}

be a vector stacking the four functions.

A GMM estimator is asymptotically normal if it is consistent and conditions (i)-(v) of  Theorem 3.4 of \citet{newey1994large} are satisfied. \medskip

The conditions are the following:
\begin{enumerate}
\item $(\boldsymbol{\beta_0,\phi_0,\beta^Y_0,}\kappa_0)$ is in the interior of $\boldsymbol{\Theta}$.
\item $\widetilde{g}(\boldsymbol{Z},\boldsymbol{\beta,\phi,\beta^Y},\kappa)$ is continuously differentiable in a neighborhood $\mathcal{N}$ of $(\boldsymbol{\beta_0,\phi_0,\beta^Y_0,}\kappa_0)$.
\item $\E[\widetilde{g}(\boldsymbol{Z},\boldsymbol{\beta_0,\phi_0,\beta^Y_0,}\kappa_0)]=0$ and $\E[||\widetilde{g}(\boldsymbol{Z},\boldsymbol{\beta_0,\phi_0,\beta^Y_0,}\kappa_0)||^2]$ is finite.
\item $\E\left[\underset{(\boldsymbol{\beta,\phi,\beta^Y,}\kappa)\in \mathcal{N}}{sup}||\triangledown\widetilde{g}(\boldsymbol{Z},\boldsymbol{\beta,\phi,\beta^Y,}\kappa)|| \right] < \infty$.
\item $\widetilde{G}'\widetilde{G}$ is non singular for $\widetilde{G}=\E\left[\triangledown_{(\boldsymbol{\beta,\phi,\beta^Y,}\kappa)}\widetilde{g}(\boldsymbol{Z},\boldsymbol{\beta_0,\phi_0,\beta^Y_0,}\kappa_0)\right]$
\end{enumerate}

Under Assumptions \ref{hyp:rand-assign}, \ref{hyp:tech-id}, and \ref{hyp:tech-cons}, the GMM estimator is  a consistent estimator of $(\boldsymbol{\beta_0,\phi_0,\beta^Y_0,}\kappa_0)$.

Condition (i) of Theorem 3.4 of \citet{newey1994large} holds  under Point 1 of Assumption  \ref{hyp:tech-cons}. Condition (ii) holds by inspection. The first part of condition (iii) is shown to hold in the consistency proof, and for the second part note that:
\begin{align*}
\E[||\widetilde{g}(\boldsymbol{Z},\boldsymbol{\beta_0,\phi_0,\beta^Y_0,}\kappa_0)||^2]=\E\left(trace\left(\widetilde{g}(\boldsymbol{Z},\boldsymbol{\beta_0,\phi_0,\beta^Y_0,}\kappa_0)\widetilde{g}(\boldsymbol{Z},\boldsymbol{\beta_0,\phi_0,\beta^Y_0,}\kappa_0)'\right)\right).
\end{align*}

Therefore for $\E[||\widetilde{g}(\boldsymbol{Z},\boldsymbol{\beta_0,\phi_0,\beta^Y_0,}\kappa_0)||^2]$ to be finite we need that:
\begin{align*}
&\E\left(\boldsymbol{\ddot{X}}_{j}'\left(\boldsymbol{\ddot{R}}^{obs}_{j}- \boldsymbol{\ddot{X}_{j}\beta_0}\right)\left(\boldsymbol{\ddot{R}}^{obs}_{j}- \boldsymbol{\ddot{X}_{j}\beta_0}\right)'\left(\boldsymbol{\ddot{R}}^{obs}_{j}- \boldsymbol{\ddot{X}_{j}\beta_0}\right)\right) < \infty \\
&\E\left(\boldsymbol{\overline{R}^{(-t)'}_{j}}\left(\boldsymbol{R}_{j}- \boldsymbol{R^{(-t)}_{j}\phi_0}\right)\left(\boldsymbol{R}_{j}- \boldsymbol{R^{(-t)}_{j}\phi_0}\right)'\boldsymbol{R^{(-t)}_{j}}\right)<\infty\\
&\E\left(\boldsymbol{\ddot{X}}_{j}'\left(\boldsymbol{\ddot{Y}}^{obs}_{j}- \boldsymbol{\ddot{X}_{j}\beta_0^Y}\right)\left(\boldsymbol{\ddot{Y}}^{obs}_{j}- \boldsymbol{\ddot{X}_{j}\beta_0^Y}\right)'\boldsymbol{\ddot{X}}_{j}\right) <\infty \\
&\E\left(\boldsymbol{\phi_0'\overline{R}^{(-t)'}_{j}}\left(\boldsymbol{Y}_{j}- \kappa_0 \boldsymbol{R^{(-t)}_{j}\phi_0} \right)\left(\boldsymbol{Y}_{j}- \kappa_0 \boldsymbol{R^{(-t)}_{j}\phi_0} \right)'\boldsymbol{R^{(-t)}_{j}\phi_0}\right)<\infty.
\end{align*}

which holds by Point 1 of Assumption  \ref{hyp:tech-norm}.

For condition (iv) we need to show that:
$\E\left[\underset{(\boldsymbol{\beta,\phi,\beta^Y,}\kappa)\in \mathcal{N}}{sup}||\triangledown\widetilde{g}(\boldsymbol{Z},\boldsymbol{\beta,\phi,\beta^Y,}\kappa)|| \right] < \infty$ \medskip

\begin{align*}
\E\left[\underset{(\boldsymbol{\beta,\phi,\beta^Y,}\kappa)\in \mathcal{N}}{sup}||\triangledown\widetilde{g}(\boldsymbol{Z},\boldsymbol{\beta,\phi,\beta^Y,}\kappa)||\right]&=\E\left(\underset{(\boldsymbol{\beta,\phi,\beta^Y,}\kappa)\in \mathcal{N}}{sup}\sqrt{trace\left(\triangledown\widetilde{g}(\boldsymbol{Z},\boldsymbol{\beta,\phi,\beta^Y,}\kappa)\triangledown\widetilde{g}(\boldsymbol{Z},\boldsymbol{\beta,\phi,\beta^Y,}\kappa)'\right)}\right)
\end{align*}

by the triangle inequality applied to the Euclidean norm, a sufficient condition for the quantity above to be finite, is that the sum of absolute values of the diagonal elements be finite.  So we need that:
\begin{align*}
& \E\left|\underset{(\boldsymbol{\beta,\phi,\beta^Y,}\kappa)\in \mathcal{N}}{sup}\left(\boldsymbol{\phi'\widetilde{R}_j^{(-t)'}\widetilde{R}_j^{(-t)}\phi}\right)\right| <\infty \\
&\E\left|\underset{(\boldsymbol{\beta,\phi,\beta^Y,}\kappa)\in \mathcal{N}}{sup}\left(\boldsymbol{\widetilde{R}_j^{(-t)'}\widetilde{R}_j^{(-t)}}\right)\right| <\infty \\
&\E\left|\left(\boldsymbol{\ddot{X}}_{j}'\boldsymbol{\ddot{X}}_{j}\right)\right| <\infty.
\end{align*}

Now by Points 1 of Assumption \ref{hyp:tech-id} we have $\E\left|\left(\boldsymbol{\ddot{X}}_{j}'\boldsymbol{\ddot{X}}_{j}\right)\right| <\infty$. The other two hold by taking the neighborhood $\mathcal{N}$ to be $\boldsymbol{\Theta}$ and then using Point 2 of Assumption \ref{hyp:tech-norm}.

Condition (v) is satisfied under Point 1 of Assumption \ref{hyp:tech-id} and Point 3 Assumption \ref{hyp:tech-norm}.\medskip

Furthermore we have $\widehat{W}=W=I$, such that $\left(\widetilde{G}'I\widetilde{G}\right)^{-1}\widetilde{G}'=\widetilde{G}^{-1}$ , therefore the asymptotic variance of the estimator becomes:
\begin{align*}
 &\left(\widetilde{G}'I\widetilde{G}\right)^{-1}\widetilde{G}'IE[\widetilde{g}(\boldsymbol{Z},\boldsymbol{\beta_0,\phi_0,\beta^Y_0,}\kappa_0)\widetilde{g}(\boldsymbol{Z},\boldsymbol{\beta_0,\phi_0,\beta^Y_0,}\kappa_0)']I\widetilde{G}\left(\widetilde{G}'I\widetilde{G}\right)^{-1}
\\&=\widetilde{G}^{-1}E[\widetilde{g}(\boldsymbol{Z},\boldsymbol{\beta_0,\phi_0,\beta^Y_0,}\kappa_0)\widetilde{g}(\boldsymbol{Z},\boldsymbol{\beta_0,\phi_0,\beta^Y_0,}\kappa_0)']\widetilde{G}^{-1'}
\end{align*}

Finally the consistency of $\widehat{\Omega}$ follows directly from Theorem 4.5 of \citet{newey1994large}. We need to only check that  $\widetilde{g}(\boldsymbol{Z},\boldsymbol{\beta,\phi,\beta^Y,}\kappa)$ is continuous in a neighborhood $\mathcal{N}$ of $(\boldsymbol{\beta_0,\phi_0,\beta^Y_0,}\kappa_0)$ with probability one, which is satisfied by inspection. And a fourth moment existence condition  $\E[sup_{(\boldsymbol{Z},\boldsymbol{\beta,\phi,\beta^Y,}\kappa) \in \mathcal{N}} ||\widetilde{g}(\boldsymbol{z},\boldsymbol{\beta,\phi,\beta^Y,}\kappa)||^2]<\infty$ which holds by Point 4 of Assumption \ref{hyp:tech-norm} and taking the neighborhood $\mathcal{N}$ to be $\boldsymbol{\Theta}$.

\textbf{Q.E.D}

\subsection{Proof of Theorem \ref{prop:kappa-dist}}
Let: \medskip

\begin{align*}
& G_{\kappa}=\E[\nabla_{\kappa}g(\boldsymbol{Z},\boldsymbol{\beta_0,\phi_0,\beta^Y_0},\kappa_0)]=\E\left[-\left(\boldsymbol{\phi_0'R_j^{(-t)'}R_j^{(-t)}\phi_0}\right)\right] \\
& G_{\beta^Y}=\E[\nabla_{\beta^Y}g(\boldsymbol{Z},\boldsymbol{\beta_0,\phi_0,\beta^Y_0},\kappa_0)]=\E\left[-\left(\boldsymbol{\phi_0'R_j^{(-t)'}X_j}\right)\right] \\
& G_{\phi}=\E[\nabla_{\phi}g(\boldsymbol{Z},\boldsymbol{\beta_0,\phi_0,\beta^Y_0},\kappa_0)]=\E\left[\left(\boldsymbol{Y_j'R_j^{(-t)}-2\kappa_0\phi_0'R_j^{(-t)'}R_j^{(-t)}}\right)\right]\\
& G_{\beta}=\E[\nabla_{\beta}g(\boldsymbol{Z},\boldsymbol{\beta_0,\phi_0,\beta^Y_0},\kappa_0)]=\E\left[-\left(\boldsymbol{Y_j'A-2\kappa_0\phi_0'R_j^{(-t)'}A}\right)\right] \\
& M_1=M_3=\E\left[-\left(\boldsymbol{\ddot{X}}_{j}'\boldsymbol{\ddot{X}}_{j}\right)\right] \ \ \ \ M_{2\phi}= \E\left[-\left(\boldsymbol{R_j^{(-t)'}R_j^{(-t)}}\right)\right] \\
&M_{2\beta}=\E[\nabla_{\beta}m_2(\boldsymbol{Z},\boldsymbol{\beta_0},\boldsymbol{\phi_0})]=\E\left[-\left(\boldsymbol{R_j^{(-t)'}X_j+\mathcal{X}_j^{(-t)}-\widetilde{\mathcal{X}}_j^{(-t)}-R_j^{(-t)'}A}\right)\right]
\end{align*} \medskip

where $\boldsymbol{Y}_{j}$ is a vector stacking $\overline{Y}_{jt}$, $R_{jt}=\overline{R}^{obs}_{jt}-\boldsymbol{\overline{X}_{jt}'\beta_0}$, $\boldsymbol{R}_{j}$ is a vector stacking all $R_{jt}$ for teacher $j$, $\boldsymbol{R^{(-t)}_{j}}$ is  a matrix stacking $T$ row vectors (of dimension $1\times(T-1)$) with each row (indexed by $t$) containing $T-1$ different $R_{jk}$ for $k\neq t$. $A$ is a $T\times K$ matrix such that each row consists of ($\phi_1\overline{X}_{jt-1}+\phi_2\overline{X}_{jt-2}+...$). $\boldsymbol{X_j^{(-t)}}$ is a $1\times(T-1)$ block matrix stacking $T\times K$ matrices of the covariates for teacher $j$ in all years except year $t$.  $\mathcal{X}_j^{(-t)}$ is a $T-1\times K$ matrix where each row consists of  $\boldsymbol{R}_j'$ multiplied by a submatrix of $\boldsymbol{X_j^{(-t)}}$. $\widetilde{\mathcal{X}}_j^{(-t)}$ is a $T-1\times K$ matrix where each row  $\boldsymbol{\phi_0'R_j^{(-t)'}}$ multiplied by a submatrix of $\boldsymbol{X_j^{(-t)}}$. 

Furthermore let:
\begin{align*}
&\psi_1(\boldsymbol{Z})=-M_1^{-1}m_1(\boldsymbol{Z},\boldsymbol{\beta_0}) \\
&\psi_2(\boldsymbol{Z})=-M_{2\phi}^{-1}m_2(\boldsymbol{Z},\boldsymbol{\beta_0},\boldsymbol{\phi_0}) \\
&\psi_3(\boldsymbol{Z})=-M_3^{-1}m_3(\boldsymbol{Z},\boldsymbol{\beta^Y_0}) \\
&g(\boldsymbol{Z})=g(\boldsymbol{Z},\boldsymbol{\beta_0,\phi_0,\beta_0^Y},\kappa_0).
\end{align*} \medskip

Then we have:

by Theorem\ref{lem:asym-norm} we have that:
\begin{align*}\sqrt{J}\left[(\boldsymbol{\widehat{\beta},\widehat{\phi},\widehat{\beta}^Y,}\widehat{\kappa})-(\boldsymbol{\beta_0,\phi_0,\beta^Y_0,}\kappa_0)\right] \rightsquigarrow N\left(0,\widetilde{G}^{-1}\E[\widetilde{g}(\boldsymbol{Z},\boldsymbol{\beta_0,\phi_0,\beta^Y_0,}\kappa_0)\widetilde{g}(\boldsymbol{Z},\boldsymbol{\beta_0,\phi_0,\beta^Y_0,}\kappa_0)']\widetilde{G}^{-1'}\right).
\end{align*}

Now note that the last row of $\widetilde{G}^{-1}$ is $G_{\kappa}^{-1}[1 \ \ G_{\beta^Y}M_3^{-1} \ \ -G_{\phi}M_{2\phi}^{-1} \ \ -G_{\beta}M_1^{-1}+G_{\phi}M_{2\phi}^{-1}M_{2\beta}M_1^{-1}]$. Multiplying that by $\widetilde{g}(\boldsymbol{Z},\boldsymbol{\beta_0,\phi_0,\beta^Y_0,}\kappa_0)$ we get:
\begin{align}
 \nonumber &G_{\kappa}^{-1}\left(g(\boldsymbol{Z})-G_{\beta^Y}M_3^{-1}m_3(\boldsymbol{Z},\boldsymbol{\beta^Y_0})-G_{\phi}M_{2\phi}^{-1}m_2(\boldsymbol{Z},\boldsymbol{\beta_0},\boldsymbol{\phi_0})-G_{\beta}M_1^{-1}m_1(\boldsymbol{Z},\boldsymbol{\beta_0})+G_{\phi}M_{2\phi}^{-1}M_{2\beta}M_1^{-1}m_1(\boldsymbol{Z},\boldsymbol{\beta_0})\right)\\
&=(G_{\kappa}^{-1})\left(g(\boldsymbol{Z})+G_{\beta^Y}\psi_3(\boldsymbol{Z})+G_{\phi}\psi_2(\boldsymbol{Z})+G_{\beta}\psi_1(\boldsymbol{Z})-G_{\phi}M_{2\phi}^{-1}M_{2\beta}\psi_1(\boldsymbol{Z})\right) \label{eq:Zob}.
\end{align}
\medskip

Now note that the asymptotic variance of $\widehat{\kappa}$ would be the lower right block of the joint variance matrix. Given that the quantity in  \eqref{eq:Zob} is a scalar, we can square it to obtain the lower right block of the joint variance matrix and then:\medskip
\begin{align}
\sqrt{J}(\widehat{\kappa}-\kappa_0)\rightsquigarrow N\left(0,\sigma^2\right)
\end{align}
where
$\sigma^2=(G_{\kappa}^{-1})^2\E\left(\left(g(\boldsymbol{Z})+G_{\beta^Y}\psi_3(\boldsymbol{Z})+G_{\phi}\psi_2(\boldsymbol{Z})+G_{\beta}\psi_1(\boldsymbol{Z})-G_{\phi}M_{2\phi}^{-1}M_{2\beta}\psi_1(\boldsymbol{Z})\right)^2\right).$

\textbf{Q.E.D}

\subsection{Proof of Lemma \ref{lem:asym-step1}}

Under Assumptions \ref{hyp:rand-assign}, \ref{hyp:tech-id}, and \ref{hyp:tech-opt}, the proof of Point 1 is similar to the proofs for Lemma \ref{prop:cons} and Theorem\ref{lem:asym-norm}. The proof of Point 2 is similar to the final part of the proof of Theorem \ref{lem:asym-norm}.

\subsection{Proof of Theorem \ref{prop:optimal}}

Let:

\begin{align}
&\E\left(m_1(\boldsymbol{Z},\boldsymbol{\beta_0})\right)=E\left(\boldsymbol{\ddot{X}}_{j}'\left(\boldsymbol{\ddot{R}}^{obs}_{j}- \boldsymbol{\ddot{X}_{j}\beta_0}\right)\right) \\&\E\left(m_3(\boldsymbol{Z},\boldsymbol{\beta^Y_0})\right)=\E\left(\boldsymbol{\ddot{X}}_{j}'\left(\boldsymbol{\ddot{Y}}^{obs}_{j}- \boldsymbol{\ddot{X}_{j}\beta_0^Y}\right)\right)\\
&\E\left(g_1(\boldsymbol{Z},\boldsymbol{\beta_0,\beta^Y_0},\kappa_0)\right)=\E\left(\boldsymbol{\overline{R}^{(-t)'}_{j}}\left(\boldsymbol{Y}_{j}-\boldsymbol{a}-\kappa_0\boldsymbol{R_j}\right)\right)
\end{align}

where $m_1(),m_3(),g_1()$ are functions. Let:

\begin{equation} \label{eq:system2}
\widetilde{g}_1(\boldsymbol{Z},\boldsymbol{\beta_0,\beta^Y_0},\kappa_0)=\left[m_1(\boldsymbol{Z},\boldsymbol{\beta_0})',m_3(\boldsymbol{Z},\boldsymbol{\beta^Y_0})',g_1(\boldsymbol{Z},\boldsymbol{\beta_0,\beta^Y_0},\kappa_0)'\right]'
\end{equation}

The proof to show that: $\sqrt{J}\left[(\boldsymbol{\widehat{\beta}^*,\widehat{\beta}^{Y*}},\widehat{\kappa}^*)-(\boldsymbol{\beta_0,\beta^Y_0,}\kappa_0)\right] \rightsquigarrow N\left(0,\Omega^*\right)$ is similar to the proof for Lemma \ref{prop:cons} and Theorem \ref{lem:asym-norm}. We need only that $\widehat{W}^* \overset{p}{\rightarrow} W^*$ and that $W^*$ be invertible, these conditions are guaranteed by  Lemma \ref{lem:asym-step1}, Point 1 of Assumption \ref{hyp:tech-id}, and Points 2 and 5 of Assumption \ref{hyp:tech-opt}. \medskip

$\Omega^* \leq \Omega_2$ by Theorem 3.2 of \citet{hansen1982large}.

\subsection{Proof of Result \ref{prop:over-id}}

The result follows from Lemmas 4.1 and 4.2 of \citet{hansen1982large}.

Note that the first two moment conditions are exactly identified. Then note that for the last moment condition we have one parameter, $\kappa_0$, and $T-1$ possible instruments such that the degrees of freedom of the $\chi^2$ distribution will be $T-1-1=T-2$. \medskip

\subsection{Proof of Result \ref{res:optimal-mom}}

We have established in Theorem \ref{prop:optimal} that the optimal GMM estimator has an asymptotic variance of $\Omega^*=\left(\widetilde{G}_1'W^*\widetilde{G}_1\right)^{-1}$ where, $W^*=\E[\widetilde{g}_1(\boldsymbol{Z},\boldsymbol{\beta_0,\beta^Y_0,}\kappa_0)\widetilde{g}(\boldsymbol{Z},\boldsymbol{\beta_0,\beta^Y_0,}\kappa_0)']^{-1}$ and $\widetilde{G}_1=\E\left[\triangledown_{(\boldsymbol{\beta,\beta^Y,}\kappa)}\widetilde{g}_1(\boldsymbol{Z},\boldsymbol{\beta_0,\beta^Y_0,}\kappa_0)\right]$. \medskip We can write our moment conditions as $\E[\boldsymbol{Q'u}]$ where $\boldsymbol{Q}$ is a block diagonal matrix with blocks $\boldsymbol{\ddot{X}_{j}}, \boldsymbol{\ddot{X}_{j}}, \boldsymbol{R^{(-t)}_{j}}$ and $\boldsymbol{u}=\begin{pmatrix} \boldsymbol{\ddot{\epsilon}_{j}} +\boldsymbol{\ddot{\mu}_{j}} \\ \boldsymbol{\ddot{\eta}_{j}} + \kappa_0\boldsymbol{\ddot{\mu}_{j}} \\  \boldsymbol{\eta_{j}} -\kappa_0 \boldsymbol{\epsilon_{j}} \end{pmatrix}$.

Note that under Assumption \ref{hyp:exog-cov} we have:

\begin{align*}
\widetilde{G}_1&=\begin{pmatrix}-\E(\boldsymbol{\ddot{X}_{j}'\ddot{X}_{j}}) & 0 &0 \\ 0& -\E(\boldsymbol{\ddot{X}_{j}'\ddot{X}_{j}}) &0 \\ -\E(\boldsymbol{B_j}+\kappa_0\boldsymbol{\overline{R}^{(-t)'}_{j}} \boldsymbol{X}_{j}) & -\E(\boldsymbol{\overline{R}^{(-t)'}_{j}}\boldsymbol{X}_{j}) & -\E(\boldsymbol{\overline{R}^{(-t)'}_{j}}\boldsymbol{R_{j}}) \end{pmatrix} \\&= \begin{pmatrix}-\E(\boldsymbol{\ddot{X}_{j}'\ddot{X}_{j}}) & 0 &0 \\ 0& -\E(\boldsymbol{\ddot{X}_{j}'\ddot{X}_{j}}) &0 \\ 0 & 0 & -\E(\boldsymbol{\overline{R}^{(-t)'}_{j}}\boldsymbol{R_{j}}) \end{pmatrix}
\end{align*}

where $\boldsymbol{B_j}$ is a matrix where each row consists of a column of $\boldsymbol{X_j^{(-t)}}$ multiplied by $\left(\boldsymbol{Y}_{j}- \kappa_0 \boldsymbol{R_{j}} \right)=(\boldsymbol{\eta}_j -\kappa_0\boldsymbol{\epsilon}_j)$. The second equality then follows from Assumption \ref{hyp:exog-cov} which makes all off-diagonal elements zero.

 Then we can write: $\widetilde{G}_1=-\E[\boldsymbol{Q'L}]=-\E[\boldsymbol{Q'}\E[\boldsymbol{L|Q}]]$ where $\boldsymbol{L}$ is a block diagonal matrix with blocks $\boldsymbol{\ddot{X}_{j}}, \boldsymbol{\ddot{X}_{j}}, \boldsymbol{R}_{j}$ and $W^*=\E[\boldsymbol{Q'uu'Q}]=\E[\boldsymbol{Q'\E[uu'|Q]Q}]$. Then the variance from the optimal GMM estimator will be:

\begin{equation*}
\Omega^*=\left(\E[\boldsymbol{Q'}\E[\boldsymbol{L|Q}]]'\E[\boldsymbol{Q'\E[uu'|Q]Q}]^{-1}\E[\boldsymbol{Q'}\E[\boldsymbol{L|Q}]]\right)^{-1}
\end{equation*}

and the optimal instrument following \citet{chamberlain1987asymptotic} are:

\begin{equation} \label{eq:optimal-inst}
\boldsymbol{Q}^{*}=\boldsymbol{\E[uu'|Q]}^{-1}\E[\boldsymbol{L|Q}]
\end{equation}

and yield an asymptotic variance of
\begin{equation*}
\Omega^*=\left(\E\left(\E[\boldsymbol{L|Q}]'\boldsymbol{\E[uu'|Q]}^{-1}\E[\boldsymbol{L|Q}]\right)\right)^{-1}
\end{equation*}

where the moment conditions will then be:

\begin{equation} \label{eq:optimal-moments}
\E[\boldsymbol{(Q^*)'u}]=\E[(\boldsymbol{\E[uu'|Q]}^{-1}\E[\boldsymbol{L|Q}])'\boldsymbol{u}]
\end{equation}

where:

\begin{equation*}
\E[\boldsymbol{L|Q}]=\begin{pmatrix} \boldsymbol{\ddot{X}_{j}} &0 & 0 \\ 0 & \boldsymbol{\ddot{X}_{j}} & 0\\ 0 & 0 & \E[\boldsymbol{R}_{j}|\boldsymbol{Q}]\end{pmatrix}
\end{equation*}

and note that $\E[\boldsymbol{R}_{j}|\boldsymbol{Q}]=\E[\boldsymbol{\mu}_{j}|\boldsymbol{R^{(-t)}_{j}}]+\E[\boldsymbol{\epsilon}_{j}|\boldsymbol{\ddot{X}_{j}}]$ because  $\boldsymbol{X}_{j}$ is independent of $\boldsymbol{\mu_j}$ by Assumption \ref{hyp:exog-cov}, and $\boldsymbol{\epsilon}_{j}$ is independent of $\boldsymbol{R^{(-t)}_{j}}$ by point 2 of Assumption \ref{hyp:exog-cov} and by point 3 of Assumption \ref{hyp:rand-assign}. Finally note that $\E[\boldsymbol{\epsilon}_{j}|\boldsymbol{\ddot{X}_{j}}]=0$ by point 3 of Assumption  \ref{hyp:exog-cov}. \medskip

\subsection{Proof of Result \ref{lem:as-outcome}}

Starting with $\E\left(\boldsymbol{\ddot{X}}_{j}'\left(\boldsymbol{\ddot{R}}^{obs}_{j}- \boldsymbol{\ddot{X}_{j}\beta_0}\right)\right)=0$: \medskip

\begin{align*}
&\E\left(\boldsymbol{\ddot{X}}_{j}'\left(\boldsymbol{\ddot{R}}^{obs}_{j}- \boldsymbol{\ddot{X}_{j}\beta_0}\right)\right)  \\
=&\E\left(\boldsymbol{\ddot{X}}_{j}'\left(\boldsymbol{\ddot{\mu}_j} + \boldsymbol{\ddot{\epsilon}_j}\right)\right)   \\
=&0
\end{align*}

where the first equality follows from  \eqref{eq:scores}, and the second equality follows from Point 3 of Assumption \ref{hyp:tech-id}. Then by Point 1 of Assumption \ref{hyp:tech-id}, we have that $\E(\boldsymbol{\ddot{X}}_{j}'\boldsymbol{\ddot{X}}_{j})$ is invertible, therefore $\boldsymbol{\beta_0}$ is uniquely identified. \medskip

Now moving to $\E\left(\boldsymbol{D_{j}}'\left(\boldsymbol{R}_j-\boldsymbol{D_{j}\alpha_0}\right)\right)=0$:
\begin{align*}
&\E\left(\boldsymbol{D_{j}}'\left(\boldsymbol{R}_j-\boldsymbol{D_{j}\alpha_0}\right)\right)  \\
=&\E\left(\boldsymbol{D_{j}}'\left(\boldsymbol{\mu_j} + \boldsymbol{\epsilon_j}-\boldsymbol{D_{j}\alpha_0}\right)\right)   \\
=&\E\left(\boldsymbol{D_{j}}'\left(\boldsymbol{\zeta_j} + \boldsymbol{\epsilon_j}\right)\right)\\
=&0
\end{align*}

where the first equality follows from  \eqref{eq:scores}, the second equality follows from  \eqref{eq:VA-outcome}. The third equality follows from the fact that $\zeta_{jt}$ is orthogonal to $D_{jt}$ by construction and by Point 2 of Assumption \ref{hyp:as-outcome}. Then by Point 1 of Assumption \ref{hyp:as-outcome}, we have that $\E(\boldsymbol{D}_{j}'\boldsymbol{D}_{j})$ is invertible, therefore $\boldsymbol{\alpha_0}$ is uniquely identified.

\subsection{Proof of Theorem \ref{lem:as-outcome-dist}}

Note that $\widehat{W}=W=I$ and consistency of the GMM estimators follows using a similar proof to Lemma \ref{prop:cons}.

Now to show asymptotic normality we start by rewriting the moments as:
\begin{align}
&\E\left(m_1(\boldsymbol{Z},\boldsymbol{\beta})\right)=E\left(\boldsymbol{\ddot{X}}_{j}'\left(\boldsymbol{\ddot{R}}^{obs}_{j}- \boldsymbol{\ddot{X}_{j}\beta}\right)\right) \\
&\E\left(m_4(\boldsymbol{Z},\boldsymbol{\beta},\boldsymbol{\alpha})\right)=\E\left(\boldsymbol{D_{j}}'\left(\boldsymbol{\widetilde{R}_j}-\boldsymbol{D_{j}\alpha} \right)\right)
\end{align}

where $\boldsymbol{Z_j}=\left(\boldsymbol{X_j,R_j^{obs},D_j}\right)$, $\boldsymbol{Z}_{2}$ stacks the $\boldsymbol{Z}_j$, and $m_1(),m_4()$ are functions. Let
\begin{equation} \label{eq:system3}
\widetilde{g}_2(\boldsymbol{Z}_{2},\boldsymbol{\beta},\boldsymbol{\alpha})=\left[m_1(\boldsymbol{Z}_{2},\boldsymbol{\beta})',m_4(\boldsymbol{Z}_{2},\boldsymbol{\beta},\boldsymbol{\alpha})'\right]'
\end{equation}

be a vector stacking the two functions. \medskip

A GMM estimator is asymptotically normal if it's consistent and conditions (i)-(v) of  Theorem 3.4 of \citet{newey1994large} are satisfied. \medskip

Condition (i) holds under Point 1 of Assumption \ref{hyp:as-outcome-tech}. Condition (ii) holds by inspection.  The first part of condition (iii) holds by Result \ref{lem:as-outcome} and the second part:
second part note that:
\begin{align*}
\E[||\widetilde{g}_{2}(\boldsymbol{Z}_{2},\boldsymbol{\beta_0},\boldsymbol{\alpha_0})||^2]=\E\left(trace\left(\widetilde{g}_{2}(\boldsymbol{Z}_{2},\boldsymbol{\beta_0},\boldsymbol{\alpha_0})\widetilde{g}_{2}(\boldsymbol{Z}_{2},\boldsymbol{\beta_0},\boldsymbol{\alpha_0})'\right)\right).
\end{align*}

Therefore for $\E[||\widetilde{g}_{2}(\boldsymbol{Z}_{2},\boldsymbol{\beta_0},\boldsymbol{\alpha_0})||^2]$ to be finite we need that:
\begin{align*}
&\E\left(\boldsymbol{\ddot{X}}_{j}'\left(\boldsymbol{\ddot{R}}^{obs}_{j}- \boldsymbol{\ddot{X}_{j}\beta_0}\right)\left(\boldsymbol{\ddot{R}}^{obs}_{j}- \boldsymbol{\ddot{X}_{j}\beta_0}\right)'\left(\boldsymbol{\ddot{R}}^{obs}_{j}- \boldsymbol{\ddot{X}_{j}\beta_0}\right)\right) < \infty \\
&\E\left(\boldsymbol{D_{j}}'\left(\boldsymbol{R}_j-\boldsymbol{D_{j}\alpha_0} \right)\left(\boldsymbol{R}_j-\boldsymbol{D_{j}\alpha_0} \right)'\boldsymbol{D_{j}}\right)  < \infty
\end{align*}

which holds by Point 4 of Assumption  \ref{hyp:as-outcome-tech}. \medskip

For condition (iv) we need to show that:
$\E\left[\underset{(\boldsymbol{\beta,\alpha})\in \mathcal{N}}{sup}||\triangledown\widetilde{g}_2(\boldsymbol{Z}_{2},\boldsymbol{\beta},\boldsymbol{\alpha})|| \right] < \infty$ for a neighborhood $\mathcal{N}$ around $(\boldsymbol{\beta_0,\alpha_0})$. \medskip

\begin{align*}
\E\left[\underset{(\boldsymbol{\beta,\alpha})\in \mathcal{N}}{sup}||\triangledown\widetilde{g}_2(\boldsymbol{Z}_{2},\boldsymbol{\beta},\boldsymbol{\alpha})||\right]&=\E\left(\underset{(\boldsymbol{\beta,\alpha})\in \mathcal{N}}{sup}\sqrt{trace\left(\triangledown\widetilde{g}_2(\boldsymbol{Z}_{2},\boldsymbol{\beta},\boldsymbol{\alpha})\triangledown\widetilde{g}_2(\boldsymbol{Z}_{2},\boldsymbol{\beta},\boldsymbol{\alpha})'\right)}\right)
\end{align*}

Since by the triangle inequality applied to the Euclidean norm, a sufficient condition for the quantity above to be finite, is that the sum of absolute values of the diagonal elements be finite.  So we need that:

\begin{align*}
&\E\left|\left(\boldsymbol{\ddot{X}}_{j}'\boldsymbol{\ddot{X}}_{j}\right)\right| <\infty \\
&\E\left|\left(\boldsymbol{D}_{j}'\boldsymbol{D}_{j}\right)\right| <\infty.
\end{align*}

these conditions are guaranteed by Point 1  of Assumption  \ref{hyp:as-outcome-tech} and Point 1 of Assumption \ref{hyp:tech-id}. \medskip

Condition (v) is satisfied under Point 1 of Assumption \ref{hyp:tech-id} and Point 1 Assumption \ref{hyp:as-outcome}.\medskip

\subsection{Proof of Theorem \ref{prop:alpha-dist}}

The asymptotic variance of $\boldsymbol{\alpha}$ can be obtained from partitioned matrix inversion. Let:

\begin{align*}
\widetilde{G}_{\alpha}&=-\E\left(\boldsymbol{D}_{j}'\boldsymbol{D}_{j}\right) \\
\widetilde{G}_{\beta}&= -\E\left(\boldsymbol{D}_{j}'\boldsymbol{X}_{j}\right) \\
\widetilde{M}_{1}&=-\E\left(\boldsymbol{\ddot{X}}_{j}'\boldsymbol{\ddot{X}}_{j}\right)
\end{align*}

Now note that the second row of $\widetilde{G}_2^{-1}$ is $[-\widetilde{G}_{\alpha}^{-1}\widetilde{G}_{\beta}\widetilde{M}_{1}^{-1},\widetilde{G}_{\alpha}^{-1}]$ which can be written as $\widetilde{G}_{\alpha}^{-1}[-\widetilde{G}_{\beta}\widetilde{M}_{1}^{-1},I]$. Multiplying that by $\widetilde{g}_{2}(\boldsymbol{Z}_{2},\boldsymbol{\beta_0},\boldsymbol{\alpha_0})$ we get:

\begin{align*}
&\widetilde{G}_{\alpha}^{-1}\left(\boldsymbol{D_{j}}'\left(\boldsymbol{R}_j-\boldsymbol{D_{j}\alpha_0}\right)-\widetilde{G}_{\beta}\widetilde{M}_{1}^{-1}\boldsymbol{\ddot{X}}_{j}'\left(\boldsymbol{\ddot{R}}^{obs}_{j}- \boldsymbol{\ddot{X}_{j}\beta_0}\right)\right) \\
=&-\E\left(\boldsymbol{D}_{j}'\boldsymbol{D}_{j}\right)^{-1}\left(\boldsymbol{D_{j}}'\left(\boldsymbol{R}_j-\boldsymbol{D_{j}\alpha_0}\right)-\E\left(\boldsymbol{D_j'X_j}\right)\E\left(\boldsymbol{\ddot{X}}_{j}'\boldsymbol{\ddot{X}}_{j}\right)^{-1}\boldsymbol{\ddot{X}}_{j}'\left(\boldsymbol{\ddot{R}}^{obs}_{j}- \boldsymbol{\ddot{X}_{j}\beta_0}\right)\right)
\end{align*}

Taking the quadratic of the above with \\ $\Gamma=\left(\boldsymbol{D_{j}}'\left(\boldsymbol{R}_j-\boldsymbol{D_{j}\alpha_0}\right)-\E\left(\boldsymbol{D_j'X_j}\right)\E\left(\boldsymbol{\ddot{X}}_{j}'\boldsymbol{\ddot{X}}_{j}\right)^{-1}\boldsymbol{\ddot{X}}_{j}'\left(\boldsymbol{\ddot{R}}^{obs}_{j}- \boldsymbol{\ddot{X}_{j}\beta_0}\right)\right)$ yields:

\begin{equation*}
V_1=\E(\boldsymbol{D}_{j}'\boldsymbol{D}_{j})^{-1}\E\left(\Gamma\Gamma'\right)\E(\boldsymbol{D}_{j}'\boldsymbol{D}_{j})^{-1'}
\end{equation*}

\medskip

\subsection{Algebra for Corollary \ref{cor:no-OLS}}
\begin{align*}
G_{\phi}&=\E\left[\left(\boldsymbol{Y_j'R_j^{(-t)}-2\kappa_0\phi_0'R_j^{(-t)'}R_j^{(-t)}}\right)\right]\\
        &=\E\left[\left(\boldsymbol{Y_j'-2\kappa_0\phi_0'R_j^{(-t)'}}\right)R_j^{(-t)}\right]\\
        &=\E\left[\left(\boldsymbol{\eta_j+\kappa_0 \theta_j -\kappa_0 \epsilon_j -\kappa_0\phi_0'R_j^{(-t)'}}\right)\boldsymbol{R_j^{(-t)}}\right]\\
        &=-\kappa_0\boldsymbol{\phi_0'}\E\left(\boldsymbol{R_j^{(-t)'}R_j^{(-t)}}\right)
\end{align*}

\subsection{Proof of Overidentification \label{section:proofid}}
To see that note that \eqref{eq:mom4} is equivalent to:

\begin{align*}
&\boldsymbol{\phi_0'}\E\left(\boldsymbol{\overline{R}^{(-t)'}_{j}}\left(\boldsymbol{Y}_{j}- \kappa_0 \boldsymbol{R^{(-t)}_{j}\phi_0} \right)\right)\\
=&\boldsymbol{\phi_0'}\E\left(\boldsymbol{\overline{R}^{(-t)'}_{j}}\left(\boldsymbol{Y}_{j}- \kappa_0\boldsymbol{R_{j}+\kappa_0\theta_{j}}  \right)\right)\\
=&\boldsymbol{\phi_0'}\E\left(\boldsymbol{\overline{R}^{(-t)'}_{j}}\left(\boldsymbol{Y}_{j}- \kappa_0\boldsymbol{R_{j}} \right)\right)=0\\
\end{align*}

where the first equality follows from \eqref{eq:decomp} and the second equality follows from the fact that $\boldsymbol{R^{(-t)}_{j}}$ and $\boldsymbol{\theta_{j}}$ are uncorrelated by construction. Notably, \eqref{eq:mom4} only requires that a linear combination of $\E\left(\boldsymbol{\overline{R}^{(-t)'}_{j}}\left(\boldsymbol{Y}_{j}- \kappa_0\boldsymbol{R_{j}}\right)\right)$   be equal to zero. However, under Assumptions \ref{hyp:rand-assign} and \ref{hyp:tech-id} we have the stronger conditions:

\begin{align*}
&\E\left(\boldsymbol{\overline{R}^{(-t)'}_{j}}\left(\boldsymbol{Y}_{j}-\kappa_0\boldsymbol{R}_{j}\right)\right) \\
=&\E\left(\boldsymbol{\overline{R}^{(-t)'}_{j}}\boldsymbol{\eta_j}- \kappa_0\boldsymbol{\overline{R}^{(-t)'}_{j}}\boldsymbol{\epsilon_j}\right) \\
=&0
\end{align*}

where the first equality follows from \eqref{eq:k-mom} and \eqref{eq:refover}. The second equality follows from the fact that:

\begin{align*}
&\E\left(\boldsymbol{R^{(-t)}_{j}}\boldsymbol{{\eta}_{j}}\right) \\
=&\E\left((\boldsymbol{\mu^{(-t)}_{j}} + \boldsymbol{\epsilon^{(-t)}_{j}})\boldsymbol{{\eta}_{j}}\right) \\
=&\E\left(\boldsymbol{\mu^{(-t)}_{j}} \boldsymbol{{\eta}_{j}}\right) + \E\left(\boldsymbol{\epsilon^{(-t)}_{j}} \boldsymbol{{\eta}_{j}}\right) \\
=&0
\end{align*}

where analogously  to $\boldsymbol{R^{(-t)}_{j}}$, each row in $\boldsymbol{\mu^{(-t)}_{j}}$  stacks $\mu_{js}$ for each $t$ with $s \neq t$ and each row in $\boldsymbol{\epsilon^{(-t)}_{j}}$  stacks $\overline{\epsilon}_{js}$ for each $t$ with $s \neq t$. The equalities then follow from  $\E\left(\mu_{js}\overline{\eta}_{jt}\right)=0$ by Assumption \ref{hyp:rand-assign} and  $\E\left(\overline{\epsilon}_{js}\overline{\eta}_{jt}\right)=0$ by Assumption \ref{hyp:rand-assign}. A similar argument shows $\E\left(\boldsymbol{\overline{R}^{(-t)'}_{j}}\boldsymbol{\epsilon_j}\right)$.\medskip
\end{document}